\documentclass[preprint]{aastex631}
\usepackage{graphicx}
\usepackage{amsmath,amssymb}
\usepackage{units}
\usepackage{color}
\newcommand{\del}{\partial}
\newcommand{\m}{\mathbf}
\newcommand{\THE}{{\boldsymbol{\theta}}}
\newcommand{\ALP}{{\boldsymbol{\alpha}}}

\newcommand{\K}{{\boldsymbol{k}}}
\newcommand{\x}{{\boldsymbol{x}}}
\newcommand{\y}{{\boldsymbol{y}}}
\newcommand{\z}{{\boldsymbol{z}}}
\newcommand{\n}{{\boldsymbol{n}}}
\newcommand{\R}{{\boldsymbol{r}}}
\newcommand{\U}{\underline}
\newcommand{\f}{\frac}
\newcommand{\T}{\tilde}
\newcommand{\N}{\nonumber}
\newcommand{\BF}{\begin{figure}\begin{center}}
\newcommand{\EF}{\end{center}\end{figure}}
\newcommand{\BE}{\begin{equation}}
\newcommand{\EE}{\end{equation}}
\newcommand{\BEA}{\begin{eqnarray}}
\newcommand{\EEA}{\end{eqnarray}}
\newcommand{\ti}{\textit}
\newcommand{\tr}{\textrm}

\def\v#1{\boldsymbol #1}
\newcommand{\ms}{M_{\odot}}

\shorttitle{ALMA Measurement of Lensing Power Spectra towards MG J0414+0534}
\shortauthors{Inoue, Minezaki, Matsushita, and Nakanishi}
\begin{document}


\title{ALMA Measurement of 10\,kpc-scale Lensing Power Spectra towards the Lensed Quasar MG J0414+0534}

\author{Kaiki Taro Inoue}
\affiliation{Faculty of Science and Engineering, 
Kindai University, Higashi-Osaka, 577-8502, Japan}

\author{Takeo Minezaki}
\affiliation{Institute of Astronomy, School of Science, University of
Tokyo, Mitaka, Tokyo 181-0015, Japan}

\author{Satoki Matsushita}
\affiliation{Institute of Astronomy and Astrophysics, Academia Sinica, \\
11F of Astronomy-Mathematics Building, AS/NTU, No.1, Sec.4, Roosevelt Rd., Taipei 10617, Taiwan, R.O.C.}

\author{Kouichiro Nakanishi}
\affiliation{National Astronomical Observatory of Japan, Mitaka, Tokyo 181-8588, Japan, \\
The Graduate University for Advanced Studies, SOKENDAI, Mitaka, Tokyo 181-8588, Japan}
\begin{abstract}
The lensing power spectra for gravitational potential, astrometric shift, and convergence perturbations are powerful probes to investigate dark matter structures on small scales. We report the first lower and upper bounds of these lensing power spectra on angular scale $\sim 1''$ towards the anomalous quadruply lensed quasar MG$\,$J0414+0534 at a redshift $z=2.639$. To obtain the spectra, we conducted observations of MG$\,$J0414+0534 using the Atacama Large Millimeter/submillimeter Array (ALMA) with high angular resolution (0\farcs 02-0\farcs 05). We developed a new partially non-parametric method in which Fourier coefficients of potential perturbation are adjusted to minimize the difference between linear combinations of weighted mean de-lensed images. Using positions of radio jet components, extended dust emission on scales $>1\,$kpc, and mid-infrared flux ratios, the range of measured convergence, astrometric shift, and potential powers at an angular scale of $\sim 1\farcs 1$ (corresponding to an angular wave number of $l=1.2\times 10^6$ or $\sim 9\,$kpc in the primary lens plane) within $1\,\sigma$ are $\varDelta_\kappa=0.021-0.028$, $\varDelta_\alpha =7-9\,$mas, and $\varDelta_\psi=1.2-1.6\,$$\textrm{mas}^2$, respectively. Our result is consistent with the predicted abundance of halos in the line of sight and subhalos in cold dark matter models. Our partially non-parametric lens models suggest a presence of a clump in the vicinity of object Y, a possible dusty dwarf galaxy and some small clumps in the vicinity of other lensed quadruple images. Although much fainter than the previous report, we detected weak continuum emission possibly from object Y with a peak flux of $\sim 100\,\mu \textrm{Jy}\, \textrm{beam} ^{-1}$ at the $\sim 4\,\sigma$ level.   
\end{abstract}


\section{Introduction}
\label{sec:1}
The cold dark matter (CDM) model has been successful in explaining structures on scales $>1\,$Mpc. However, on scales $<1\,$Mpc, discrepancies between theory and observation remain. In particular, the observed number of dwarf galaxies inside a Milky Way (MW)-sized galaxy is far less than the theoretically predicted number of subhalos that would host dwarf galaxies\citep{kauffmann1993,klypin1999,moore1999}. Recent hydrodynamical simulations with baryonic feedback and reionization \citep{wetzel2016, brooks2017, fielder2019} indicate that MW no longer has the 'missing satellite problem' if the detection efficiency of sky survey is taken into account \citep{kim2018}. However, it remains uncertain whether this resolution applies beyond our MW \citep{nashimoto2022}. Moreover, directly counting the number of dwarfs cannot trace completely dark halos with a mass below $\sim 10^8 \ms$. To overcome this limitation, gravitational lensing offers a powerful approach to directly probe such low-mass dark halos residing in far universe.

It has been known that some quadruply lensed quasars show anomalies in the flux ratios of lensed images. Although the relative positions of lensed images can be fitted with a smooth gravitational potential on angular scales of a few arcseconds, the flux ratios deviate from the prediction by typically $10-40$ percent. Some theoretical works claimed that such anomalies in the flux ratios can be caused by dwarf galaxy-sized subhalos residing in a lensing galaxy halo \citep{mao1998,metcalf2001,chiba2002,dalal-kochanek2002,keeton2003,inoue-chiba2003,xu2009,xu2010}. Radio observations \citep{kochanek2004, metcalf2004, mckean2007, more2009}, mid-infrared (MIR) observations \citep{chiba2005,sugai2007, minezaki2009,macleod2013}, and near-infrared observations \citep{fadely2012} supported the claim\footnote{Anomalies in flux ratios can be caused by microlensing due to stars in the lensing galaxy\citep{zimmer2011}. However, the effects of microlensing is negligible in MIR or radio observations due to the large size of source.}.
However, the scenario is not so simple. Any small-mass halos in sight lines of lensed images can also change the flux ratios and the relative positions of lensed images \citep{metcalf2005a, xu2012}. In CDM models, it has been argued that the major cause of anomalies in flux ratios is small-mass halos residing in the intergalactic space rather than subhalos \citep{inoue-takahashi2012, takahashi-inoue2014, inoue-minezaki2016, ritondale2019} though some pointed out that the anomalies might be explained by a complex potential of the primary lens \citep{evans2003, oguri2005, gilman2017, hsueh2017, hsueh2018}. 

Based on $N$-body simulations, \citet{inoue2016} pointed out that the total contribution of line-of-sight (LOS) structures in changing the flux ratios of lensed images for lens source redshift $z_\textrm{s}>1$ amounts to 60 to 80 percent. A subsequent analysis obtained a similar result \citep{despali2018}. Frameworks for modelling 'LOS lensing' have been explored in the literature \citep{erdl1993, rennan1996, mccully2014, mccully2017, birrer2017, fleury2021}. LOS lensing has been used to constrain warm, mixed, and other dark matter models \citep{inoue-etal2015, kamada2016, kamada2017, gilman2018, gilman2020, hsueh2020, enzi2021}.

\begin{figure}
\epsscale{0.6}
\plotone{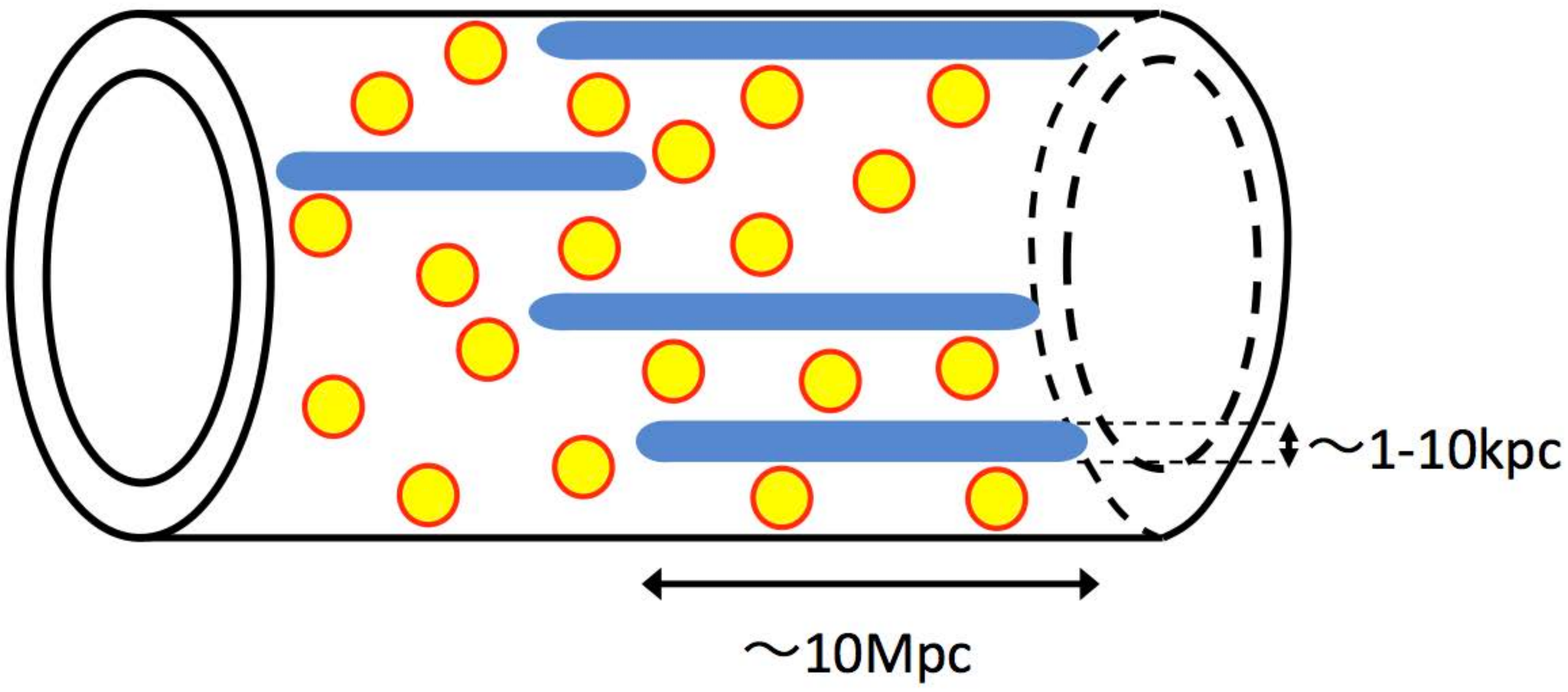}
\vspace{-0.8cm}
\caption{Schematic picture of halos (yellow disks) and troughs (blue bars) in sight lines. The cylinder represents bundles of light rays that pass the vicinity of an Einstein ring in the lens plane.  }
\label{fig:trough.pdf}
\end{figure}

Numerical simulations suggest that potential fluctuations due to LOS structures include positive and negative perturbations: they consist of small-mass halos and voids aligned in sight lines. The density contrast of voids with a radius of $\sim 10\,$Mpc at the present time is $\sim -1$. Thus the mass deficient in a void at $z\sim 1$ is equal to the cosmological matter density $\sim 10^{11}\,\ms \,\textrm{Mpc}^{-3}$. \citet{takahashi-inoue2014} demonstrated that 1) the amplitude of typical convergence perturbation due to LOS structures for an anomalous lens system with a source redshift $z_\textrm{s} \sim 3$ and a lens redshift $z_\textrm{l} \sim 1$ is $\delta \kappa =10^8\,\ms \textrm{arcsec}^{-2} \sim 10^6\,\ms \textrm{kpc}^{-2}$ and 2) the typical amplitudes of negative mass components on a scale of the Einstein radius of the primary lens are approximately equal to those of positive mass counterparts. Therefore, the total length of the aligned voids or a \textit{trough} \citep{gruen2016} is expected to be $\sim 10\,$Mpc, which is equal to the comoving radius of a single void at the present time. Note that a trough structure may consist of a number of small troughs separated in a sight line. In other words, a negative projected density region that corresponds to a trough can be interpreted as sight lines that do not \textit{scatter} with small halos in the intergalactic space (Figure \ref{fig:trough.pdf}).  

To \ti{directly} measure potential fluctuations due to LOS structures and subhalos, it is necessary to measure relative astrometric shifts of lensed extended images \citep{treu-koopmans2004, koopmans2005, vegetti2009, vegetti2010, chantry2010, vegetti2014}. Typical astrometric shifts due to LOS structures are of the order of a few milli-arcseconds \citep{takahashi-inoue2014} but they can be significantly enhanced by the strong lensing effect \citep{inoue2005b}. Indeed, such anomalous astrometric shifts corresponding to a small-mass clump have been observed in a near-infrared (NIR) image of the lensed quasar B1938+666 with a source redshift $z_\textrm{s}=0.881$ \citep{vegetti2012}. A similar result has been obtained for submillimeter continuum and line images of the lensed submillimeter galaxy SDP.81 with a source redshift $z_\textrm{s}=3.042$ \citep{inoue-minezaki2016, hezaveh2016a}.

The origin of the detected mass clumps is not known. Our theoretical analysis based on $N$-body simulations suggests that they probably reside in the intergalactic space rather than in the primary lensing galaxy \citep{inoue2016}, particularly for SDP.81 with a relatively high source redshift. However, determining the distance to the clumps is difficult because of lack of brightness of the host galaxy. Therefore, constraining CDM models using such results without assuming the nature of the clumps is also difficult. In order to constrain CDM models on scales of $\lesssim 10\,$kpc, we need to measure potential fluctuations due to LOS structures that can distinguish CDM models from other dark matter models. 

So far, in large weak lensing surveys, the convergence power spectrum has been measured on scales of $>1\,$Mpc. Although the direct measurement of weak lensing effects on scales of $<1\,$Mpc is a difficult task, the enhancement of weak lensing effects via the strong lensing effects enables us to measure the convergence power spectrum \citep{hezaveh2016b, chatterjee2018, bayer2018, cagan2020, bayer2023} on scales of $<1\,$Mpc with currently available telescopes such as ALMA. 

In this paper, we develop a new formalism based on source plane $\chi^2$ evaluation to measure lensing power spectra for potential, astrometric shift, and convergence (see Appendix A for definition). Subsequently, we conduct a mock analysis and apply the formalism to the submillimeter data of the lensed quasar MG\,J0414+0534, which were observed using ALMA at 340\, GHz \citep{inoue2017,inoue2020}. It has an anomaly in the flux ratio in the MIR band and the positions of jets in the low frequency radio band are measured very accurately. Combining these multi-wavelength data, we would be able to measure the lensing power spectra with unprecedented accuracy. In Section 2, we briefly review the previous observations and lens models of MG\,J0414+0534. In Section 3, we briefly describe our ALMA observations of MG\,J0414+0534. In Section 4, we explain our new formalism. In Section 5, we present our results on mock simulations. In Section 6, we provide our results on the reconstructed perturbations, the source intensity, power spectra, test with visibility fitting, and object Y obtained from our ALMA observations. In Section 7, we discuss about the consistency with CDM models. In Section 8, we conclude and discuss the robustness of our formalism. 

In what follows, we adopt a \ti{Planck} 2018 cosmology with matter density of $\Omega_{m,0}=0.315$, energy density of cosmological constant $\Omega_{\Lambda,0}=0.685$, and Hubble constant
$H_0=67.4$\,km/s \citep{planck2018}.

\section{Review of MG\,J0414+0534}
\label{sec:2}
MG\,J0414+0534 \citep{hewitt1992} is a quadruply 
lensed radio-loud quasar with an anomaly in the flux ratios. 
As shown in Figure \ref{fig:ALMA-con-cy24}, it has four lensed quasar core images: A1, A2, B, and C. The quasar at redshift $z_\ti{S}=2.639$ \citep{lawrence1995} is lensed by an elliptical galaxy G at redshift $z_{\ti{L}}=0.9584$ \citep{tonry1999} and object X, which may be a companion galaxy less massive than G \citep{schechter1993}. The lensed images of the quasar, galaxy G and object X were observed using the Hubble Space Telescope (HST) WFPC2/PC1 in the near infrared (NIR)/ optical (OPT) band \citep{falco1997}. The accuracy of relative positions of the lensed quadruple images is $3\,$mas. Very Long Baseline Array (VLBA) observations at 5\,GHz \citep{trotter2000} and 8.4\,GHz \citep{ros2000} resolved small-scale ($\lesssim 1\,$kpc) radio jet components p, q, r, and s. The flux ratio of image A2 to image A1 (A2/A1) indicated an anomaly in the mid-infrared (MIR) band \citep{minezaki2009, macleod2013}. The MIR flux ratios suggest the presence of a small-mass dark clump near the secondary brightest lensed image A2 \citep{macleod2013}. 
 
In order to model the gravitational potential of the lensing objects in MG\,J0414+0534, \citet{trotter2000} used a Taylor expansion for the potential with $m=3$ and $m=4$ multipole moments of the mass that is exterior and interior to the Einstein ring radius. However, the best-fitted model cannot explain the observed anomaly in the MIR flux ratio A2/A1. Since the radial size of the lensed jet components are significantly smaller than the tangential size, constraining the radial profile of the potential perturbation is difficult. 

To measure the anomaly in the flux ratio, \citet{minezaki2009} used a smooth potential with a singular isothermal ellipsoid (SIE) for G, a singular isothermal sphere (SIS) for X, an external shear (ES) for clusters and other large scale structures. \citet{macleod2013} added an SIS to explain the VLBA positions of jet components, and the MIR flux ratios. Although the model explained the VLBA positions and MIR flux ratios, the assumed mass profiles are \textit{ad hoc} and possible gravitational perturbations from multiple objects were not considered. More realistic models for multiple dark objects with an arbitrary mass profile are required.

To probe the origin, as part of ALMA Cycle 2 program, we performed observations of MG\,J0414+0534(Project ID: 2013.1.01110.S, PI: K.T. Inoue). Using our ALMA Cycle 2 data, we discovered a faint continuum emission in the vicinity of image A2. Assuming that the emission is coming from object Y, a possible dusty dwarf galaxy, we can explain the anomaly in the flux ratios and the differential dust extinction observed in optical to NIR bands \citep{inoue2017}. However, \citet{stacey2018} pointed out that the faint continuum signal disappeared after self-calibration of visibilities and argued that the identification was spurious (we will discuss about the robustness of the faint emission in Section \ref{sec:6.5}). As part of ALMA Cycle 4 program, we performed high-resolution observations of MG\,J0414+0534(Project ID: 2016.1.00281.S, PI: S. Matsushita). To obtain more realistic models, we combined our Cycle 2 and Cycle 4 data of MG\,J0414+0534. Then we performed imaging of the lensed extended source with high resolution ($0\farcs 02-0\farcs 05$). Because of the complexity of the source intensity in the submillimeter band, we were able to use more complex models to fit simultaneously the ALMA data, the VLBA positions, and MIR flux ratios. Using our ALMA Cycle 2 and Cycle 4 data, we discovered a possible interaction between the quasar jets and interstellar medium \citep{inoue2020}.

\section{ALMA Observations}
\label{sec:3}
Our Cycle 2 and Cycle 4 observations of MG\,J0414+0534 were performed on June 13 and August 14, 2015 and on November 1, 8, 10, and 11, 2017, respectively. For the Cycle 2 and Cycle 4 observations, the maximum and minimum baselines were 1.574\,km and 15\,m, and 13.894\,km and 113\,m, and the angular resolutions and maximum recoverable scales were $\sim 0\farcs 2$ and $2''$, and $\sim 0\farcs 02$ and $\sim 0\farcs 4$, and the rms noise were $\sim 20\,\mu$Jy/beam and $\sim 30\,\mu$Jy/beam, respectively (for details, refer to \citet{inoue2017,inoue2020}).  

\begin{figure}
\epsscale{0.5}
\hspace{-0.3cm}
\plotone{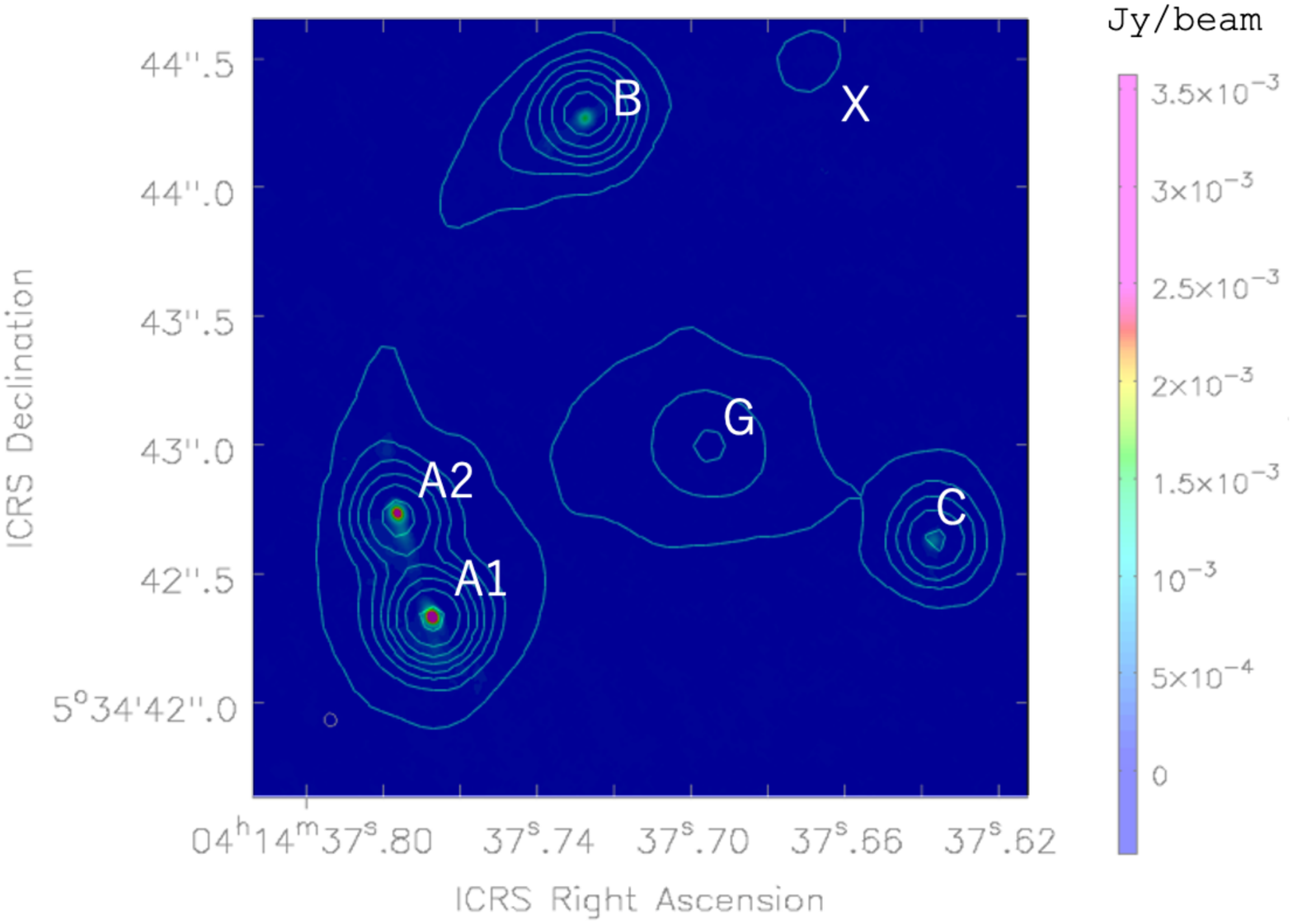}
\vspace{-0cm}
\caption{ALMA 0.88\,mm (Band 7 340\,GHz) continuum image of MG\,J0414+0434 overlaid with the HST near-infrared/optical (F814W+F675W) image (in contours)\citep{falco1999}. G is the primary lensing galaxy and X is an 'object X', a possible companion galaxy. The ALMA image was obtained from our ALMA Cycle 2 (project ID: 2013.1.01110.S, PI: K.T. Inoue) and Cycle 4 (project ID: 2016.1.00281.S, PI: S. Matsushita) observations. The imaging was conducted with a Briggs weighting of $robust=0.5$.  The synthesized beam size is ${0\farcs 051\times 0\farcs 047}$ and the PA is $25\fdg 7$. The rms background noise is $23\,\mu \textrm{Jy}\,\textrm{beam}^{-1}$. } 
\label{fig:ALMA-con-cy24}
\end{figure}

After performing phase-only self-calibration for both the Cycle2 and Cycle 4 data, we combined them with weights inversely proportional to the variance of errors. Note that before combining the data, we relabeled the position reference frame of the Cycle 2 data as ICRS\footnote{In ALMA Cycle 2, the position reference frame in Measurement Sets (storing data of visibilities and information of observation) were given as J2000. However, the actual position reference frame of the phase calibrators used in Cycle 2 was ICRS, which causes systematic errors of $\sim 0\farcs01$ in the positions.}. In order to assess possible systematic differences in amplitude of visibilities, we compared both continuum data with a common uv-range between 140\,m and 1400\,m. The difference was observed to be $\lesssim 10\%$, which is comparable to the typical error value in amplitude of visibilities in ALMA observations. Then we performed the CLEAN imaging with a Briggs weighting of $robust=0,0.5$ to measure the power spectra using a CASA task \textit{tclean}(Figure \ref{fig:ALMA-con-cy24}). Imaging with $robust=-1$ was also used to fit positions of lensed images with data in other wavelengths. In the subsequent analysis, we used only continuum data because the line data \citep{stacey2018, inoue2020} had insufficient S/N to constrain the lens potential. 

\section{Method of Analysis}
\subsection{Overall Procedure}
\label{sec:4.1}
\begin{figure}
\epsscale{0.9}
\hspace{-0.3cm}
\plotone{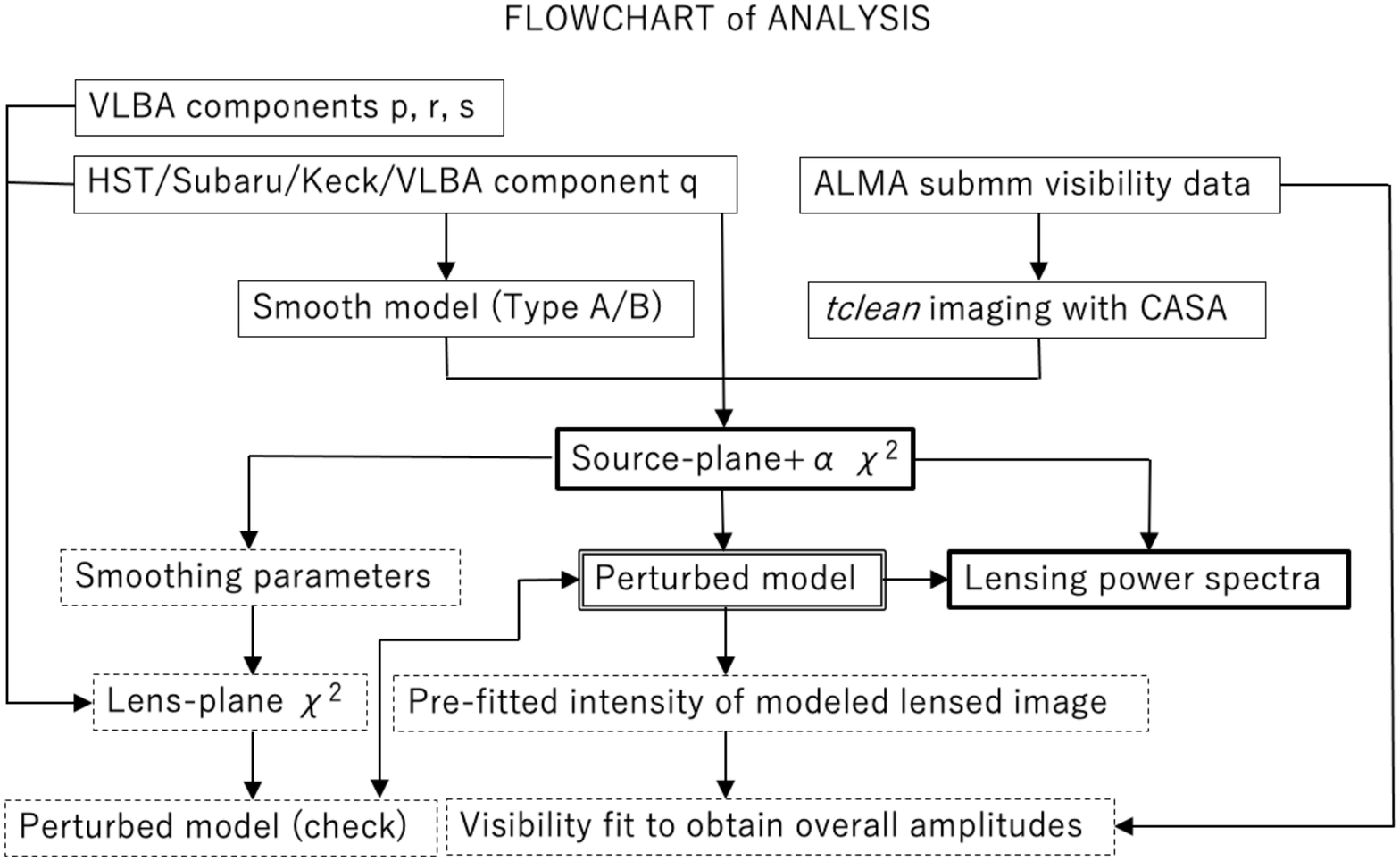}
\vspace{-0.5cm}
\caption{The master flowchart of overall procedure. Items marked in thin, thick, and dotted lines represent the preprocess, main process, and postprocess, respectively. 'Source plane+$\alpha ~\chi^2$' is described in equation (1). 'Visibility fit' is described in section 6.3 and 'lens-plane $\chi^2$' is described in section 6.4. The MIR data are obtained with Subaru \citep{minezaki2009} and Keck \citep{macleod2013}. } 
\label{fig:flowchart}
\end{figure} 
The overall procedure of our model fitting (see Figure \ref{fig:flowchart}) using the ALMA data of lensed images of an extended source is described as follows: (1) We derive a best-fitted smooth model that consists of galaxy halos with smooth gravitational potentials. (2) Using the obtained smooth model, we calculate a fiducial model source intensity for a particular weighting. (3) For a given potential perturbation, we calculate the shifts of each de-lensed extended images and their effect on the positions and fluxes of the lensed images of the quasar core. (4) Based on (3), we try to minimize $\chi^2$ (we call it 'source plane$+\alpha\,\, \chi^2$') defined as 
\BE 
\chi^2=\chi^2_{\textrm{ext}}+\chi^2_{\textrm{core}}+\chi^2_{\textrm{flux}}.
\label{eq:chi2}
\EE 
RHS of equation (\ref{eq:chi2}) consists of three terms: The first term $\chi^2_{\textrm{ext}}$ constrains the differences between the intensities of de-lensed images of an extended source in the source plane. The second term $\chi^2_{\textrm{core}}$ constrains the predicted positions of lensed images of a quasar core in the source plane. The third term $\chi^2_{\textrm{flux}}$ constrains the predicted flux ratios of the lensed images of a quasar core. In what follows, we use our ALMA data, the HST/VLBA data, and the MIR data 
for calculating $\chi^2_{\textrm{ext}}$, $\chi^2_{\textrm{core}}$, and $\chi^2_{\textrm{flux}}$, respectively.

Because of limited observable sky area of lensed images (i.e., thin arcs) compared to the sky area inside the arcs, we need to minimize a regularized $\chi^2_{\textrm{reg}}$ instead of $\chi^2$. More detailed definitions are described in the following subsections. 

To apply the above procedure to MG\,J0414+0534, we first model the primary lensing galaxy G using an SIE and a possible companion galaxy X using a cored isothermal sphere (CIS) that can account for an absence of a bright spot in the vicinity of X. As conducted in \citet{macleod2013}, possible effects from neighboring clusters and other large scale structures are modeled as an ES centered at the SIE. First, we perturb our background Type A models (SIE-ES-CIS) in which object Y (see Section \ref{sec:6.5}) is not explicitly modeled by discrete Fourier modes (see Section \ref{sec:4.6} for details) defined in the interior of a square that encompasses the lensed arcs\footnote{We did not include flexion (third derivative of potential) terms in describing potential fluctuations of massive objects located far from the lens \citep{okura2007, mccully2017, fleury2021} as they can be considered as low frequency Fourier modes. Similarly, we did not consider the third or fourth order multipoles $m=3$ and $m=4$ explicitly.}. Then, we perturb Type B models in which object Y is modeled by an SIE (SIE-ES-CIS-SIE) by the Fourier modes. From these results, we check whether these two procedures give a similar power spectra or not. In order to obtain the smooth models, we only use the HST or VLBA data for the positions of quadruply lensed images and galaxies (objects) and our Subaru and the Keck MIR data for the flux ratios of quadruply lensed images.

We use our combined ALMA Cycle 2 and Cycle 4 data as well as those used to obtain the smooth models to calculate $\chi^2_{\textrm{ext}}$, $\chi^2_{\textrm{core}}$, and $\chi^2_{\textrm{flux}}$, respectively. We do not constrain the position of G and X in the final fitting procedure because we found that such constraints do not affect the fitting much. Our Fourier decomposition of gravitational perturbation can describe perturbed gravitational potentials of the primary lensing galaxy G, object X, possible substructures, and LOS structures. 

\subsection{Regularization of Fitting}
\label{sec:4.2}
Although we can obtain a model that minimizes $\chi^2$ defined in equation $\ref{eq:chi2}$, the best-fitted model tends to yield a very large amplitude of perturbations at regions outside the lensed image because of limited sky area of lensed images. Since we aim to reconstruct a potential perturbation $\delta \psi$ defined in a region R that extends beyond the whole lensed image, we need to add a smoothing term $R_s$ to $\chi^2$, which functions as a penalty or regularization term \citep{treu-koopmans2004,koopmans2005}. In order to ensure the smoothness of the potential perturbation $\delta \psi$ and the first and second derivatives\footnote{\citet{treu-koopmans2004} and \citet{koopmans2005} use only the second derivative (curvature) of the source and potential perturbations as a regularizer. In our formalism, we add the squared potential perturbation and the squared first derivative to the regularizer but the source is not regularized.}, we adopt a smoothing term defined as 
\BE
R_s =\dfrac{\langle (\delta \psi)^2 \rangle}{(\delta \psi_0)^2}+\dfrac{\langle (\delta \alpha) ^2 \rangle}{(\delta \alpha_0)^2}+\dfrac{\langle (\delta \kappa)^2 \rangle}
{(\delta \kappa_0)^2},
\EE
where $\delta \psi$ is the potential perturbation projected on the primary lens plane, $\delta \alpha $ is the astrometric shift, $\delta \kappa$ is the convergence perturbation, and $\delta \psi_0, \delta \alpha_0 $ and $\delta \kappa_0$ are the smoothing parameters. $\langle \rangle$ denotes ensemble averaging over region R. For a given set of smoothing parameters $\delta \psi_0, \delta \alpha_0$, and $\delta \kappa_0$, we minimize a regularized $\chi^2_{\textrm{reg}}$ defined as 
\BE
 \chi^2_{\textrm{reg}} = \chi^2+R_s,
\EE
where $\chi^2$ is given by equation (\ref{eq:chi2}). From the obtained set 
of solutions, we select one that satisfies a $\chi^2/{\textrm{dof}}\sim 1$. Thus a set of the best-fitted model parameters and smoothing parameters, which determines the best-fitted potential perturbation $\delta \psi$ is determined. Note that we do not change the parameters of the smooth model in the fitting procedure.

\subsection{Source Plane $\chi^2_{\textrm{ext}}$ for Extended Sources}
\label{sec:4.3}
In order to fit perturbed lensing models to the observed data of lensed images of an extended source, we adopt a fit in which '$\chi^2$' fit is performed in the source plane. The reason is as follows:  First, the 
bias for estimating the potential perturbation in the 'source plane $\chi^2$' fit is expected to be smaller than that in the
traditional 'lens plane $\chi^2$' fit. Owing to the strong lensing effect from 
a primary lens, the weak lensing effect from a potential perturbation is significantly enhanced in the lens plane. Therefore, astrometric shifts due to a potential perturbation become very anisotropic and inhomogeneous in the lens plane, which would result in a biased estimate.  Second, we do not need to assume any functional forms for the source intensity. It can be obtained $\ti{a posteriori}$ rather than $\ti{a priori}$. As we shall describe later, the source intensity of a lensed quasar can have a very complex structure and the dynamical range can be very large. In such lens systems, the ensemble of source 
intensity does not necessarily obey homogeneous and isotropic gaussian statistics, which is often assumed in the traditional 'lens plane $\chi^2$' fit.  
Third, systematic errors due to 
coupling between noises at different regions or coupling between signals and noises
can be significantly suppressed in the source plane. Since multiple images are generated from a single image, we expect a gain factor of $\sim \sqrt{N}$ for a de-lensed image generated from $N$ multiple images. Moreover, de-lensed PSFs with a high magnification factor are significantly contracted (typically by a factor of $\sim 1/10$) in one direction. Therefore, the areas of de-lensed sidelobes in the source plane are reduced by $\sim 1/10$, leading to a suppression of systematic errors.

We now derive astrometric shifts due to a potential perturbation. 
The lens equation for an unperturbed (background) primary lens system with a deflection angle $\v{\alpha}_0$ is
\BE
\y=\x-\v{\alpha}_0(\x),
\EE
where $\x$ and $\y$ are the coordinates in the primary lens plane and the source plane, respectively. If the deflection angle is perturbed by $\delta \v{\alpha}$,
then the lens equation is given by
\BE
\tilde{\y}=\tilde{\x}-\v{\alpha}_0(\tilde{\x})-\delta \v{\alpha}(\tilde{\x}),
\EE 
where  
\BEA
\tilde{\x}&=&\x+\delta \x,
\label{eq:gauge1}
\\
\tilde{\y}&=&\y+\delta \y,
\label{eq:gauge2}
\EEA
and $\delta \v{\alpha}=\nabla \delta \psi$\footnote{If a massive object resides in regions outside the lens plane of the primary lens, a rotational (magnetic) component of astrometric shift may not be negligible.}. Equations (\ref{eq:gauge1}) and (\ref{eq:gauge2}) represent a relationship between the unperturbed and perturbed coordinates. Assuming that the second order terms due to coupling between the perturbation of the deflection angle and the astrometric shifts in the primary lens plane ($=d \delta \ALP$) are negligible, the
lens equation for the perturbation of the coordinates is 
\BE
\delta \y = \biggl(\v{1}-\frac{\del \v{\alpha}_0 }{\del \x}\biggr)\delta \x -\delta \ALP,
\EE   
where $\v{1}$ is a unit matrix. The choice of unperturbed coordinates $\x$ and $\y$ are arbitrary. Therefore, we need to fix a 'gauge' to determine the unperturbed coordinates. In what follows, we consider gauges in which the positions of 
lensed images are given by a sum of singular/cored isothermal ellipsoids 
with an external shear, representing galaxies, and a nearby cluster, respectively. If we choose a set of unperturbed coordinates with $\delta \y=\v{0}$,
then the lens equation yields astrometric shifts in the primary lens plane, 
\BE
\delta \x=\frac{\del \x}{\del \y}\delta \ALP=M \delta \ALP,
\EE 
where $M$ is the magnification matrix. We call this the 'lens plane gauge'.
If we choose unperturbed coordinates with $\delta \x=\v{0}$,
then the lens equation yields astrometric shifts in the source plane, 
\BE
\delta \y=-\delta \ALP.
\EE 
We call this the 'source plane gauge'. Note that the above formalism applies only for the weak lensing regime: the perturbation does not change the number of lensed images as $d \delta \ALP$ is sufficiently small. 

In terms of $n$-lensed images at $\x_i$ in the primary lens plane ($i=1,\cdots,n$), the source intensity at $\y=\y(\x_i)$ can be estimated using a linear combination of intensities $I_i$ observed in the primary lensed plane with weightings $w_i$,

\BE
\hat{I}(\y)=\f{\displaystyle{\sum_{i=1}^{n}} w_i I_i(\x_i(\y))}
{\displaystyle{\sum_{i=1}^{n}} w_i}. 
\label{eq:sigma_y}
\EE   

In what follows, we consider only two particular types of \ti{a priori} weighting choices for de-lensing: 
the homogeneous weighting in which the weighting is unity ($w_i=1$), and the magnification weighting in which the weighting is an absolute magnification ($w_i=|\mu_i|$). However, analogous to the weighting scheme in interferometry, one may also 
introduce the 'robust' weighting defined as 
\BE
w_i=\dfrac{|\mu_i|+1}{2}+\dfrac{r(|\mu_i|-1)}{4},
\EE
where $\mu_{\textrm{tot}}=\sum_i |\mu_i|$ is the total magnification and $r$
is the 'robust' parameter. $r=-2$ and $r=2$ correspond to the homogeneous weighting and the magnification weighting, respectively. The robust parameters $-2<r<2$ represent weightings between the homogeneous and magnification weightings. 

\begin{figure}
\epsscale{0.7}
\plotone{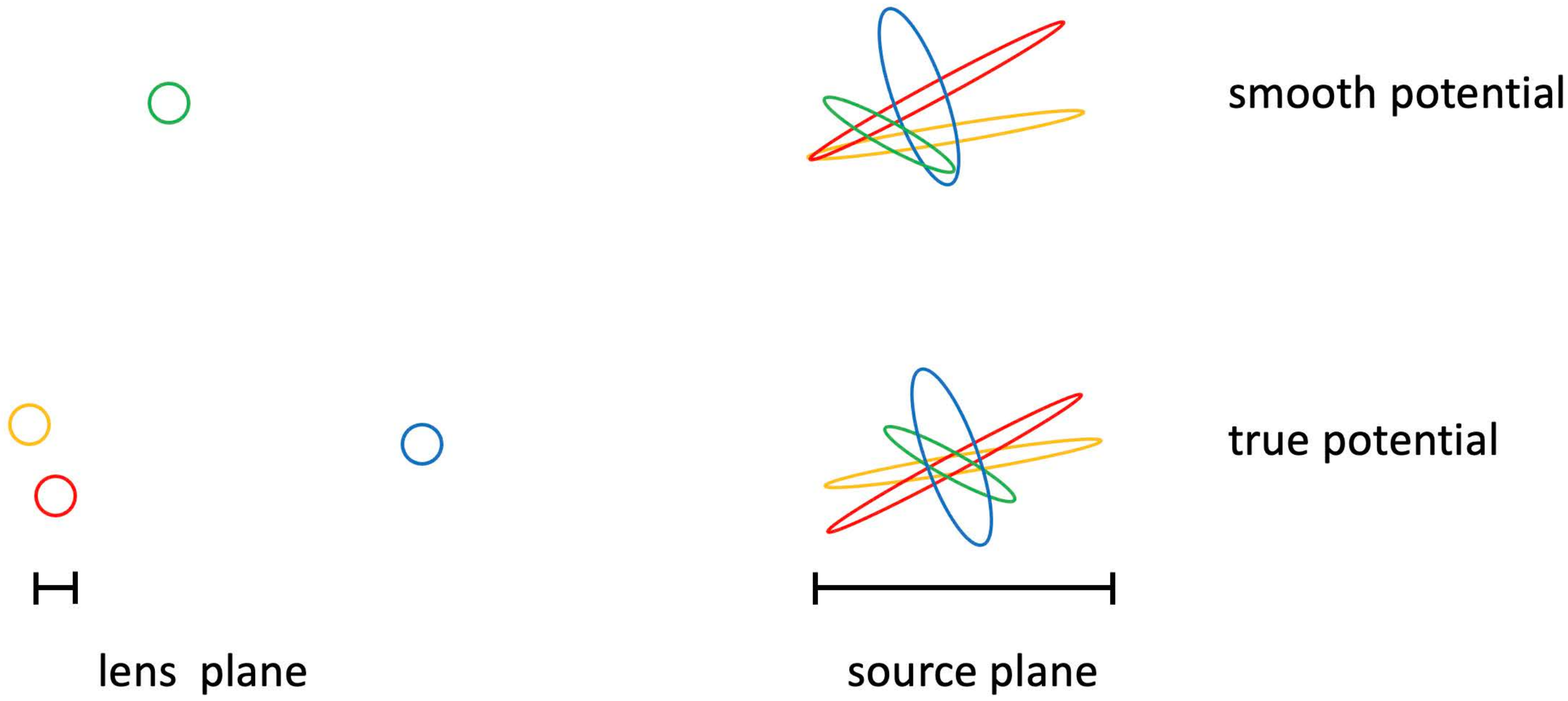}
\vspace{-1cm}
\caption{Super-resolution achieved by de-lensing. Each colored circle shows a lensed image of a point source convolved with a PSF in the lens plane (left). The corresponding de-lensed (zoomed up) images in the source plane are shown with ellipses with the same color (right). The horizontal bars in both the figures indicate scale bars with the same angular size. Due to errors in the potential, de-lensed PSFs are not aligned for models with the background smooth potential (top right) but aligned for those with a perturbed true potential (bottom right).   }
\label{fig:PSF}
\end{figure}

Owing to large magnification in strong lenses, the effective angular resolution of a source image can be significantly improved. Let us consider a system in which a point source is quadruply lensed and the point spread function (PSF) of lensed images is circularly symmetric and homogeneous in size and shape. As shown in Figure \ref{fig:PSF},  in the source plane, 
the de-lensed PSFs are shrunk compared to the original PSFs.
Therefore, the synthesized de-lensed PSF made from a linear combination of the 
de-lensed PSFs is significantly smaller than the original PSF in the lens plane,
though the shape is complicated in the tail. If the fitted gravitational potential is smooth and not  
perfectly correct, de-lensed images would become blurred due to residual astrometric shifts of PSFs (top right in Figure \ref{fig:PSF}). In order to find a true gravitational potential, it is necessary to align the de-lensed PSFs on all relevant pixels in the source plane (bottom right in Figure \ref{fig:PSF}).    

For fold or cusp caustics, the choice of weighting scheme yields noticeable differences: the 'synthesized' de-lensed PSF for the homogeneous weighting is more isotropic but larger than that for the magnification weighting. As shown in Figure \ref{fig:PSF}, the de-lensed PSF shrinks more for a lensed image with larger magnification. Therefore, in the source plane, the effective angular resolution of a de-lensed image given by the magnification weighting is expected to be much better than that given by the homogeneous weighting. Moreover, the S/N of de-lensed image is expected to be significantly better for the magnification weighting because it corresponds to the inverse-variance weighted average provided that the observational noise dominates. However, the de-lensed PSF is more anisotropic because of the small effective number of de-lensed PFSs and the information of pixels with small magnification is partially lost. Thus, it is not clear whether the magnification weighting is an optimal choice for estimating lensing power spectra.  

If the lens model is perfect and the S/N is sufficiently large, then the intensity of each lensed images must be equal. However, owing to finite angular resolution, errors in the intensity and the measured gravitational potential of lensing objects, the estimated source intensity on each pixel defined in equation (\ref{eq:sigma_y}) differ from the true intensity regardless of the weighting scheme. 

By minimizing the difference in the intensity at pixels 
in the source plane, one can obtain a
more accurate gravitational perturbation towards lensed images.    
Suppose that a sufficiently bright region in the source plane 
consists of $N$ pixels. For a given position $\y_j$ in the source plane,
the positions of multiple lensed images are $\x_1(\y_j),\x_2(\y_j),\cdots,\x_n(\y_j)$,
where $n(\y_j)$ is the total number (even) of the lensed images. The 
range of subscript $1$ to $n(\y_j)/2$ corresponds to images with a 
positive parity and that of $n(\y_j)/2$ to $n(\y_j)$ corresponds to images
with a negative parity. Although a statistic that can measure the difference in intensity between lensed images can have many definitions, we adopt a weighted mean of all the lensed images with a positive parity subtracted by a weighted mean of 
all the lensed images with a negative parity. The reason is as follows: First, the collective patterns of astrometric shifts
due to perturbations depend on the parity\citep{inoue2005a}. If the de-lensed images with different parity are synthesized,
the signal of a source smaller than the PSF beam size may be weaken due to cancellation of perturbation. Second, for quadruple lenses with cusp or fold caustics, 
a pair of images with a different parity can have a significantly larger magnification than the other pair, leading to loss of information. Grouping with parity avoids such selections. Third, grouping with parity is easy to implement as the parity can be directly calculated from magnification matrices. Other selections of grouping need calculation of the boundaries at which the sign of parity changes, which results in an increase in computation time. Assuming that the correlation in flux errors at different pixels is negligible\footnote{In the source plane, correlation between pixels due to side lobes is significantly suppressed as side lobe patterns do not correlate with the shape of lensed images.}, 
$\chi^2_{\textrm{ext}}$ defined in the 'lens plane gauge' ($\delta \y=\v{0}$) is
given by
\BEA
\chi^2_{\textrm{ext}} &=&
\displaystyle{\sum_{j=1}^{N_\textrm{p}}}\frac{1}{\varepsilon_\textrm{dif} (\y_j)^2}\left(\f{\displaystyle{\sum_{i=1}^{n(\y_j)/2}} w_i I(\x_i(\y_j)+\delta\x_i(\y_j))}
{\displaystyle{\sum_{i=1}^{n(\y_j)/2}} w_i} \right.
\nonumber
\\
&-&
\left.  \f{\displaystyle{\sum_{i=n(\y_j)/2+1}^{n(\y_j)}} w_i I(\x_i(\y_j)
+\delta\x_i(\y_j))}
{\displaystyle{\sum_{i=n(\y_j)/2+1}^{n(\y_j)}} w_i} \right)^2,
\label{eq:chisq}
\EEA
where $I$ is the observed intensity, $\varepsilon_\textrm{dif}$ is the error between de-lensed weighted 'mean' images with different parities, 
$\delta \psi$ is the potential perturbation projected on the primary lens plane, 
$\delta \alpha $ is the 
strength of deflection angle and $\delta \kappa$ is the convergence perturbation and $N_\textrm{p}$ is the total number of pixels. 
  
\subsection{Constraint on Positions of Core}
\label{sec:4.4}
To constrain the HST/VLBA positions of lensed images of a quasar core, we add the following constraining term defined in the 'source plane gauge' ($\delta \x =\v{0}$) as
\BE
 \chi^2_{\textrm{core}}=\displaystyle{\sum_{(\textrm{J},\textrm{K})}} \dfrac{|\delta \ALP(\x_\textrm{J}(\y_{\textrm{core}}))-\delta \ALP(\x_\textrm{K}(\y_{\textrm{core}}))|^2}{\langle |\delta \x_\textrm{J}-\delta \x_\textrm{K}|^2 \rangle}, 
\label{eq:const_pos}
\EE
where $\y_{\textrm{core}}$ is the fitted source position in the source plane, $\delta \x_\textrm{J}$ and $\delta \x_\textrm{K}$ are the astrometric shifts at the positions $\textrm{J}$ and $\textrm{K}$ of a
lensed core due to perturbation, respectively. The ensemble average of the shift difference $\langle |\delta \x_\textrm{J}-\delta \x_\textrm{K} |^2\rangle$ in a lens plane can be estimated using the residual errors in the positions of the lensed images in the best-fitted smooth model. We may consider that the shift differences in the denominator should be replaced with those
in the source plane $\langle |(M^{-1} \delta \x_\textrm{J})(\y_{\textrm{core}})- (M^{-1}\delta \x_\textrm{K})(\y_{\textrm{core}}) |^2\rangle$. However, we found that such a choice is too restrictive: by changing the fitted position of the quasar core $\y_{\textrm{core}}$, the errors in the source plane can be increased if the added potential perturbation on brightest lensed images gives a similar astrometric shift in the source plane (see also \citet{takahashi-inoue2014}). The sum in equation (\ref{eq:const_pos}) is taken over four sets of closest pairs of lensed images. For MG\,J0414+0534, the closest pairs $(\textrm{J},\textrm{K})$ are (A1,A2), (A2,B), (B,C), and (C,A1).

\subsection{Constraint on Flux Ratios of Core}
\label{sec:4.5}
To constrain the MIR flux ratios of lensed images of a quasar core, we add the following constraining term:
\BE
\chi^2_{\textrm{flux}}=\displaystyle{\sum_{\textrm{J}}}\biggl(\frac{\mu_\textrm{J}+\delta \mu_\textrm{J}}{\mu_\textrm{A1}+\delta \mu_\textrm{A1}}-\frac{\mu_\textrm{J}}{\mu_\textrm{A1}} \biggr)^2 /(\sigma_{\mu_\textrm{J}/\mu_\textrm{A1}})^2, 
\label{eq:const_flux}
\EE
where $\mu_\textrm{J}$ is the magnification factor at lensed image J ($\textrm{J}=\textrm{A2},\textrm{B},\textrm{C}$) of a core (or best-fitted image position of J) in the smooth model and 
$\delta \mu_\textrm{J}$ is the perturbation and $\sigma_{\mu_\textrm{J}/\mu_\textrm{A1}}$ is the observational error in the magnification ratio $\mu_\textrm{J}/\mu_\textrm{A1}$.

\subsection{Fourier Mode Expansion}
\label{sec:4.6}
A gravitational potential perturbation $\delta \psi$ due to halos and voids in the vicinity of photon paths of lensed images is expanded in terms of Fourier modes. In what follows, to discretise the potential, we impose a Dirichlet boundary condition \footnote{Instead we may impose the periodic boundary condition. However, in that case, we must consider a square with twice the side length.} $\delta \psi=0$ at the boundary of a square R with a side length of $L$ centered at $(x_{c1}, x_{c2})$ in the (primary) lens plane. The parameters are selected to satisfy that the entire lensed image is contained in the square. The potential perturbation $\delta \psi$ is set to zero outside the boundary. In our model, the contribution from masses inside the square is modeled as Fourier modes and the contribution from masses outside the square is modeled as an external shear in the smooth model and low (spatial) frequency Fourier modes inside the square. We can express the core structure or distortion of the primary lens as well as halos and voids in sight lines inside the square. To take into account the gravitational effect from masses near the boundary, we need to adjust the size of the square such that the distance between the lensed arc and boundary is larger than the half of the minimum angular wavelength in the real Fourier modes.    

The potential perturbation at $(x_1,x_2)$ in the lens plane can be decomposed as, 
\BEA
&\delta \psi&\!\!\!\!\!\!\!(x_1,x_2) =\sum_{m,n > 0}
\biggl[\Psi_{mn}^{++} \cos{k_m^+\tilde{x}_1}\cos{k_n^+\tilde{x}_2} \nonumber
\\
&+&\!\!\!\!\!\!\! \Psi_{mn}^{+-} \cos{k_m^+\tilde{x}_1}\sin{k_n^-\tilde{x}_2}
+
\Psi_{mn}^{-+} \sin{k_m^-\tilde{x}_1}\cos{k_n^+\tilde{x}_2} \nonumber
\\
&+&\!\!\!\!\!\!\!
\Psi_{mn}^{++} \sin{k_m^-\tilde{x}_1}\sin{k_n^-\tilde{x}_2} \biggr],
\label{eq:decomposition}
\EEA  
where $\tilde{x}_1=x_1-x_{c1}$, $\tilde{x}_2=x_2-x_{c2}$ are the relative positions in the lens plane of the primary lens, $k_m^{+}= (m~\textrm{mod}~2) m \pi/L $ and $k_m^{-}=((m+1)~ \textrm{mod}~2) m \pi/L $ are the angular wave numbers, $m$ and $n$ are non-zero positive integers, and $\Psi_{mn}^{++},\Psi_{mn}^{+-},\Psi_{mn}^{-+}$, and $\Psi_{mn}^{--}$ are expansion coefficients. Note that the relation between real and complex Fourier coefficients is described in Appendix B. Subsequently, the mean squared potential perturbation $\delta \psi$ is given by
\BEA
\langle (\delta \psi)^2 \rangle &=&\frac{1}{4}\sum_{m,n > 0}
\biggl[(\Psi_{mn}^{++})^2+(\Psi_{mn}^{+-})^2 \nonumber \\
&+&\!\!\!(\Psi_{mn}^{-+})^2 
+ (\Psi_{mn}^{++})^2 \biggr].
\EEA 
Similarly, the mean squared 
astrometric shift $\langle (\delta \alpha)^2 \rangle $
and mean squared convergence $\langle (\delta \kappa)^2 \rangle $, 
which is equal to the mean squared shear $\langle (\delta \gamma)^2 \rangle $, 
can be written in terms of the expansion coefficients as
\BEA
\langle (\delta \alpha)^2 \rangle &=&\frac{1}{4}\sum_{m,n > 0}
\biggl[(k_m^{+2}+ k_n^{+2} )(\Psi_{mn}^{++})^2
\nonumber
\\
&+&\!\!\!(k_m^{+2}+ k_n^{-2})(\Psi_{mn}^{+-})^2 
\nonumber
\\
&+&\!\!\!(k_m^{-2}+ k_n^{+2})(\Psi_{mn}^{-+})^2
\nonumber
\\ 
&+&\!\!\!(k_m^{-2}+ k_n^{-2})(\Psi_{mn}^{++})^2 \biggr],
\EEA 
and
\BEA
\langle (\delta \kappa)^2 \rangle
  &=&\frac{1}{4}\sum_{m,n > 0}
\biggl[(k_m^{+2}+ k_n^{+2} )^2(\Psi_{mn}^{++})^2
\nonumber
\\
&+&\!\!\!(k_m^{+2}+ k_n^{-2})^2(\Psi_{mn}^{+-})^2 
\nonumber
\\
&+&\!\!\!(k_m^{-2}+ k_n^{+2})^2(\Psi_{mn}^{-+})^2
\nonumber
\\ 
&+&\!\!\!(k_m^{-2}+ k_n^{-2})^2(\Psi_{mn}^{++})^2 \biggr].
\EEA 

\section{Mock Simulation}
\subsection{Generation of Mock Data}
\label{sec:5.1}
Before performing mock simulations, we prepared 
a fiducial unperturbed smooth model 
based on our ALMA observations, and previous HST and VLBA observations of MG J0414+0534. The procedure is as follows: 

First, we used a Type A model consisting of an SIE, an ES, and a CIS. SIE, ES, and CIS model the primary lensing galaxy G, a large-scale external shear, and object X, respectively. Using the SIE-ES-CIS model, we fitted the positions of quadruple images of a quasar core and the centroid of G and X observed in the OPT/NIR band in the CASTLES database. The assumed HST position errors are $0 \farcs 003$ for lensed images and the centroid of G and $0 \farcs 2$ for the centroid of X. The error for X was relaxed because X may be a lensed image of the quasar host galaxy rather than that of a companion galaxy (see \citet{inoue2017} for details of model parameters). The size of the core of CIS was selected to best fit the observed parameters but constrained to not have an additional pair of images of a quasar core, which has not been observed in any radio bands. We did not consider any relative fluxes of lensed images for parameter fitting.

Second, we carried out continuum imaging with a Briggs weighting of $\ti{robust}=-1$ using the Measurement Set from mock 'Cycle 4' observations. Then we rotated and translated the image to fit the positions of the lensed quasar images of the quasar core in the OPT/NIR band. We also subtracted off the lensed images of the lensed peaks using the best-fitted synthesized elliptical Gaussian beam in the lens plane. The peak intensity of the elliptical Gaussian beam at image A2 was fixed to 95 percent of the peak intensity at image A2. The 5 percent reduction is due to fluxes from an extended region. The peak intensities at other lensed images were given by the MIR flux ratios.

\begin{figure}
\epsscale{0.7}
\plotone{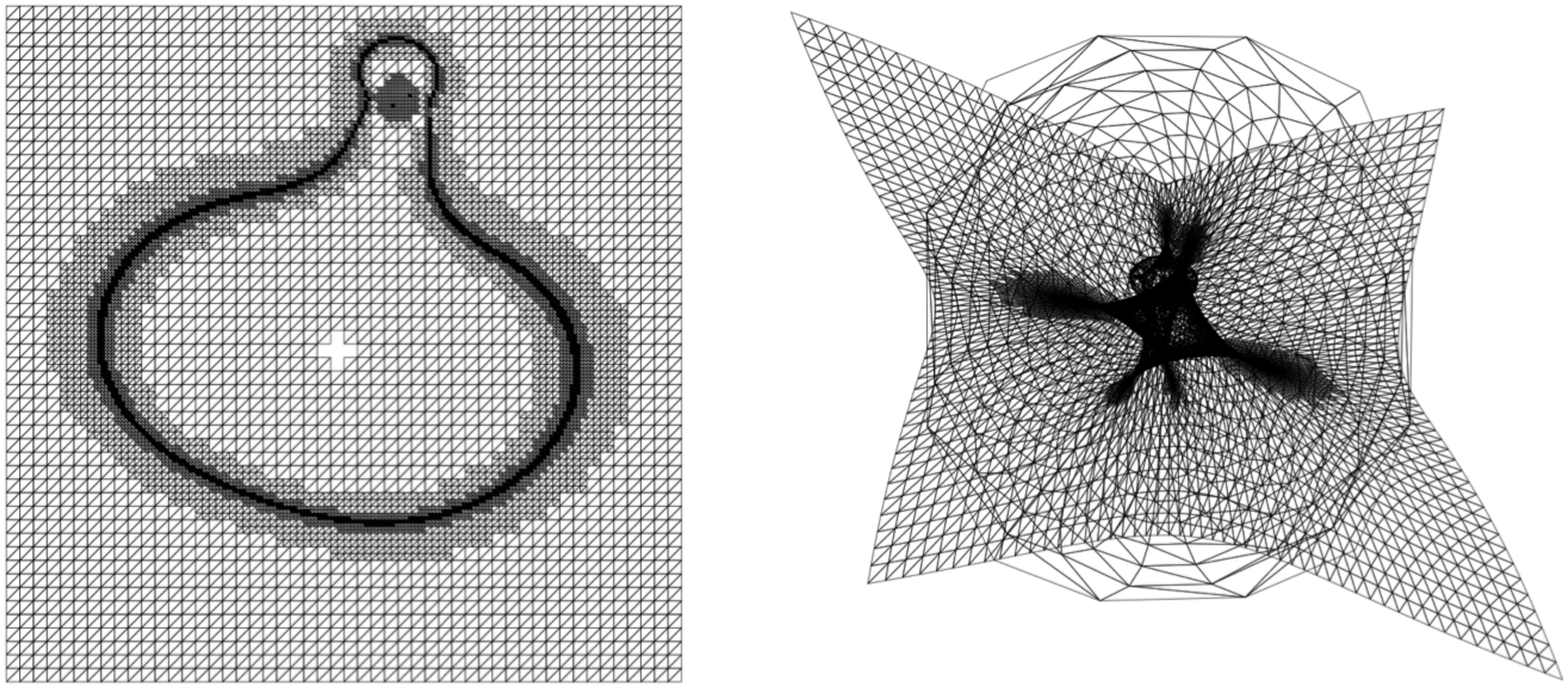}
\vspace{-1cm}
\caption{Meshes in the lens plane (left) and source plane (right) for the fiducial smooth model. }
\label{fig:meshes}
\end{figure}

Third, we reconstructed the source image from the continuum image of the Cycle 4 observations, in which imaging was carried out using the natural weighting. We used the homogeneous weighting for de-lensing. To carry out de-lensing, as shown in  (Figure\ref{fig:meshes}, left), a square region with a side length of $2$ arcsec centered at the center of the best-fitted SIE is covered with $50 \times 50$ square meshes. Then we selected all the meshes at which the absolute magnification is larger than $\mu_1=6$ and we subdivided these meshes into four square meshes. Similarly, we iterated this process with thresholds $\mu_2=18$ and $\mu_3=65$. This choice resulted in a similar number of meshes in each tier and the final mesh size being the same as the pixel size of the original image. We also subdivided
meshes in a circular region around the center of object X into square meshes with a side length of $0\farcs 1$ to resolve a small closed critical curve. Moreover, we omitted a region within a radius of $0\farcs 1$ centered at the center of an SIE to avoid a singularity. Each square mesh was divided into two right triangles and their vertices were mapped into the corresponding vertices in the source plane (Figure\ref{fig:meshes}, right). The source plane was covered with square meshes and at the center of each mesh, the number of triangles that include the center and the mapped triangles in the lens plane were computed. For a given point in the source plane, with the mesh that contains the point, the number of lensed image and the first guess of the corresponding points in the lens plane were computed. Using the first guess values, the corresponding accurate points in the lens plane was computed using Newton's method. Thus for a given point in the source plane, the corresponding lensed points in the lens plane can be numerically obtained with shorter CPU time. For brevity, we used the homogeneous weighting for de-lensing the continuum image. We found that the source (after subtracting bright spots) consists of two bright spots and a surrounding extended structure with a core \citep{inoue2020}. 
\begin{figure}
\epsscale{0.7}
\plotone{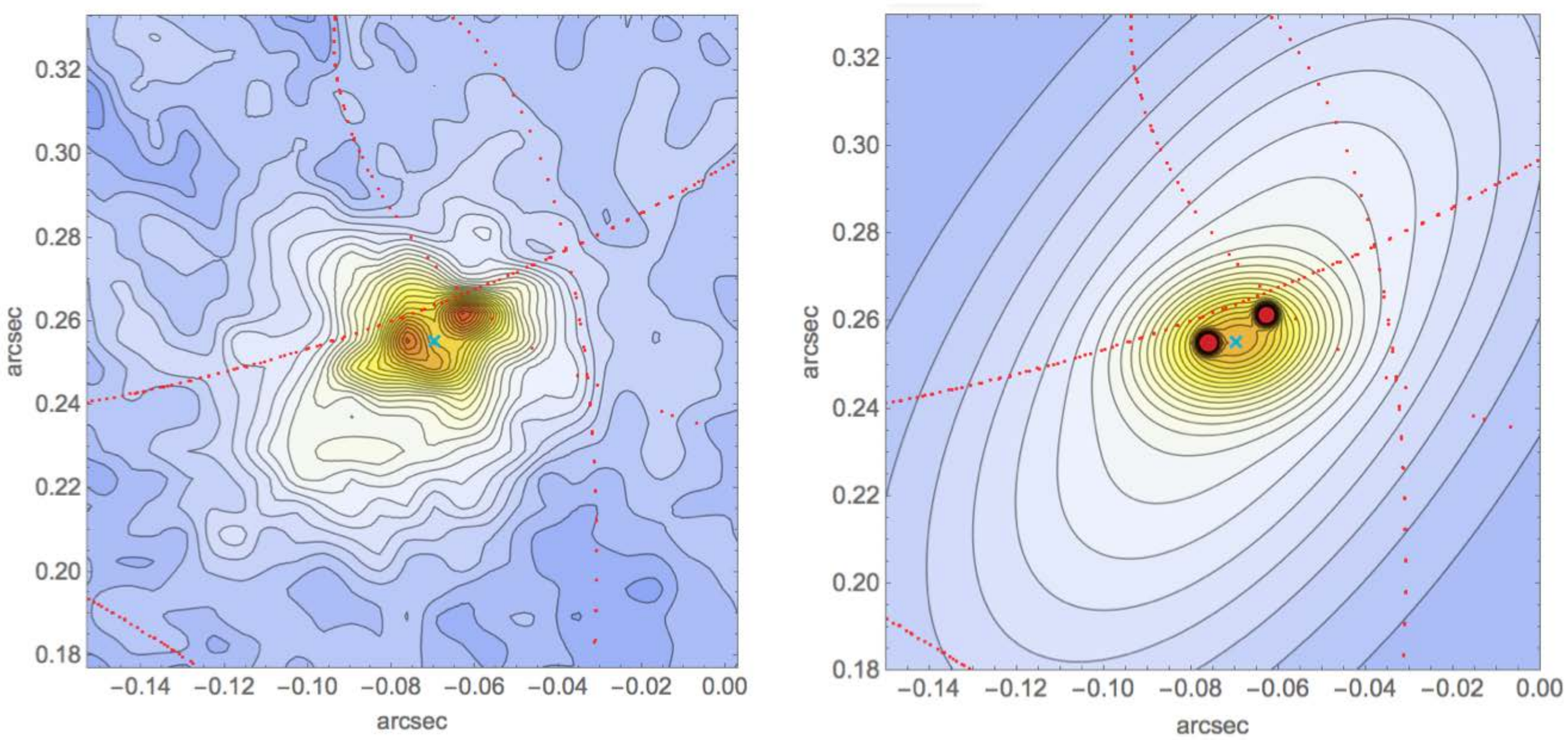}
\vspace{-1cm}
\caption{Contour map of a source image made from ALMA continuum Cycle 4 image 
(left) and fitted model image (right). The quasar core component is subtracted off in both the maps. The color images and contours show the intensity of MG J0414+0534. Potential perturbation is not taken into account. Red points show the positions of caustics. The green X shows the position of the fitted quasar core.}
\label{fig:mock-source}
\end{figure}

Finally, we made a 'true' source image and added a random potential perturbation 
to the obtained background lens potential: First, we fitted the two bright spots and extended structures observed in the Cycle 4 core-subtracted source image  (Figure \ref{fig:mock-source}, left) with two identical spherical Gaussian functions with a full width at half maximum (FWHM) of 12\,pc and two-component concentric elliptical 
Gaussian functions, respectively (Figure \ref{fig:mock-source}, right). The two bright spots and extended structures represent jet/core components and cold/warm dust emissions, respectively. The amplitudes, ellipticity and axis directions of these Gaussian functions were obtained from a 
$\chi^2$ fit to the Cycle 4 core-subtracted source image on pixels with $>3 \sigma$. Second, we added a quasar core component represented by a spherical Gaussian function with FWHM of 12\,pc in the source plane. The amplitude was adjusted to recover fluxes of bright spots observed in the lens plane. Third, we added a random Gaussian perturbation consisting of $4\times 3^2=36$ discrete modes to the potential of the background lens model. The side length of a square region was set to $3\farcs 6$ and the center was set at $(-0\farcs 4,0\farcs 3)$ in which the centroid of G is at $(0,0)$. The selected parameters satisfy the condition that the distance between the boundary and the lensed arcs, objects X and Y are longer than the half of the shortest angular wavelength of $1\farcs 2$. The 'true' potential perturbation was assumed to vanish at the boundary and the outside of the square. The coefficients of the potential perturbation were assumed to obey a Gaussian distribution with a zero mean and a standard deviation of $0.57\,\textrm{mas}^2$. We made random $10^5$ realizations and selected one set that satisfied the constraints on the relative astrometric shifts in the OPT/NIR bands (equation (\ref{eq:const_pos})) and the MIR flux ratios of the lensed quasar cores (equation (\ref{eq:const_flux})).

\subsection{Mock Observation}
\label{sec:5.2}
\begin{figure}
\hspace{-0.3cm}
\epsscale{0.6}
\plotone{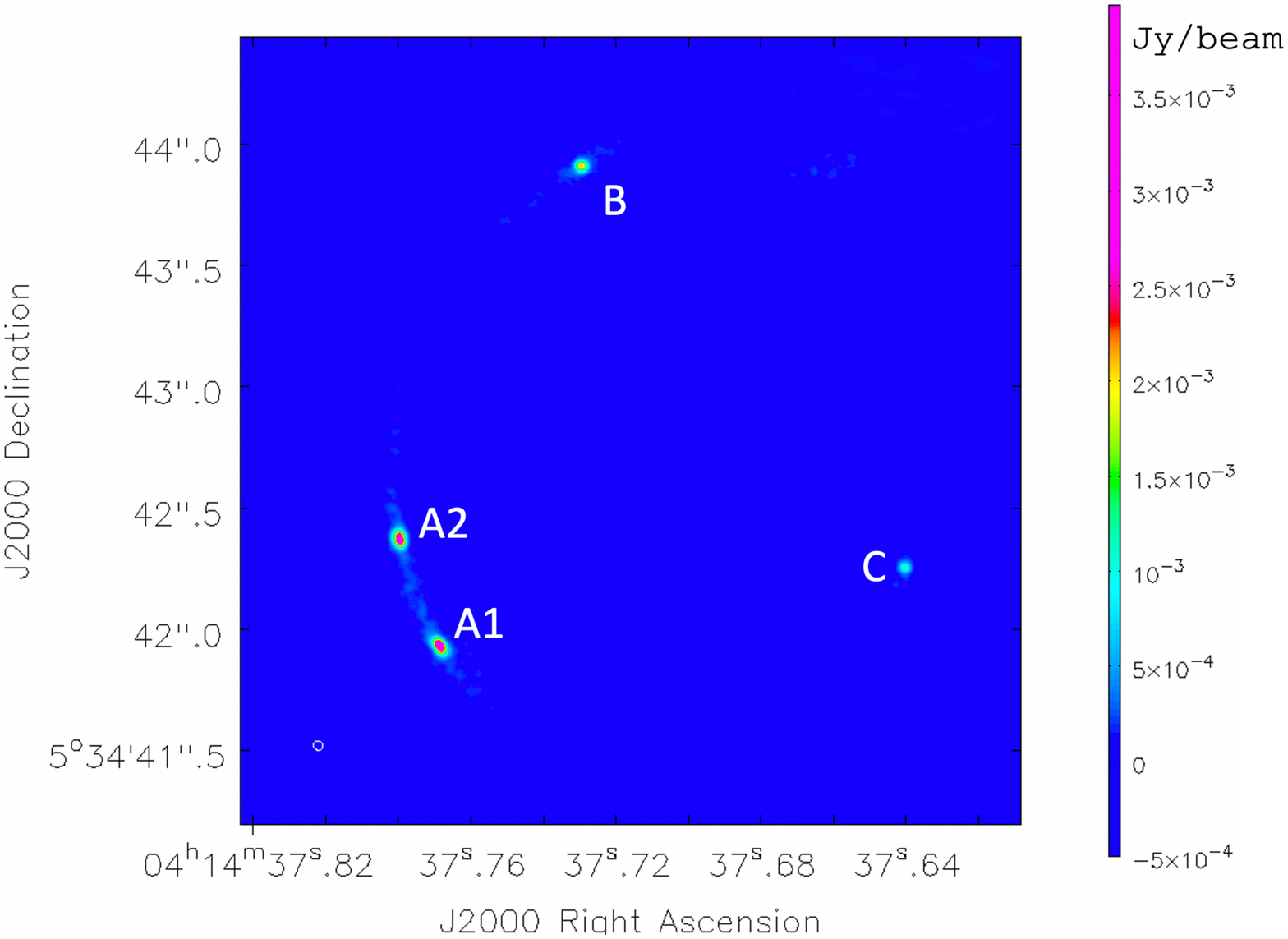}
\vspace{-0.5cm}
\caption{Mock ALMA continuum image (simulated Cycle 2 + Cycle 4 observations) of MG J0414+0434. The imaging was carried out with a Briggs weighting of $robust=0.5$. The synthesized beam size is ${0\farcs 041\times 0\farcs 037}$ and the PA is $54\fdg 2$. The background rms noise is $20\,\mu \textrm{Jy}\,\textrm{beam}^{-1}$.} 
\label{fig:mock-lens-image}
\end{figure} 

We carried out ALMA mock continuum observations of the lensed 'true' source using the $simobserve$ 
task in CASA. We used the same observation dates, antenna configuration, precipitable water vapor (PWV) and integration time as used in our actual Cycle 2 and Cycle 4 observations. We added thermal noise for a ground temperature of $271\,$K. The line-free band width was set to 4189\,MHz, approximately equal to the one for the actual observations. After the mock 'Cycle 2' and 'Cycle 4' observations, we concatenated the obtained Measurement Sets as was conducted for our actual data. Then continuum imaging was performed using the CLEAN algorithm (\textit{tclean} in CASA) with a Briggs weighting of $\ti{robust}=0.5$ (Figure \ref{fig:mock-lens-image}). To measure the positions of quadruple images of a quasar core, we also performed continuum imaging with a Briggs weighting of $\ti{robust}=-1$ using the Measurement Set from mock 'Cycle 4' observations. We also subtracted off the lensed images of the mock 'quasar core' using the best-fitted elliptical Gaussian beam. The peak intensity of the elliptical Gaussian beam at image A2 was set to the 95 percent of the peak intensity at image A2. The 5 percent reduction is due to fluxes from an extended region. The peak intensities at other lensed images were given by the 'true' flux ratios. We observed that our CLEANed image gives a $\sim 60$ per cent increase in intensities except for the brightest spot with S/N$\gtrsim 50$ that corresponds to the mock quasar core. Therefore, to test our mock analysis, we multiplied the mock observed intensities for peak-subtracted lensed images by a constant of $1/1.6$. Such a uniform change in intensity does not much affect the reconstruction of perturbation as the scale of potential perturbation is much smaller than the whole lensed image.

\begin{figure}
\hspace{-1.5cm}
\epsscale{0.6}
\plotone{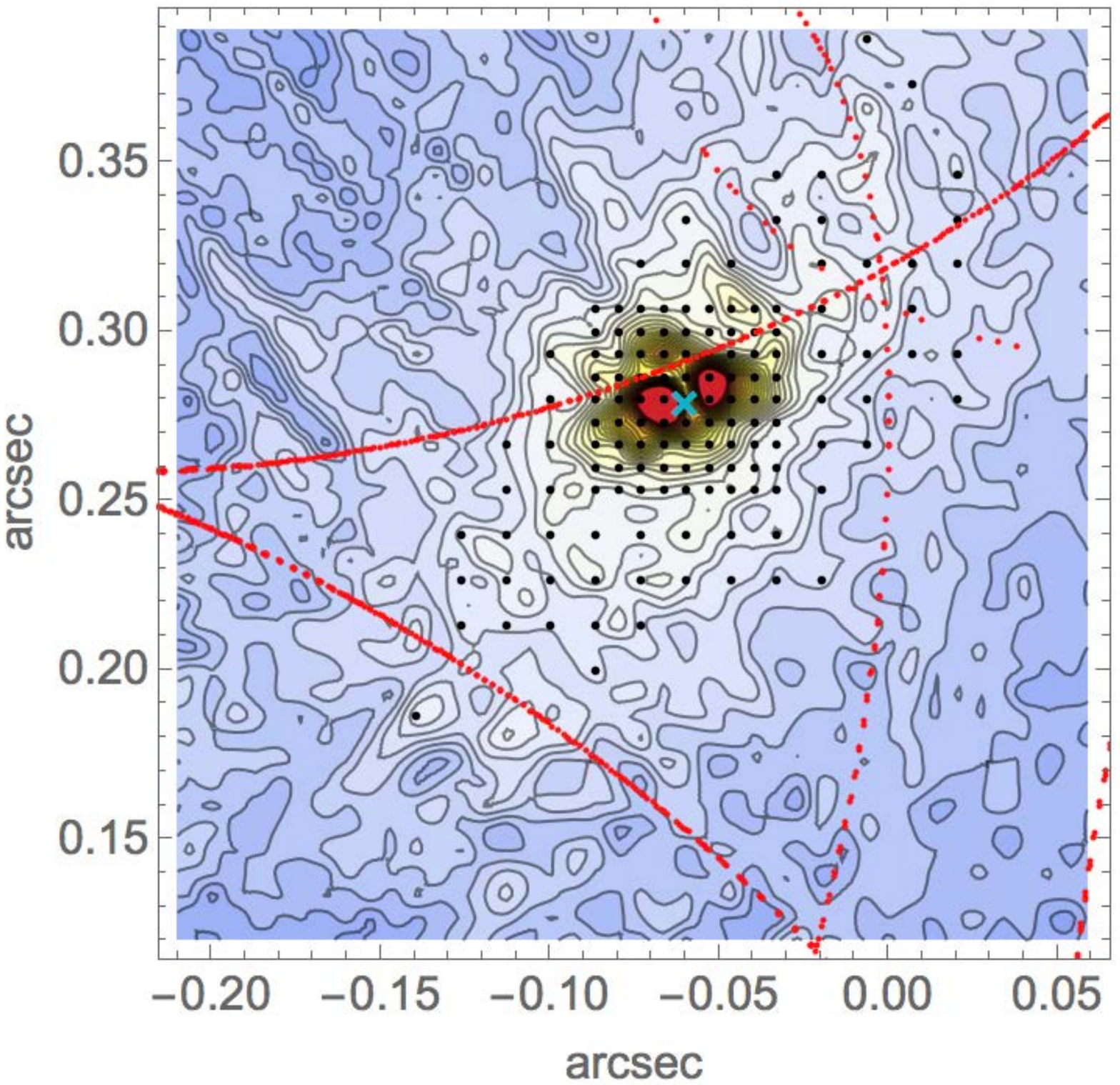}
\vspace{-0.5cm}
\caption{Contour map of a de-lensed image made from simulated ALMA observations of a core-subtracted mock source. The color images and contours show the intensity. The black points show the centers of meshes. The red points show the positions of caustics. The green X shows the position of the fitted quasar core.  }
\label{fig:mock-source-mesh}
\end{figure} 

\subsection{Mock Analysis}
\label{sec:5.3}
We used the SIE-ES-CIS (Type A) model to fit
the positions of mock quadruple images of quasar cores and the centroids of the primary lensing galaxy G and object X (see \citet{inoue2017} for details). Note that the positions of the centroid of G and X were also perturbed by the added potential perturbation. We assumed that the errors in the positions in the lens plane are equivalent to the values in the OPT/NIR data. 

Then we performed $\chi^2$ minimization numerically to obtain the best-fitted parameters for the mock smooth model. As described in Sec.5.1, a square region with a side length of $2$ arcsec centered at the center of the best-fitted SIE was covered with $50 \times 50$ square meshes, which were subdivided iteratively. The source plane at which the estimated signal in the source plane is larger than $5\,\sigma$ was covered with meshes with a side length of 6.66\,mas or 13.3\,mas. The meshes cover the central and surrounding region of the core-subtracted mock source (Figure \ref{fig:mock-source-mesh}). The size of small meshes is smaller than the two bright spots. Since the mesh sizes are larger than the de-lensed PSF size of $\sim 4\,$mas, the effect of spatial correlation of errors is expected to be small. The mesh size is larger in the outer region since the curvature of the de-lensed intensity is smaller in the outer region.         

In order to avoid regions at which our weak lensing formalism breaks \footnote{If absolute magnification is excessively large, the second order effect due to coupling between the perturbations on the parameters in the background and foreground cannot be neglected.}, we excluded meshes at which the maximum absolute magnification is larger than 30. We also excluded meshes that yielded an odd number of images, due to crossing over the caustic. For each pixel, $1\,\sigma$ errors $\varepsilon_I$ in the reconstructed source and $\varepsilon_\textrm{dif}$ in the difference in the weighted mean de-lensed images for a positive/negative parity were measured from random translations of the mock lensed image. In order to suppress contributions from signals, we subtracted off fluxes larger than $4\,\sigma$ in the lens plane to estimate the errors in the source plane.

If the scale of the fluctuation in the de-lensed intensity is smaller than the fitted synthesized beam, 
the reconstructed intensity differs (typically smaller for brighter regions) from the true value due to the contribution 
within the beam. Moreover, the fitted synthesized beam may differ from the 'true' PSF due to systematic errors. 
To take into account such effects, we modeled the 
'true' errors in the weighted mean de-lensed images at a pixel centered at $\y_j$ as
\BE
\varepsilon_\textrm{dif}(\y_j)=\varepsilon_\textrm{dif}(\y_j;\textrm{fid})|\hat{I}(\y_j)/\varepsilon_I(\y_j)|^{\beta},
\EE
where $\varepsilon_\textrm{dif}(\y_j;\textrm{fid})$ is the nominal value obtained from random 
translations of a de-lensed image and $\beta $ is a constant parameter. If the contamination is described by a Poisson process, we expect $\beta=0.5$. In what follows, we adjusted $\beta$ before $\chi^2$ minimization to yield a best-fitted reduced $\chi^2$ of $\sim 0.8 - 1.5$.

\begin{figure}
\vspace{0.3cm}
\hspace{0.5cm}
\epsscale{0.6}
\plotone{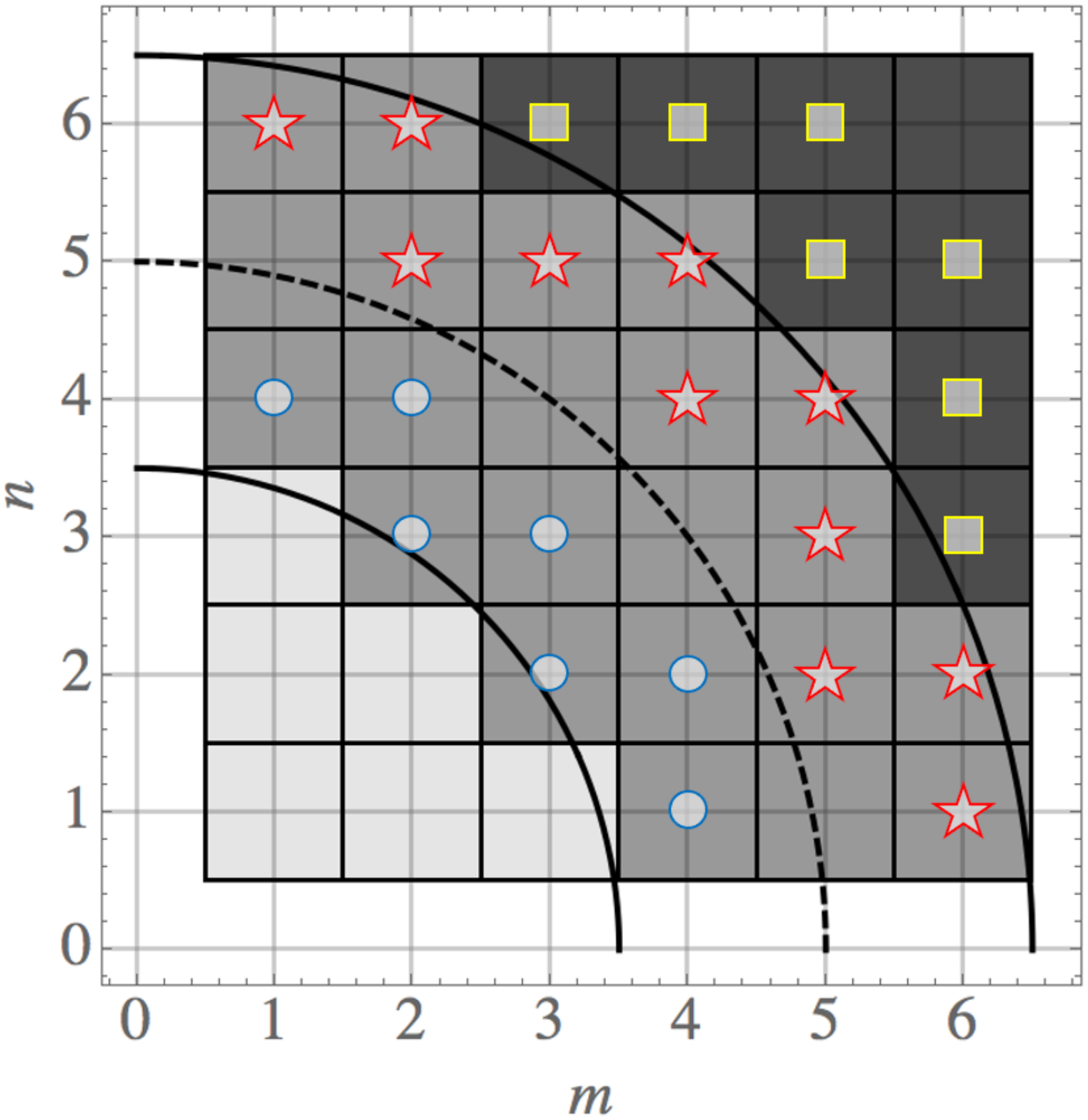}
\vspace{0.cm}
\caption{Fourier modes used in our analysis. $m,n$ are Fourier mode numbers that specify a Fourier mode function $\delta \psi$ defined in equation (\ref{eq:decomposition}). A potential perturbation is decomposed into 36 modes, which include 6 low-frequency modes (light gray), 22 intermediate frequency modes (middle gray) and 8 high-frequency modes (dark gray). We also use three frequency bins: 7 low-to-intermediate frequency modes (circle), 11 intermediate-to-high frequency modes (star), and 7 high-frequency modes (square).}
\label{fig:fourier-modes}
\end{figure} 
If a potential perturbation is decomposed into $N\times N$ modes, we use modes with $3.5<\sqrt{m^2+n^2}<N+0.5$ to analyze the mock perturbation due to LOS structures or subhalos. For instance, for $N=6$, we use 22 intermediate-frequency modes for analysis of perturbation (Figure \ref{fig:fourier-modes}) in our mock simulations. This ensures an approximate rotational symmetry of the correlation and avoids degeneracy with low multipole (=monopole, dipole and quadrupole) contributions from the galaxy of the primary lens.

In this mock analysis, we used the same square boundary with a side length of $L=3\farcs 6$ and $N=6$. Therefore, our mock analysis is limited to systems in which the gravity of halos at the boundary does not significantly affect the lensed arcs. For comparison, we adjusted $\beta$ to satisfy the condition $\chi^2/\textrm{dof}=2.0$ in the initial model without any perturbation.
\subsection{Mock Result}
\label{sec:5.4}
\begin{table}
\hspace{-4cm}
\caption{Parameters for the mock best-fitted models. }
\label{tab:1}
\hspace{1cm}
\setlength{\tabcolsep}{2pt}
\begin{tabular}{lccccccc}
\hline
\hline
weighting & $\delta \psi_0$ & $\delta \alpha_0$ &  $\delta \kappa_0$ & $\beta$ &$\chi^2/\textrm{dof}$ & $n_{\textrm{obs}}$ &dof
\\
\hline
magnification & $0.0011$ & $0.0016$ & $0.0073$ & $0.383$ & $0.76(2.0)$ & $138+7$ & 106
\\
\hline
homogeneous &  $0.0014$ & $0.0049$ & $0.0061$ & $0.661$ & $1.51(2.0)$ & $163+7$ & 131
\\
\hline
\end{tabular}
\flushleft{Note.  $\delta \psi_0 $,  $\delta \alpha_0 $, and $\delta \kappa_0 $ are smoothing parameters
that control the smoothness of the rms values of potential, astrometric shift, and convergence perturbations.
$\beta$ is an index that describes the increase of errors in bright compact regions. $n_{\textrm{obs}}$ is the number of meshes plus the number of constraints for the lensed images of a quasar core. 
The unit of $\delta \psi_0 $ is $\textrm{arcsec}^2$ and that of 
$\delta \alpha_0 $ is $\textrm{arcsec}$. }

\end{table}
As shown in Table \ref{tab:1}, the 
magnification weighting gives smaller $\beta$ than the homogeneous weighting. This is expected because the magnification weighting corresponds to the inverse-variance weighted average which maximizes the S/N. 

\begin{figure}
\vspace{-3.0cm}
\hspace{-0.3cm}
\epsscale{1}
\plotone{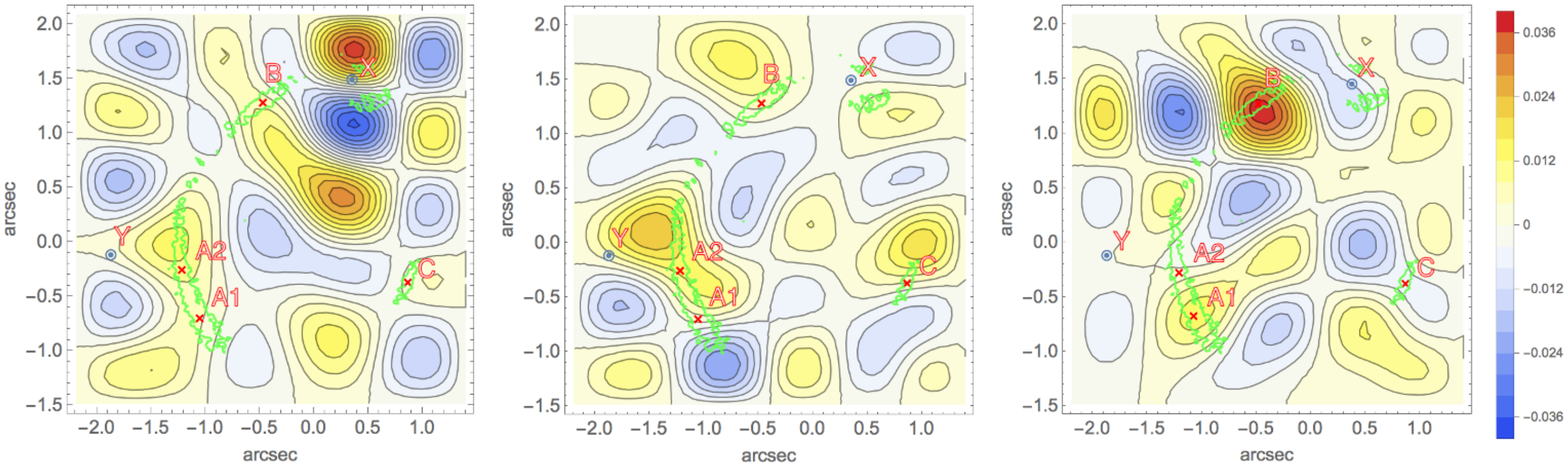}
\vspace{-2.8cm}
\caption{Mock simulation of reconstruction of convergence perturbation $\delta \kappa$. The plotted images are contour maps for the original values (left), the best-fitted reconstructed values obtained with the magnification weighting (middle) and the homogeneous weighting (right). The plots were obtained from the 36 modes that best fit the mock data. Inside the green curves, the 'observed' intensity is larger than $4\,\sigma$. The contour spacing is 0.002. red X's show the original positions of the lensed quasar core. Blue circled dots show the positions of object X (upper right) and Y (lower left, see Section \ref{sec:6.5}) in each panel. At the boundary of each panel, the potential perturbation is set to zero and the center of the coordinates is the centroid of observed G.    }
\label{fig:conv-mock}
\end{figure} 

\begin{figure}
\vspace{-2cm}
\hspace{-0.3cm}
\epsscale{1.0}
\plotone{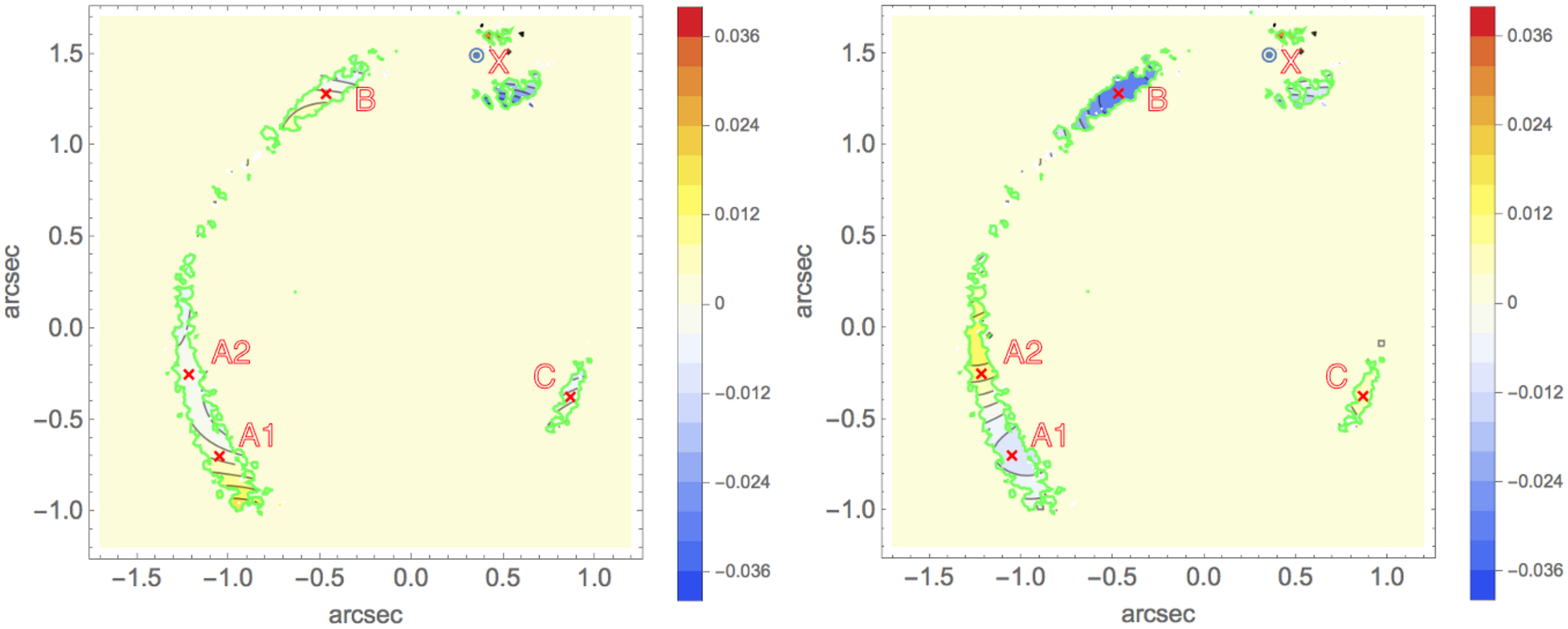}
\vspace{-2.5cm}
\caption{Contour maps of differences between the original and best-fitted convergence perturbation $\delta \kappa$ obtained with the magnification weighting (left) and homogeneous weighting (right). The contour spacing is 0.002. The red and blue symbols and green curves are the same as in Figure \ref{fig:conv-mock}. The value are set to be zero in which the 'observed' intensity is smaller than $4\,\sigma$ for illustrative purposes.  }
\label{fig:conv-dif-mock}
\end{figure} 

We plotted the original and best-fitted convergence perturbation $\delta \kappa$ in Figure \ref{fig:conv-mock} and
their differences in Figure \ref{fig:conv-dif-mock}. We can see in these figures that the fidelity of convergence perturbation in the vicinity of images A1, A2, and B are better for the magnification weighting. The result is not surprising as these images have large magnification and thus the de-lensed images of have large weighting. On the other hand, the fidelity of convergence perturbation in the vicinity of X and image C are better for the homogeneous weighting though the fidelity in the vicinity of images A1, A2, and B are worse. The result is again not surprising as the homogeneous weighting weighs each lensed image equally. Fluctuations that are reconstructed with the magnification weighting seem to be more anisotropic than those reconstructed with the homogeneous weighting. This implies that the magnification weighting may not be an optimal choice for the purpose of reconstructing correlation functions and lensing power spectra though the fidelity of perturbation in real space (i.e., lens plane) is better than that obtained with the homogeneous weighting.

\begin{figure}
\hspace{0.cm}
\epsscale{1.2}
\plotone{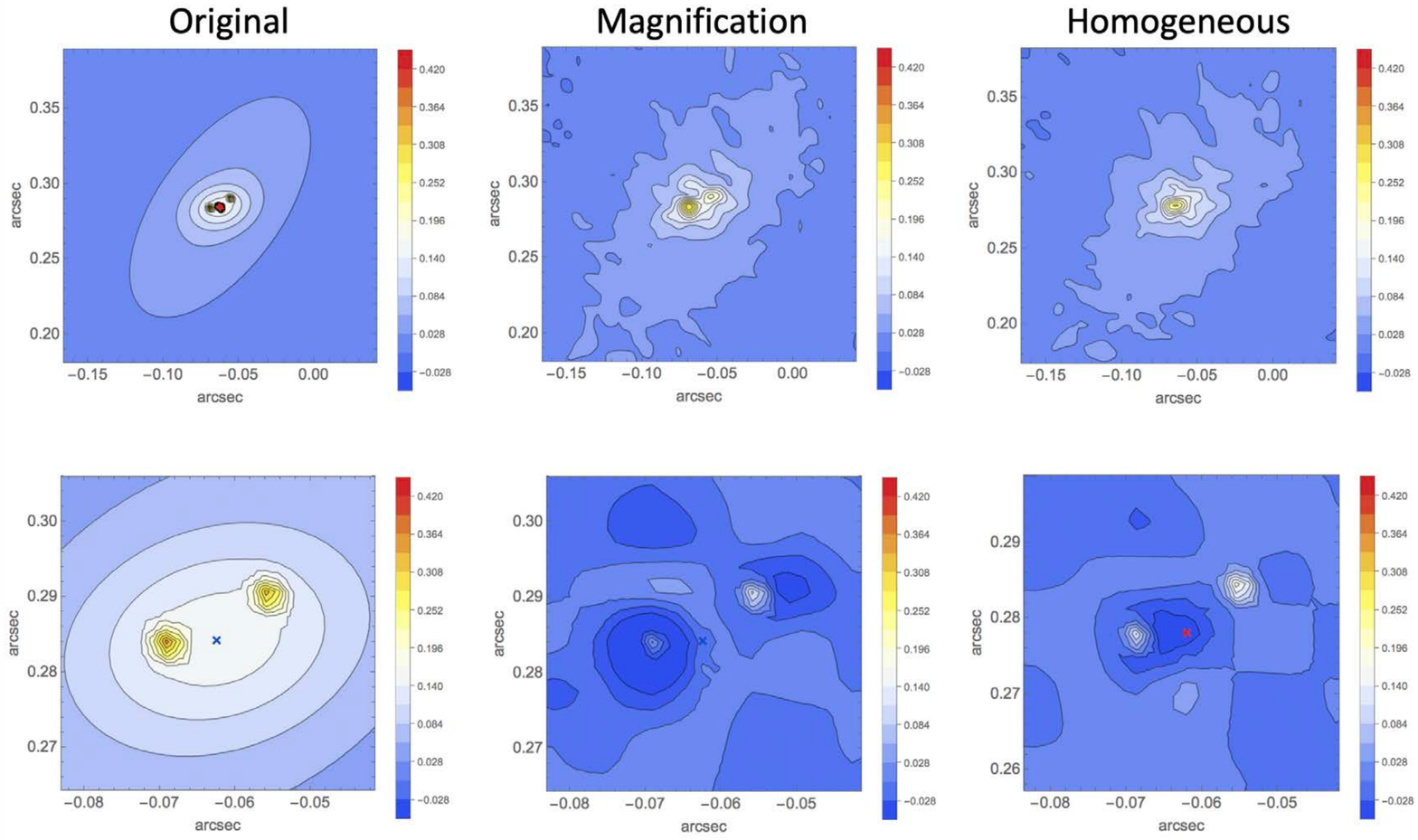}
\vspace{-1.1cm}
\caption{Continuum intensity for the original and de-lensed mock sources. Except for the original image on the top left, the bright quasar components (at the position of a blue or red X) were subtracted off. The colors show intensity in units of $\textrm{Jy}\, \textrm{arcsec}^{-2}$. The plotted images are the original source (top left), de-lensed source reconstructed with the magnification weighting (top middle), de-lensed source reconstructed with the homogeneous weighting (top right), zoomed up images of 
the original source (bottom left), differences between the de-lensed source reconstructed with the magnification weighting and the original one (bottom middle), and difference between the de-lensed source reconstructed with the homogeneous weighting and the original one (bottom right).     }
\label{fig:source-mock}
\end{figure} 

\begin{figure}
\vspace{-2cm}
\hspace{-0.5cm}
\epsscale{0.8}
\plotone{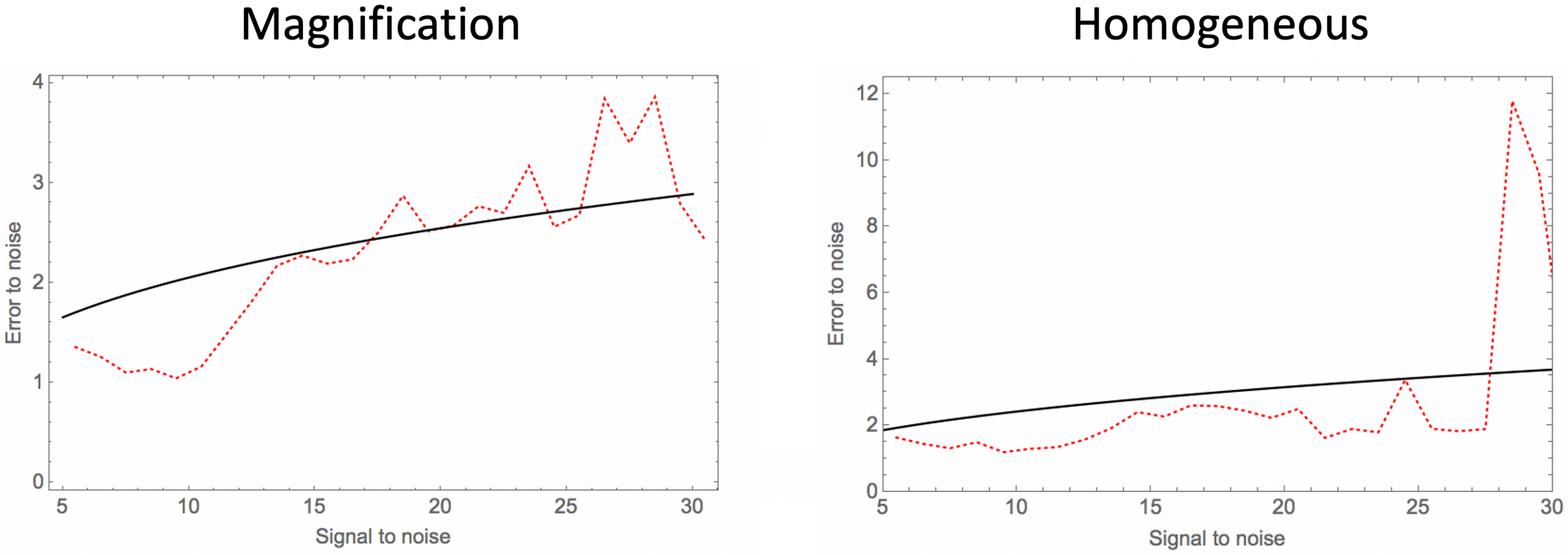}
\vspace{-2.cm}
\caption{Residual error in the reconstructed mock source intensity as a function of S/N. The pixel size is $0\farcs 0066$. The red dotted curves show the mean (over a bin width of $\varDelta (S/N)=1$) ratio of the 
residual error in mock source intensity in units of the $1\,\sigma $ nominal noise as a function of the ratio of the signal to the $1\,\sigma $ nominal noise at a given pixel. The black full curves show the best-fitted polynomial functions $(S/N)^\beta $ with $\beta=0.319\,(\textrm{left})$ and $\beta=0.383\,(\textrm{right})$ for the magnification and homogeneous weightings, respectively. }
\label{fig:error-mock}
\end{figure} 

The magnification weighting also gives a better fidelity of intensity
in the source plane. As shown in Figure \ref{fig:source-mock},
two bright spots with a separation of $0\farcs 014$, which is one third of the beam size were resolved for the magnification weighting. Thus 'super-resolution' was achieved. However, the homogeneous weighting failed to resolve the two spots. Since de-lensed PSFs obtained with the homogeneous weighting are larger than peak structures, large residual errors remain for peaks with high S/N. The residual errors were larger than the nominal errors for larger S/N because brighter pixels affect the neighboring pixels much larger than fainter pixels. Such effects were observed in both the weightings, but the difference was more prominent for the homogeneous weighting due to the sizes of the de-lensed PSFs. As shown in Figure \ref{fig:error-mock}, the residual error in the reconstructed mock source intensity shows a significant deviation from the best-fitted polynomial function in pixels with $S/N\sim 30$ in the homogeneous weighting. Thus, the mock residual errors support our assumption on errors in intensity difference between de-lensed images: Stronger S/N dependency for the homogeneous weighting than for the magnification weighting. 

\begin{table}
\hspace{-6.5cm}
\caption{Simulated mock lensing power spectra (see Appendix A) for 3 bins of angular wavenumbers centered at $l=0.74\times 10^6, 1.08\times 10^6, 1.30\times 10^6$. They were calculated using the fitted 7 low-to-intermediate, 11 intermediate-to-high, and 7 high frequency modes shown in Figure \ref{fig:fourier-modes}, respectively. The measured lensing power spectra within a range of $1\,\sigma $ error obtained with the magnification weighting and those with the homogeneous weighting are shown in the 4th and 8th rows, respectively.    }
\vspace{0.1cm}
\label{tab:2}
\setlength{\tabcolsep}{2pt}
\hspace{-2.1cm}
\scriptsize{
\begin{tabular}{lccccccccc}
\hline
\hline
   & $\varDelta_\psi[\textrm{arcsec}^2]$ & $\varDelta_\alpha[\textrm{arcsec}]$ &  $\varDelta_\kappa$ & $\varDelta_\psi[\textrm{arcsec}^2]$& $\varDelta_\alpha[\textrm{arcsec}]$&$\varDelta_\kappa$&$\varDelta_\psi[\textrm{arcsec}^2]$&$\varDelta_\alpha[\textrm{arcsec}]$ &$\varDelta_\kappa$
\\
\hline
$l  [10^6]$ & $0.74\pm 0.06$  & $0.74\pm 0.06$ & $0.74\pm 0.06$ &$1.08\pm 0.07$&$1.08\pm 0.07$ &$1.08\pm 0.07$ &$1.30\pm 0.08$&$1.30\pm 0.08$ & $1.30\pm 0.08$
\\
\hline
'true' values & $0.000644$ & $0.00214$ & $0.00362$  &$0.00101$ &$0.00526$ &$0.0138$ &$0.00111$ &$0.00688$ & $0.0217$
\\
\hline
magnification  & $0.00099$ & $0.00360$ & $0.00656$ & $0.000637$ &$0.00329$ &$0.0087$ &$0.000658$ &$0.00410$ & $0.0128$
\\
   & $\pm 0.00012$ & $\pm 0.00045$ & $\pm 0.00083$ & $\pm 0.00011$ & $\pm 0.00056$ &$\pm0.0015$ &$\pm0.00014$ &$\pm0.00092$ & $\pm0.0029$
\\ 
absolute error [$1\sigma$]   &$2.9$ & $3.3$ & $3.6$ &$3.5$ & $3.5$ & $3.4$ & $2.9$ & $3.0$ & $3.1$
\\ 
relative error [\%]   &$54$ & $68$ & $83$ & $37$ & $37$ & $53$ &$40$ & $41$ & $41$
\\ 
\hline
homogeneous  &  $0.000881$ & $0.00303$ & $0.00531$ &$0.000851$ &$0.00431$ &$0.0110$ & $0.000730$ & $0.00456$ & $0.0143$
\\
   & $\pm 0.00018$ & $\pm 0.00061$ & $\pm 0.00110$ & $\pm 0.00014$ & $\pm 0.00070$ & $\pm 0.0018 $ & $\pm 0.00017$ & $\pm 0.00104$ & $\pm 0.0033$
\\
absolute error [$1\sigma$]  &$1.3$ & $1.4$ & $1.5$ & $1.1$& $1.4$ & $1.6$ & $2.2$ & $2.3$ & $2.4$
\\ 
relative error [\%]   &$37$ & $42$ & $47$ & $16$& $18$ & $20$ & $34$ & $34$ & $34$
\\ 
\hline
\end{tabular}
}
\end{table}
In contrast, the homogeneous weighting gave much better results than 
the magnification weighting in reconstructing the lensing power spectra (Table \ref{tab:2}). 
The relative errors \footnote{$1\,\sigma$ errors are calculated by considering $>10^6$ sets of random Gaussian potentials that give $\varDelta \chi \le 1$ with the smoothing term smaller than that for the best-fitted model. The best-fitted potential was selected as the 'center' of the random Gaussian potentials. } between the mean value of the reconstructed power spectra are
$40$ to $80$ percent for the magnification weighting and $20$ to $50$ percent for the homogeneous weighting for 3 bins of angular wavenumbers centered at $l=0.74\times 10^6, 1.08\times 10^6, 1.30\times 10^6$. In other words, our results suggest that the magnification weighting results in significantly large systematic errors compared with the homogeneous weighting when estimating the powers. Since the homogeneous weighting weighs the multiple lensed images equally, it is probable that the information loss of fluctuations in regions beyond the lensed images and the bias in the estimated powers were reduced. We found that the relative errors in potential and astrometric shift perturbations are much smaller than those of convergence perturbation in intermediate scales regardless of weighting scheme.  
\begin{figure}
\hspace{0.cm}

\plotone{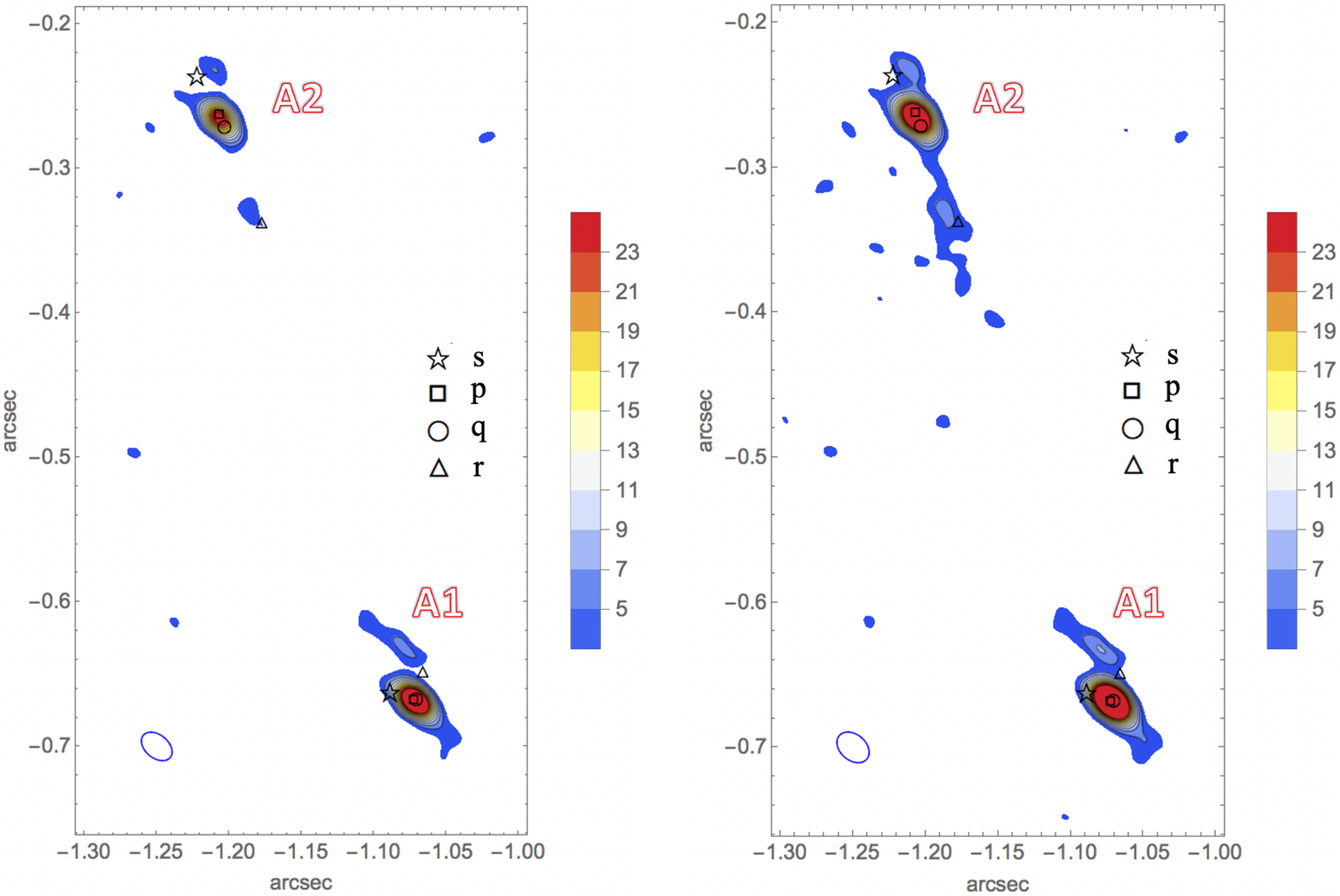}
\vspace{0.cm}
\caption{Zoomed up images of ALMA (Cycle 2 and Cycle 4) 0.88\,mm (Band 7 340\,GHz) continuum intensity of MG\,J0414+0434 showing A1 and A2. The contours and colors show intensity larger than $3\,\sigma$. The images were obtained from the data of Cycle 4 observations (left) and the combined data of Cycle 2 and Cycle 4 observations (right). The imaging was carried out with a Briggs weighting of $robust=-1$. The values in the legends show intensity in units of $1\,\sigma$. The contours start from a $3\,\sigma$ level and increase with a step of $2\,\sigma$. The small ellipses in the bottom left corners show the synthesized beam sizes $0\farcs024\times 0\farcs016$ with PA $50\fdg 4$ (left) and $0\farcs025\times 0\farcs018$ with PA $49\fdg 5$ (right). The $1\,\sigma$ errors are $94\,\mu \textrm{Jy}\, \textrm{beam}^{-1}$ (left) and $66\,\mu \textrm{Jy}\, \textrm{beam}^{-1}$ (right). Boxes, circles, triangles, and stars correspond to jet components p, q, r, and s at 5\,GHz, respectively. The coordinates are J2000 and centered at the centroid of the primary lensing galaxy G.   }
\label{fig:alma-lens-image}
\end{figure} 

\begin{figure}
\hspace*{-0.6cm}
\epsscale{1.2}
\plotone{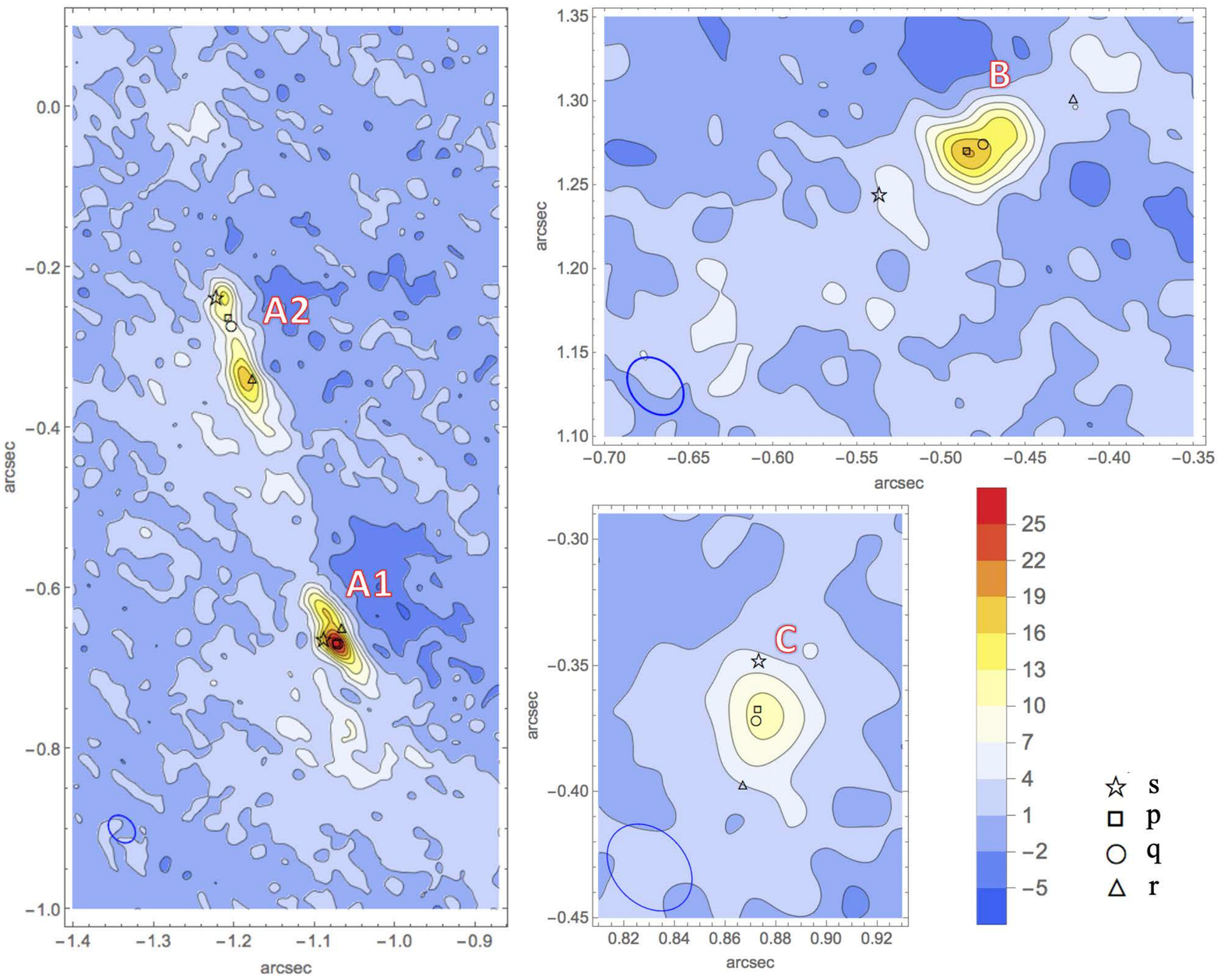}
\vspace{0.cm}
\caption{Peak-subtracted ALMA (Cycle 2 and Cycle 4) 0.88\,mm (Band 7, 340\,GHz) continuum intensity of MG\,J0414+0434. Point-like sources at radio-jet/core components p and q were subtracted off with parameters $c_{\textrm{p}}=0.626, c_{\textrm{q}}=0.324$. The contours and colors show intensity. The imaging was carried out with a Briggs weighting of $robust=0$. The ticks in the legend show intensity in units of the background noise of $1\,\sigma=30\,\mu \textrm{Jy}\,\textrm{beam}^{-1}$. The contours start from a $-5\,\sigma$ level and increase with a step of $3\,\sigma$. The small ellipses in the bottom left corners in each panel show the synthesized beam size $0\farcs038\times 0\farcs029$ with PA $42\fdg 5$. The symbols represent p, q, r, and s at 5\,GHz as in Figure \ref{fig:alma-lens-image}. The coordinates are J2000 centered at the centroid of the primary lensing galaxy G.   }
\label{fig:alma-lens-image-core-subtracted}
\end{figure} 

% Parameters from chisqfit1N6
\begin{table*}
\hspace{-2cm}
\caption{Fiducial smooth model parameters, $\chi^2$, and flux ratios in models best-fitted to the MIR flux ratios, HST positions of galaxies, and VLBA positions of jet component q. }
\label{tab:3}
\hspace{-2.4cm}
\setlength{\tabcolsep}{2pt}
{\footnotesize
\begin{tabular}{lccccccccccc}
\hline
\hline
Type  &  $b_\textrm{G}$  & ($y_{s1}, y_{s2})$ & $e_\textrm{G}$ & $\phi_{\textrm{G}}$
& $\gamma$ & $\phi_{\gamma}$ & ($x_{\textrm{G}1}, x_{\textrm{G}2})$ &$b_\textrm{X}$ & ($x_{\textrm{X}1}, x_{\textrm{X}2})$ &$r_\textrm{X}$ 
\\
\hline
A & $1\farcs 094$ & $(-0\farcs 0562, 0\farcs2654)$ &0.307 &$-87\fdg9$ & 0.0918 & 
$47\fdg5$ & $(0\farcs 006,-0\farcs 009)$  &$0\farcs 219$  & $(0\farcs 445,1\farcs 483)$ & $0\farcs 018$ 
\\ 
\hline
B & $1\farcs 107$ & $(-0\farcs 0981, 0\farcs 2254)$ &0.280 &$-87\fdg9$ 
& 0.0790 & $47\fdg 5 $ & $(-0\farcs 001,-0\farcs 002)$  &$0\farcs 160$  & 
$(0\farcs 316,1\farcs 490)$ & $0\farcs 019$ 
\\
\hline
\hline
Type & $z_\textrm{Y}$ & $b_\textrm{Y}$ & $e_\textrm{Y}$ & $\phi_{\textrm{Y}}$ & ($x_{\textrm{Y}1}, x_{\textrm{Y}2})$ & & &  &  &  
\\
\hline
A & & & & & &  &  & &  & 
\\
\hline
B & $0.661$ & $0\farcs 028$ & 0.604 &$79\fdg 5$ &$(-1\farcs 850,-0\farcs 042)$ &  & &  & & 
\\
\hline
Type & $\chi^2_{\tr{flux}}$ & $\chi^2_{\tr{posG}}$ &  $\chi^2_{\tr{posX}}$ & $\chi^2_{\tr{posY}}$ & $\chi^2_{\tr{posq}}$  & $\chi^2_{e} $ & $\chi^2_{\tr{tot}}/$dof & A2/A1 & B/A1 & C/A1 &
\\
\hline
A & 24.09 & 4.58 & 0.45 & &  6.58 & & 35.7/3 & 1.007 & 0.350 & 0.175 & 
\\
\hline
B & 7.014 & 0.264 & 0.555 & 0.645 &  0.018 & 0.427 & 8.92/1 & 0.928 & 0.358 & 0.174 & 
\\
\hline 

\end{tabular}
}
\flushleft{Note. In Type A models, object Y is not modeled explicitly but in Type B models, Y is modeled explicitly in the smooth model. $b_\textrm{G}$ is the effective Einstein radius of the primary lensing galaxy G, ($y_{s1}, y_{s2}$) is a set of source coordinates of the jet component q, $e_\textrm{G}$ is the ellipticity of G, $\phi_{\textrm{G}}$ is the direction of the major axis of G, $\gamma$ is the amplitude of the external shear, $\phi_{\gamma}$ is the direction of the external shear, ($x_{\textrm{G}1}, x_{\textrm{G}2}$) is a set of coordinates of the centroid of G, $b_\textrm{X}$ is the Einstein radius of object X, ($x_{\textrm{X}1}, x_{\textrm{X}2}$) is a set of coordinates of the centroid of X, and $r_\textrm{X}$ is the assumed core radius of X (see \citet{inoue2017} for details). For simplicity, $r_\textrm{X}$ is fixed in fitting.
$z_\textrm{Y}$ is the redshift of object Y, $b_\textrm{Y}$ is the effective Einstein radius of object Y, 
($x_{\textrm{Y}1}, x_{\textrm{Y}2}$) is a set of coordinates of the peak position of Y, $e_\textrm{Y}$ is the ellipticity of Y, $\phi_{\textrm{Y}}$ is the direction of the major axis of Y.  $\chi^2_{\tr{tot}}$ is the sum of contributions from the flux ratios $\chi^2_{\tr{flux}}$,  the positions of the lensed images of the jet component q $\chi^2_{\tr{posq}}$, lensing galaxy
G $\chi^2_{\tr{posG}}$, and object X $\chi^2_{\tr{posX}}$. The coordinates are
centered at the centroid of G (CASTLES database) \citep{falco1997}. The assumed errors are $2\,\tr{mas}$ for the VLBA positions, $5\,\tr{mas}$, $100 \,\tr{mas}$, and $100 \,\tr{mas}$
for the HST positions of the centroids of G, X, and ALMA position of Y. Here the error in the HST position of G includes the systematic difference of $\sim 2\,\tr{mas}$ between the HST and VLBA maps.  We assume that the central position of object Y is $(-1\farcs 865, -0\farcs116)$ (see Section \ref{sec:6.5}). }
\label{tab:best-fitted smooth model parameters}
\end{table*}

\section{Results of ALMA Observations}
\subsection{Smooth Model}
\label{sec:6.1}
To obtain background lensing models from our ALMA data, we used lensed images of a compact radio component q in the VLBA map at 5\,GHz \citep{trotter2000} instead of the HST position of the quasar core. In a continuum map of the Cycle 4 observation in which imaging was performed with ({\it{robust}}$\,=-1$), the positions of lensed images of q are shown with circles (Figure\ref{fig:alma-lens-image}). 

The reason is as follows: First, the accuracy of the VLBA positions ($\lesssim 2\,\textrm{mas}$) are better than that of the HST positions ($\sim 3\,\textrm{mas}$). Second, as the position of the radio emission at 5\, GHz is very close to the OPT/NIR emission of the quasar core ($<2\,$mas), we can consider component q as the quasar core rather than a jet \citep{inoue2020}. 

In order to match the VLBA positions to the HST positions, we translated and rotated the coordinates of the VLBA map to best-fit the positions of lensed q to the HST positions in the OPT/NIR band. The obtained rotation angle is $1.76^\circ$ (East of North). The residual differences between the lensed q and the HST positions, A1, A2, B, and C are $0.0, 2.4, 2.0$, and $1.3\,\textrm{mas}$.  

Then our ALMA maps were rotated by the same angle (assuming a perfect alignment with the coordinates in the VLBA image) and the position of a local peak in image A1 was placed on the brightest image A1 of the lensed gravitational center of p and q in the corresponding VLBA data (see Figure\ref{fig:alma-lens-image}). The differences between local peaks in the ALMA image and the gravitational centers of p and q for A2, B, and C were 2.3, 4.0, and  $1.4\,\textrm{mas}$, which are smaller than the pixel size of $5\,$mas\footnote{The pixel size is approximately equal to the smallest observable astrometric shift in the source plane, which is given by the synthesized beam size $\sim 0\farcs 04$ divided by a mean magnification factor of the quadruple images \citep{inoue2005a,inoue2005b}.} used in the continuum maps. We assumed that the errors in the VLBA positions are $2\,\textrm{mas}$, which is the mean size of the synthesized beam. Although the positions of emission at 340\,GHz may slightly differ from those at 5\,GHz, the differences in the positions in the lens plane were found to be smaller than $4\,$mas. The differences in the source plane are expected to be much smaller due to demagnification. 

As conducted in our mock analysis, we used the HST positions of the centroids of the primary lensing galaxy G, object X observed in the OPT/NIR band, and the MIR flux ratios observed with the Subaru and Keck telescopes \citep{minezaki2009, macleod2013}. Considering a possible misalignment of $\sim 2\,$mas, the positional error of the centroid of G was assumed to be $5\,$mas. The positional error of X and Y were assumed to be $0\farcs1$. 

To obtain the best-fitted Type B model, we also used the central position of object Y observed with ALMA (see Section \ref{sec:6.5} for details), and assumed observational constraints for the ellipticities of G and Y. For G, the mean ellipticity of isophotes in the HST $I$ band was measured as $e_\textrm{G}=0.20\pm 0.02$ \citep{falco1997}. Taking into account the misalignment between the baryonic and dark matter components and halo flattening, we adopt a conservative error value $\delta e_\textrm{G}=0.20$ for the ellipticity of the projected total matter (baryon + dark matter) in the halo of G. As the expected ellipticity of the halo of Y, we adopt a mean ellipticity $\sim 0.5$ of projected dark matter halo measured in cluster scales \citep{kathinka2009, oguri2010, okabe2020}. We assume a conservative error value $\delta e_\textrm{Y}=0.20$ for the ellipticity of Y. To include these constraints in fitting, we add a term        
\BE
\chi^2_e=\biggl(\frac{0.2-e_\textrm{G}}{0.2} \biggr)^2+\biggl(\frac{0.5-e_\textrm{Y}}{0.2} \biggr)^2,
\EE
to $\chi^2_{\textrm{tot}}$ in modelling Type B. For simplicity, we assume the same ratio between the core size $r_X$ of X and the effective Einstein radius $b_\textrm{G}$ of G as used in Type A. Therefore, $r_\textrm{X}$ is fixed in Type B models. We also parameterized the redshift $z_\textrm{Y}$ of object Y. The best-fitted parameters of the smooth models of Type A and B are shown in Table \ref{tab:3}.

\subsection{Source Plane Fit to ALMA image}
\label{sec:6.2}
To subtract lensed images of bright core components p and q from a continuum image, we used the best-fitted elliptical Gaussian beam obtained from CASA. We considered p and q as two point-like sources, whose intensities are described by an elliptical Gaussian beam synthesized beam) multiplied by a constant. The intensities $c_\textrm{p}$ and $c_\textrm{q}$ (in units of the observed peak intensity of q at the fitted position) at the fitted positions of p and q in image A2 were selected as free parameters but constrained so as not to become negative. $c_\textrm{p}+c_\textrm{q}$ was also constrained to be 0.95\footnote{Although image A1 is slightly brighter than image A2, the angular separation between p and q in image A1 is far smaller than the separation in image A2 (Figure\ref{fig:alma-lens-image-core-subtracted}). That is why we chose A2 as the fiducial image.}. The 5 percent decrease accounts for emission from extended dust regions (Figure \ref{fig:alma-lens-image-core-subtracted}). Note that $c_\textrm{p}$ and $c_\textrm{q}$ were not fit independently as our numerical analysis showed that such fits tend to give a solution with a negative flux. 

The intensities at the positions of lensed p and q in image A1, B, and C were given by the MIR flux ratios of the quasar core and intensities at the positions of lensed p and q in image A2. If we allow independent change in both $c_\textrm{p}$ and $c_\textrm{q}$, $\chi^2$ minimization would give a solution with a negative hole in the peak-subtracted image due to over subtraction. Therefore, we fixed $c_\textrm{p}+c_\textrm{q}$ to be a constant.   

We constructed multi-scale meshes in the source plane in a similar manner to our mock analysis but the mesh sizes was fixed to be $6.66$\,mas. The mesh size was determined from the fluctuation scale of the de-lensed image in the brightest region.

We used only pixels with $>3.8\,\sigma$ in the source plane. In order to estimate possible enhancement in the errors in the source plane due to sidelobes, we subtracted off fluxes larger than $4\,\sigma$ in the lens plane and carried out random translations of the lens plane around the lensed image. Using the data set, we found that the absolute values of the non-diagonal components in the covariance matrix of the difference in the weighted sum of the de-lensed source intensity is $\lesssim 10$ per cent of the diagonal components. Therefore, the effect of spatial correlation of errors in the pixels in the source plane is expected to be small.  

Before performing $\chi^2$ analysis, we adjusted $\beta$ to satisfy the condition $\chi^2/\textrm{dof} \sim 3$ for an unperturbed model. As a fiducial value, we set $c_\textrm{q}=0.20$ and $c_\textrm{p}=0.75$. Then we minimized $\chi^2$ using de-lensed images reconstructed with the magnification or homogeneous weighting as was conducted in the mock analysis. 
\begin{table*}
\hspace{-2cm}
\caption{Results of source plane fits : assumed parameters, minimized $\chi^2$, flux ratios, and rms perturbations in models best-fitted to the ALMA data, MIR flux ratios, HST positions of galaxies, and VLBA positions of jet component q. }
\hspace{-3.2cm}
\setlength{\tabcolsep}{2pt}
{\scriptsize
\begin{tabular}{llccccccccccccccccccl}
\hline
\hline
Type &  & $N^2$ & $L$ & $\beta$ & $\delta \psi_0$ & $\delta \alpha_0$ &  $\delta \kappa_0$ & $c_{\textrm{q}}$&    $\chi^2/\textrm{dof}$ & $N_\textrm{p}$ & $n_\textrm{obs}$ & dof & A2/A1 & B/A1 & C/A1 & $\chi^2_{\textrm{pos}}/\textrm{dof}_{\textrm{pos}}$ & $\textrm{dof}_{\textrm{pos}}$ &  $\delta \kappa_{\textrm{rms}}^q$ & $\delta \gamma_{\textrm{rms}}^q$
\\
\hline
A & mag. & 36 & $3\farcs6$ & $0.67$ &  $0.0014$ & $0.0053$ & $0.0090$ & $0.324$ & $0.80(3.0)$ & $46$ &7  & $13$ & 0.943 & 0.374 & 0.146  &0.67(0.73) & 24  & 0.015 & 0.012
\\
 \hline
A & mag. & 36 & $4\farcs2$  & $0.67$ &  $0.00096$ & $0.0060$ & $0.0057$ & $0.333$ & $0.83(3.0)$ & $46$& 7 & $13$ & 0.905 & 0.349 & 0.153 & 0.68(0.73) & 24 & 0.011 & 0.010
\\
\hline  
A & hom.  & 36 & $3\farcs6$ & $0.77$ & $0.00065$ & $0.0085$ & $0.0125$ & $0.452$ & $1.2(2.8)$ & $51$ & 7  & $18$ & 0.958 & 0.346 & 0.131 &1.0(0.73) & 24 & 0.018 & 0.011
\\
\hline
A & hom. & 36 & $4\farcs2$ & $0.77$ &  $0.00075$ & $0.0065$ & $0.0340$ & $0.400$ & $1.2(2.8)$ & $51$ &7 & $18$ &0.952 & 0.369 & 0.132 &1.0(0.73) & 24 & 0.025 & 0.007
\\
\hline
B & mag. & 36 & $3\farcs6$ & $0.48$ &  $0.00070$ & $0.0064$ & $0.0058$ & $0.229$ & $1.0(2.4)$ & $48$ &7 & $15$ &0.924 & 0.354 & 0.152 &0.28(0.32) & 24 & 0.010 & 0.0076
\\
\hline
B & mag. & 36 & $4\farcs2$ & $0.48$ &  $0.00053$ & $0.0053$ & $0.0070$ & $0.134$ & $1.0(2.4)$ & $48$ &7 & $15$ &0.918 & 0.359 & 0.152 &0.29(0.32) & 24 & 0.013 & 0.0091
\\
\hline
B & hom. & 36 & $3\farcs6$ & $0.77$ &  $0.00089$ & $0.0059$ & $0.0066$ & $0.306$ & $1.0(1.7)$ & $57$ &7 & $24$ &0.943 & 0.339 & 0.131 &0.65(0.32) & 24 & 0.019 & 0.011
\\
\hline
B & hom. & 36 & $4\farcs2$ & $0.77$ &  $0.00080$ & $0.0050$ & $0.0130$ & $0.080$ & $1.0(1.7)$ & $57$ &7 & $24$ &0.894 & 0.344 & 0.126 &0.60(0.32) & 24 & 0.026 & 0.011
\\
\hline
MIR &  & & & & & & & & & & & & $0.919$ & $0.347$ 
& $0.139$ & &  & &
\\
&  & & & & & & & & & & & & $\pm 0.021$ & $\pm 0.013$ 
& $\pm 0.014 $ & & &  &

\\
\hline
\label{tab:parameters-for-best-fit}
\end{tabular}
}
\vspace{-0.8cm}
\flushleft{Note. 'mag.' and 'hom.' represent the magnification and  
homogeneous weighting, respectively. $N^2$ is the number of mode functions. $L$ is the side length of a square at which the Dirichlet condition is imposed. $\beta$ is the power index of the expected error as the function of signal in the source plane. The units of smoothing parameters $\delta \psi_0$ and $\delta \alpha_0$ are $\textrm{arcsec}^2$ and $\textrm{arcsec}$, respectively. $c_\textrm{q}$ is the ratio of the flux of $q$ to the peak flux at the lensed image A2 of $q$. $\chi^2/\textrm{dof}$ is the reduced $\chi^2$ for the ALMA image ($\textit{robust}=0$), VLBA positions of q, and MIR flux ratios. $N_\textrm{p}$ is the total number of pixels. $n_\textrm{obs}$ is the total number of constraints for the relative angular distance ($=4$) and the MIR flux ratios ($=3$) of the lensed images of the quasar core (assumed to be q). $\chi^2_{\textrm{pos}}/\textrm{dof}_{\textrm{pos}}$ is the reduced $\chi^2$ for the positions of four VLBA jet components p, q, r, and s in which only the source positions are adjusted while the Fourier modes are fixed. Numbers in parentheses are the values for the corresponding unperturbed smooth model. The last column shows the observed MIR flux ratios \citep{minezaki2009, macleod2013}. }
\end{table*}

\begin{figure}
\vspace*{-1.5cm}
\epsscale{0.7}
\plotone{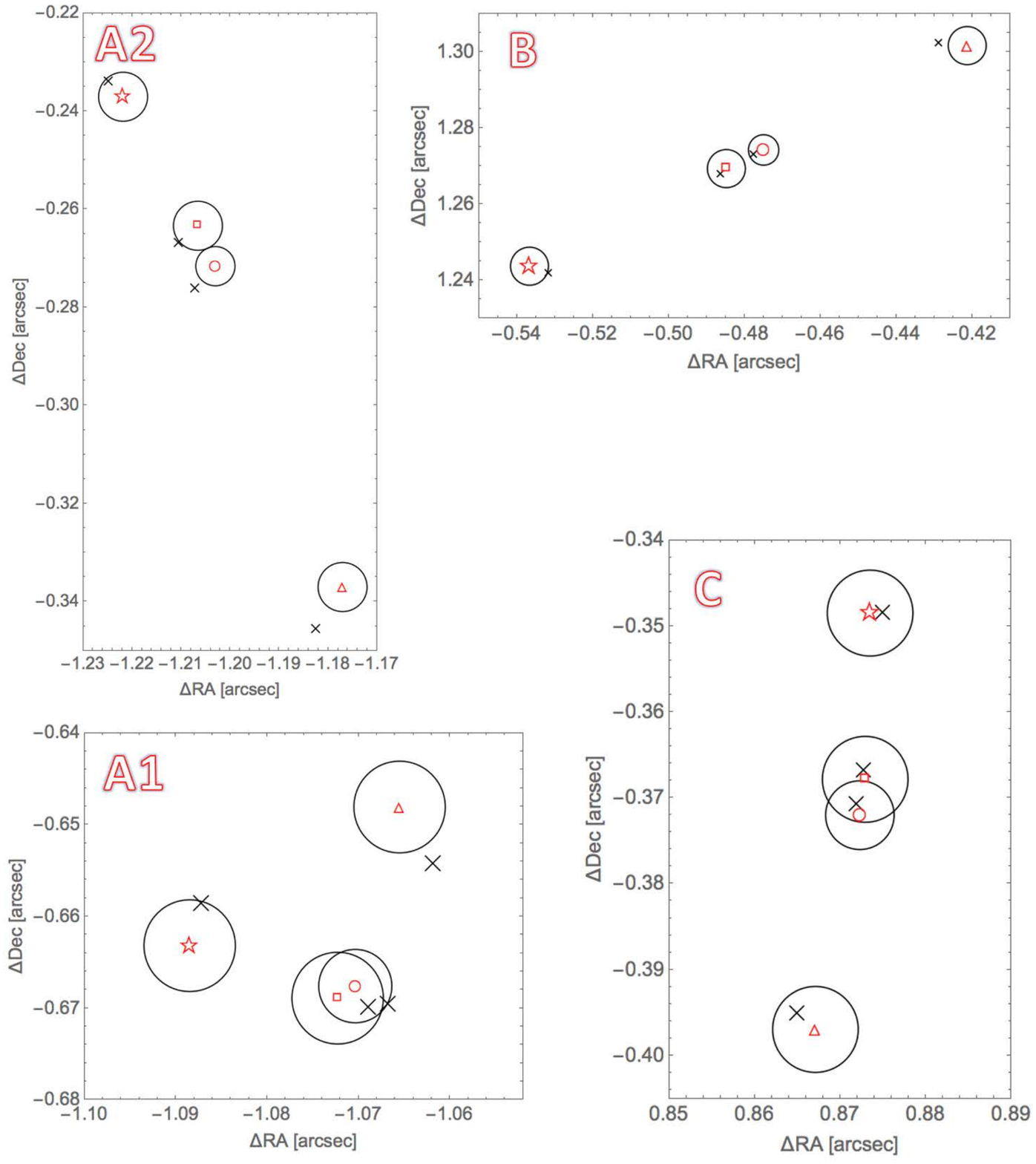}
\vspace*{-1.5cm}
\caption{The best-fitted positions of four VLBA jet components in the lens plane using the ALMA data, MIR flux ratios, HST positions of galaxies, and VLBA positions of lensed jet component q. The red symbols represent the positions of p, q, r, and s observed at 5\, GHz as in Figure \ref{fig:alma-lens-image} and circles indicate their errors. The black X's represent the best-fitted positions in a Type A model obtained with the magnification weighting with $L=3\farcs 6$ and $c_\textrm{q}+c_\textrm{p}=0.95$. The coordinates are J2000 centered at the centroid of the primary lensing galaxy G. North is up and east is left.}
\label{fig:VLBA-fit}
\end{figure} 

\begin{figure}
\vspace*{-2.cm}
\hspace*{-0.4cm}
\epsscale{1.2}
\plotone{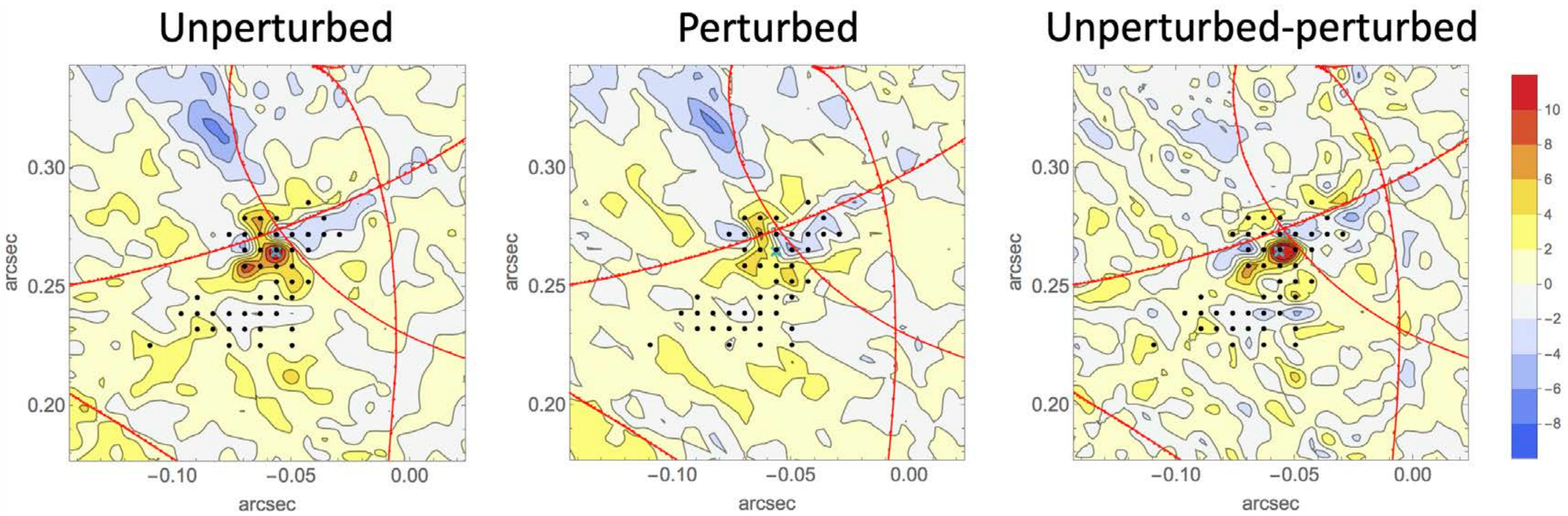}
\vspace{-3.cm}
\caption{Difference in de-lensed mean peak-subtracted images in the source plane. The left panel shows the difference in intensity of de-lensed mean peak-subtracted images with the same parity in the unperturbed smooth model. The middle panel shows the difference in intensity of de-lensed mean peak-subtracted images with the same parity in a Type A model obtained with the magnification weighting, $L=3\farcs 6$, and $c_\textrm{q}+c_\textrm{p}=0.95$. The right panel shows the change in the difference in intensity between the unperturbed (left) and perturbed (middle) models. The contours begin from a $-8\,\sigma$ level and increase with a step of $2\,\sigma$. The black points show the positions of meshes. The red curves show the caustics. The green X's show the position of the best-fitted core component q.  }
\label{fig:dif-source-plane}
\end{figure} 

\begin{figure}
\vspace*{-3.5cm}
\epsscale{1.2}
\plotone{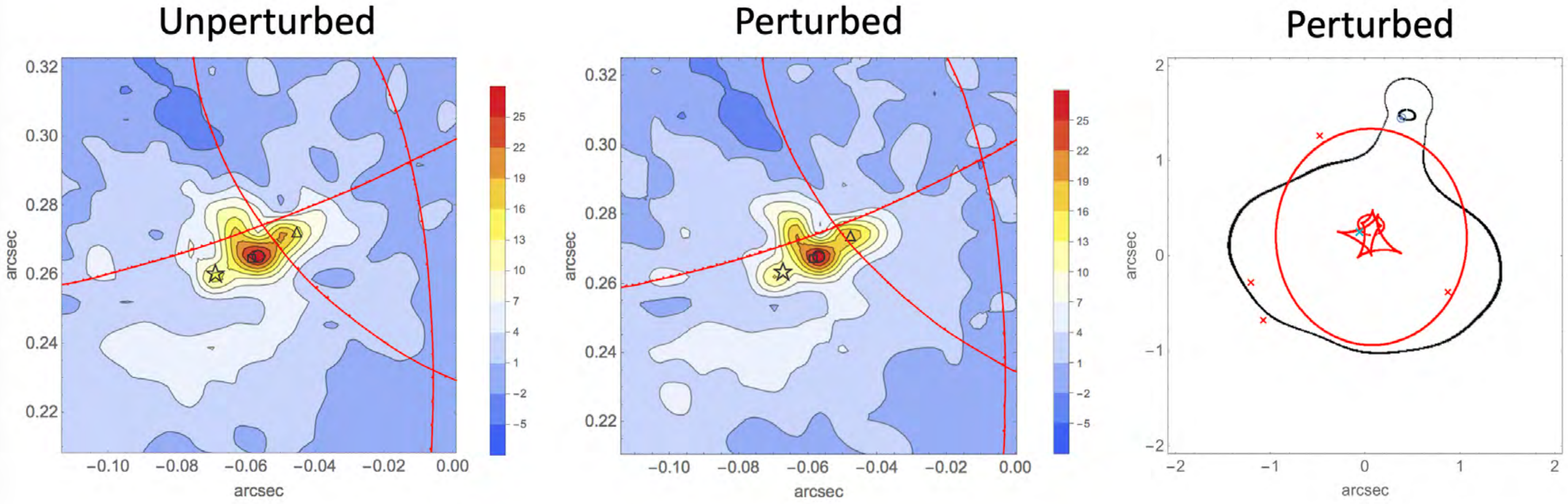}
\vspace*{-3.5cm}
\caption{De-lensed peak-subtracted source images for unperturbed and perturbed models, and caustics overlaid with critical curves in the perturbed model.  The intensity of de-lensed peak-subtracted images is shown in contours  
for an unperturbed Type A model (left) and a perturbed (middle) Type A model obtained with the magnification weighting with $L=3\farcs 6$ and $c_\textrm{q}+c_\textrm{p}=0.95$. The red curves in all the panels 
show the caustics in the corresponding model. In the left and middle panels, the black symbols represent the positions of p, q, r, and s observed at 5\, GHz as in Figure \ref{fig:alma-lens-image}. In the right panel, the caustics and critical curves in the perturbed best-fitted model are shown: The red curves are the caustics, the black curves are the critical curve, the blue circled dot is the centroid of object X, the red X's are the best-fit positions of q in the lens plane and the green X is the best-fitted position of q in the source plane. }
\label{fig:model-best-fit}
\end{figure} 

\begin{figure}
\hspace*{-0.5cm}
\epsscale{0.9}
\plotone{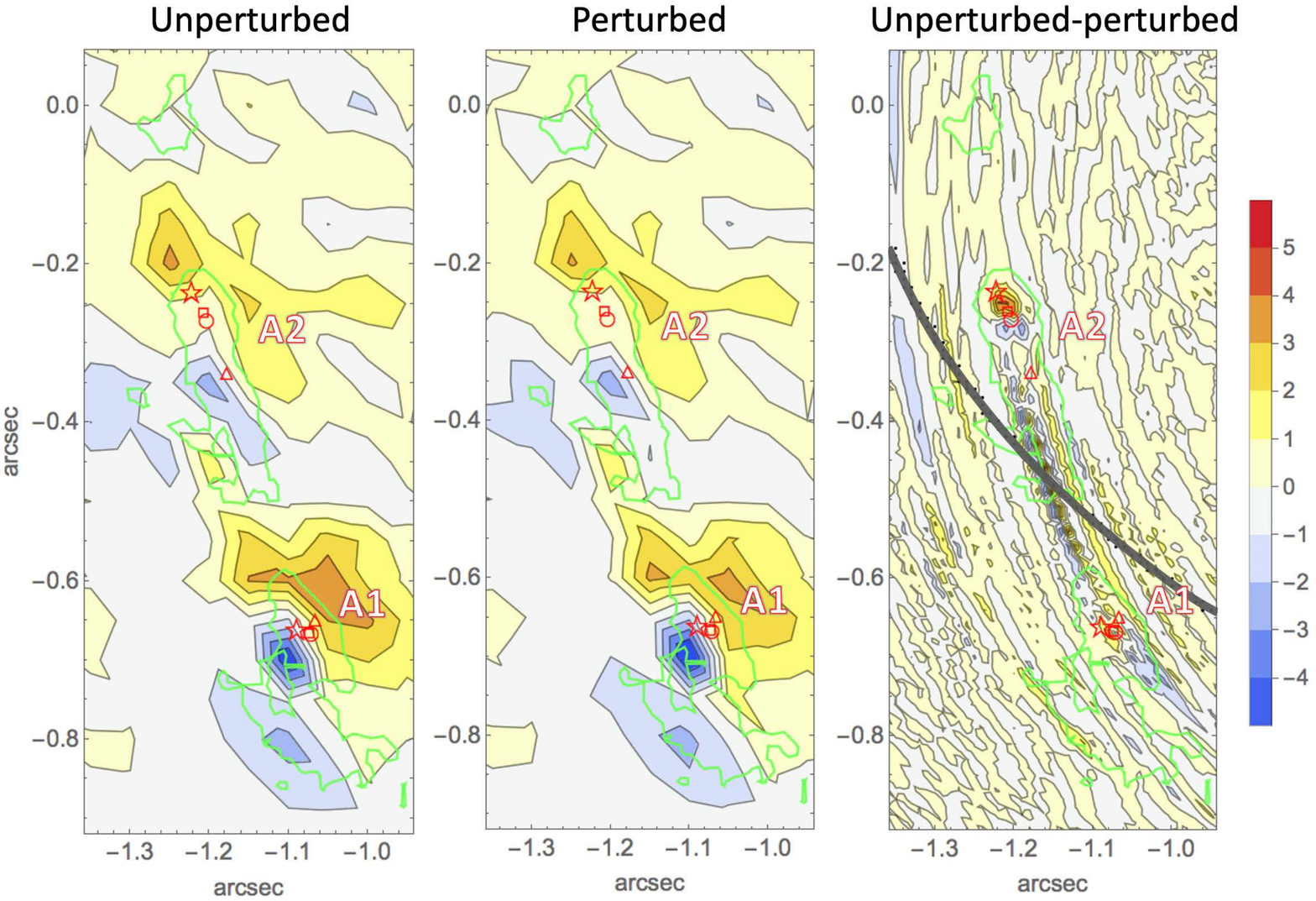}
\vspace{0.cm}
\caption{Image residual in the lens plane. The left panel shows the contour map of the residual intensity: smoothed (corresponding to a pixel size of $0\farcs 05$ ) intensity of an unperturbed background Type A model is subtracted by the intensity of the ALMA image with $\textit{robust}=0$. The middle panel shows the contour map of the residual intensity: smoothed intensity of a perturbed Type A model obtained with the magnification weighting with $L=3\farcs 6$ and $c_\textrm{q}+c_\textrm{p}=0.95$ is subtracted by the intensity of the ALMA image with $\textit{robust}=0$. The right panel shows the intensity of the unperturbed background Type A model subtracted by the intensity of the perturbed Type A model. The red symbols represent the positions of p, q, r, and s observed at 5\, GHz as in Figure \ref{fig:alma-lens-image}.  The contours start from a $-4\,\sigma$ level and increase with a step of $1\,\sigma$. Green curves denote the boundary inside which the signal is $>3\,\sigma$. The thick black curve in the right panel shows the critical curve.  }
\label{fig:dif-lens-plane}
\end{figure}

\begin{figure}
\hspace*{-0.8cm}
\epsscale{1.1}
\plotone{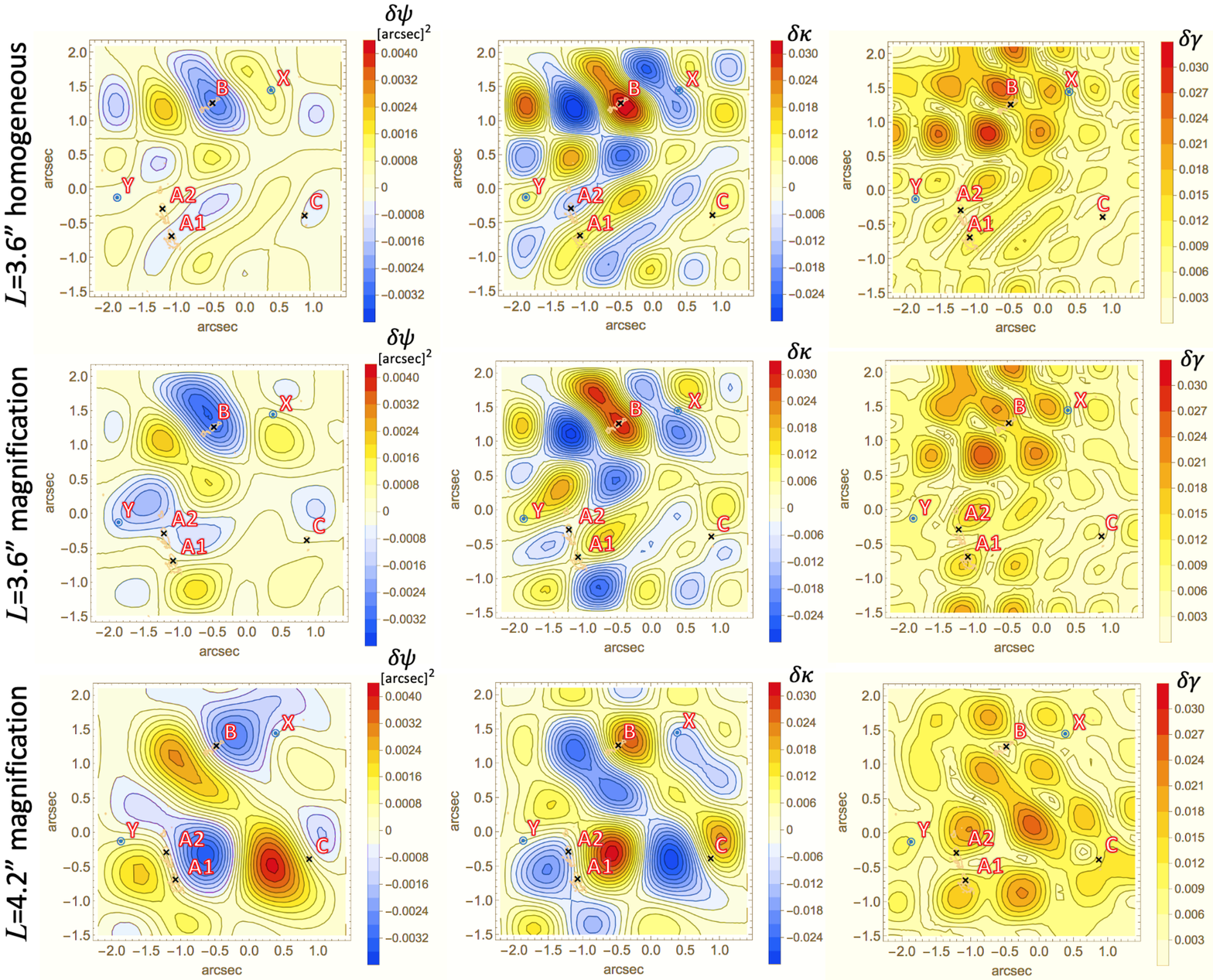}
\vspace{-0.2cm}
\caption{Contour maps of perturbations (36 modes) in Type A models best-fitted to the ALMA data, MIR flux ratios, and VLBA positions of a core component q. The plotted maps show the potential perturbation $\delta \psi$, convergence perturbation $\delta \kappa $, and shear perturbation $\delta \gamma$ reconstructed with $L=3\farcs6$ and the homogeneous weighting (top column), $L=3\farcs6$ and the magnification weighting (middle column), $L=4\farcs2$ and the magnification weighting (bottom column). Light brown curves denote the boundaries of regions in which the signals are larger than $3\,\sigma$. The centers of the square region are $(-0.4,0.3)$ (top),  $(-0.4,0.3)$ (middle),  and $(-0.7,0.3)$, (bottom). The coordinates are J2000 centered at the centroid of the primary lensing galaxy G. The slight difference between the best-fitted potential wells around A1, A2, and B images of the quasar core reconstructed with $L=3\farcs6$ and the magnification weighting (middle left) and that with $L=4\farcs2$ and the magnification weighting (bottom left) may be due to the slight difference in the angular scales of the Fourier modes (see also Figure \ref{fig:powerspectrum-result.pdf}). }
\label{fig:real-space-best-fit-full}
\end{figure} 

\begin{figure}
\hspace*{-0.8cm}
\epsscale{1.1}
\plotone{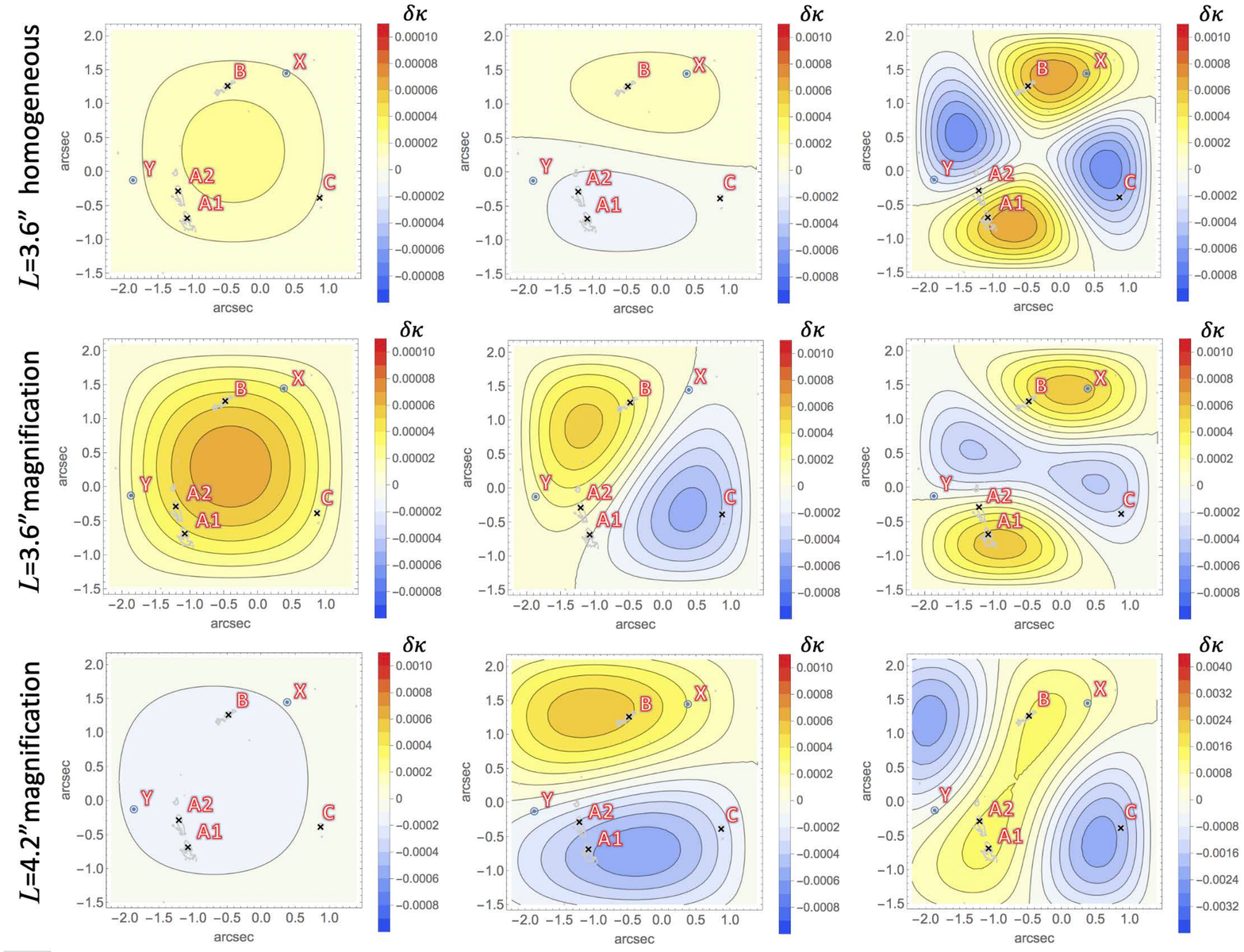}
\vspace{-0.2cm}
\caption{Contour maps of low frequency modes of the best-fitted convergence perturbations for the 3 Type A models shown in Figure \ref{fig:real-space-best-fit-full}. The plotted maps show the lowest mode (monopole) (left), the sum of the second and third lowest 2 modes (dipole) (middle), and the sum of the fourth to sixth lowest 3 modes (quadrupole) (right). The symbols are the same as Figure \ref{fig:real-space-best-fit-full}.   }
\label{fig:real-space-best-fit-low}
\end{figure} 

\begin{figure}
\vspace*{-0.8cm}
\hspace*{-0.5cm}
\epsscale{0.8}
\plotone{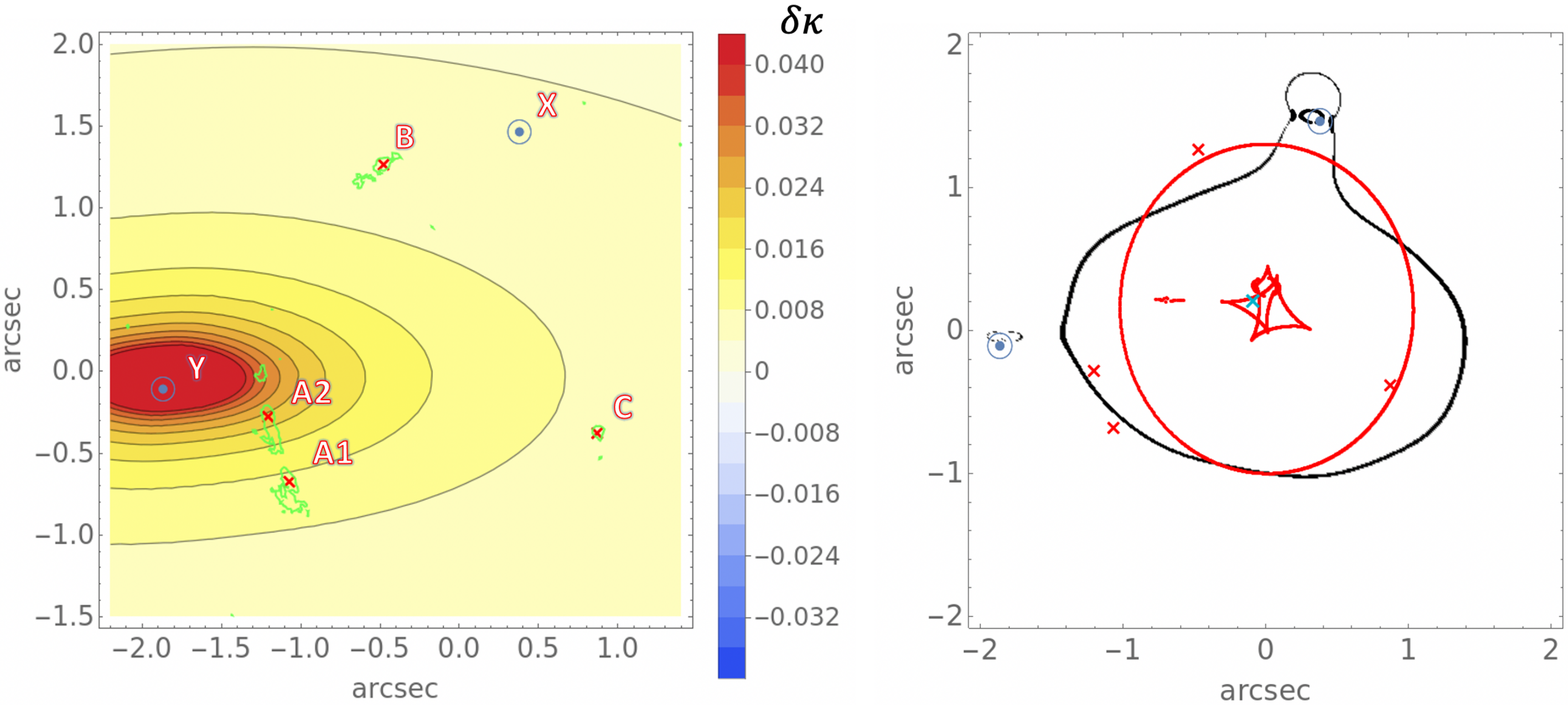}
\vspace*{-1.2cm}
\caption{Convergence of object Y (left) and caustics overlaid with critical curves in a perturbed Type B model in which object Y is explicitly modeled (right). The left panel shows a contour map of the convergence of a modeled object Y in the background Type B model. Green curves denote the boundary inside which the signal is $> 3\, \sigma$. The right panel shows the caustics, cuts (black), and critical curves (red) in a perturbed Type B model with the magnification weight and $L=3\farcs6$. The red X's are the best-fitted positions of q in the lens plane and the green X is the best-fitted position of q in the source plane. Blue circled dots are the centroid of object X and the position of the peak intensity of object Y.     }
\label{fig:convergence-objY.pdf}
\end{figure} 

\begin{figure}
\vspace*{-1cm}
\hspace*{0.cm}
\epsscale{0.7}
\plotone{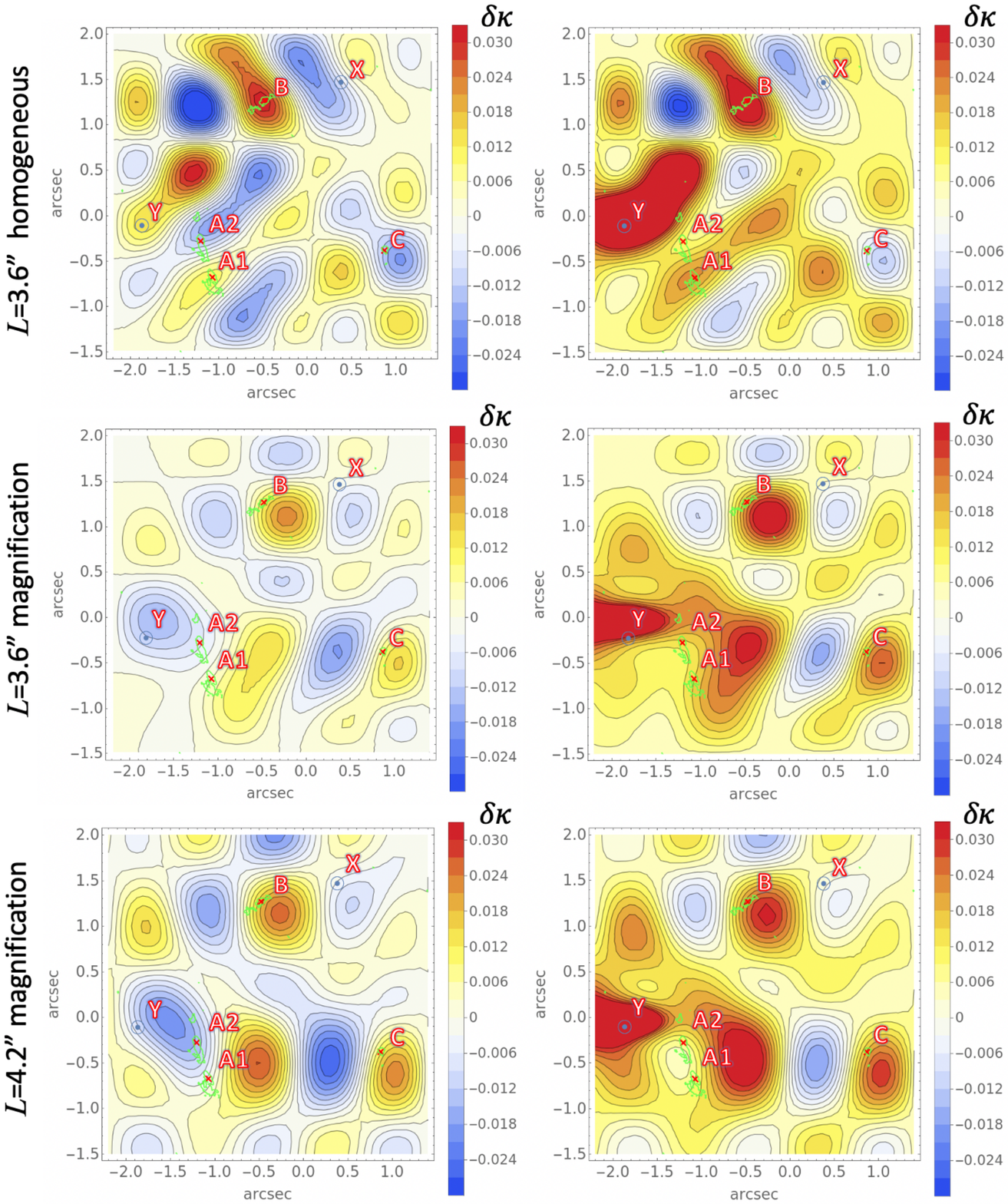}
\vspace*{-1.2cm}
\caption{Contour maps of convergence perturbations $\delta \kappa $ (36 modes) in Type B models in which object Y is explicitly modeled. The left row shows the convergence perturbations in which contribution from object Y is subtracted. The right row represents the total convergence perturbations (object Y plus Fourier modes). The model was obtained with the source plane fits with $L=3\farcs6$ and the homogeneous weighting (top column), $L=3\farcs6$ and the magnification weighting (middle column), $L=4\farcs2$ and the homogeneous weighting (bottom column). Green curves denote the boundaries of regions in which the signals are larger than $3\,\sigma$. The centers of the square region that determines the Dirichlet boundary condition are $(-0.4,0.3)$ (top),  $(-0.4,0.3)$ (middle),  and $(-0.7,0.3)$, (bottom). The coordinates are J2000 centered at the centroid of the primary lensing galaxy G.}
\label{fig:object-Y-perturbation.pdf}
\end{figure} 

\begin{table*}
\hspace{-4cm}
\caption{Model parameters and the lensing power spectra obtained from the 22 intermediate frequency modes. They were obtained from the 
36 modes that were best-fitted to the ALMA data, MIR flux ratios, HST positions of galaxies, and VLBA positions of a jet component q.}
\vspace{0.1cm}
\hspace{-3.32cm}
\setlength{\tabcolsep}{2pt}
\footnotesize{
\begin{tabular}{llcccccccccc}
\hline
\hline
Type & weighting & $N^2$ & $L$ & $l[10^5]$ & $\delta \theta $ & $\varDelta_\psi[\textrm{arcsec}^2]$ & $\varDelta_\alpha [\textrm{arcsec}]$ &  $\varDelta_\kappa$ & $\chi^2/\textrm{dof}$ &  $R_s/\textrm{dof}$ & dof
\\
\hline
A & hom. & 36 & $3\farcs6$ &$ 8.6\pm 2.6$ & $1\farcs5^{+0\farcs 64}_{-0\farcs 35}$  & $0.00070\pm0.00004$ & $0.0033\pm0.0002$ & $0.0079\pm0.0005$  & 1.2 &0.12 & 18
\\
\hline
A & hom. & 36 & $4\farcs2$ & $ 7.4 \pm 2.2$ &  $1\farcs 75^{+0\farcs 75}_{-0\farcs 40}$  &  $0.00072\pm 0.00003$ & $0.0030\pm0.0002$ & $0.0066\pm0.0003$ & 1.2 & 0.12 &18
\\
\hline  
\label{tab:power-parameters-ALMA}
\end{tabular}
}
\vspace{-0.7cm}
\small{
\flushleft{Note.  $N^2$ is the number of mode functions. $L$ is the side length of a square at which the Dirichlet condition is imposed. $l$ is the angular wave number in degree and $\delta \theta$ is the corresponding angular wavelength. To obtain de-lensed images, the homogeneous weighting was used.}}
\end{table*}

\begin{figure*}
\vspace{-2cm}
\hspace*{0.cm}
\vspace{-3.0cm}
\epsscale{1.1}
\plotone{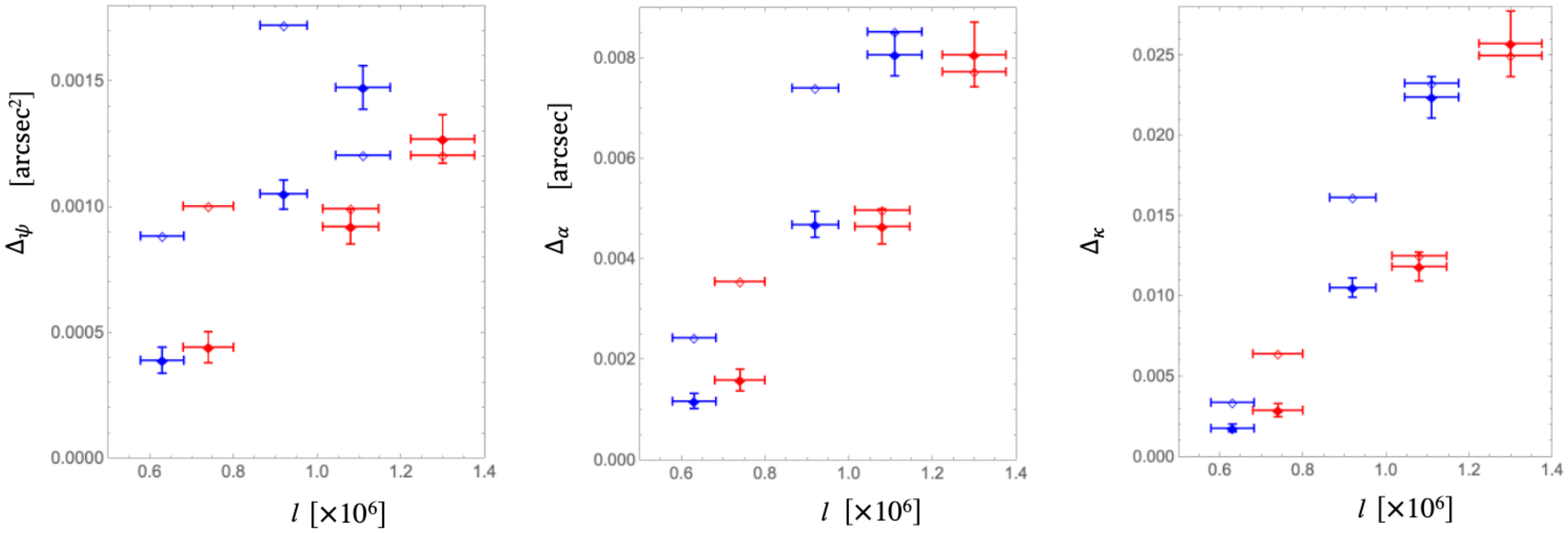}
\caption{Lensing power spectra for potential (left), astrometric shift (middle), and convergence (right) perturbations for small bins. The measured values were obtained from the ALMA image, the MIR flux ratios, the HST positions of galaxies, and the VLBA positions of lensed images of a core component q. Blue (red) vertical and horizontal bars centered at filled diamonds show $\sim 1 \sigma$ errors in the square root of power and the bin range of angular wavenumber $l$ used for a Type A model obtained with the homogeneous weighting and $L=4\farcs2$ ($L=3\farcs 6$), respectively.  Blue (red) horizontal bars centered at empty diamonds show the bin range of angular wavenumber $l$ and the best-fitted values for a Type B model obtained with the homogeneous weighting and $L=4\farcs2$ ($L=3\farcs 6$), respectively. The power spectra at the smallest, intermediate, largest angular  bins were obtained from 7 low-to-intermediate frequency modes, 11 intermediate-to-high frequency modes, and 7 high frequency modes. The highest one ($(m,n)=(6,6)$) was not included (see Figure 9). Note that the errors in the three (potential, shift, and convergence) lensing power spectra are not independent as they are given by common Fourier modes with an angular scale of $360^\circ/l$. }   
\label{fig:powerspectrum-result.pdf}
\end{figure*} 

In our algorithm, the phases of Fourier modes are fixed. Therefore, the position of the square boundary at which the gravitational potential vanishes may affect the reconstruction of potential. Moreover, lensing power spectra may depend on the scale of fluctuation. To take into account these ambiguities, we considered two types of models to describe the potential perturbation due to subhalos and LOS structures: 36 modes with $L=3\farcs 6$ centered at $(-0\farcs 4,0\farcs 3)$. To test the effect of possible masses in the vicinity of object Y, we also considered another model with 36 modes with $L=4\farcs 2$ centered at $(-0\farcs 7,0\farcs 3)$. In this model, the east (left) boundary of a square region was shifted towards the east by the half of the shortest wavelength ($0\farcs6$) while the west (right) boundary was fixed. The number of modes, the size and position of the square region satisfy the following conditions: 1) the number of modes, a squared integer should be smaller than the number of pixels in the source plane. 2) the distance between the boundary of the square region and the lensed quasar core should be larger than the half of the shortest wavelength in the Fourier modes. 3) the square region includes the central positions of object X and Y. 

The center of the coordinates was located at the centroid of the primary lensing galaxy G. For each model, a set of parameters that gives the minimum of $\chi^2$ for the image obtained with $\textit{robust}=0$ is given in Table \ref{tab:parameters-for-best-fit}. To analyze the perturbation in the real space, we used the magnification weighting and to analyze the lensing power spectra, we used the homogeneous weighting for de-lensing.

To check the consistency with the VLBA map at 5 \,GHz, we further fitted the positions of lensed VLBA jet components p, q, r, and s as well as the centroids of the primary lensing galaxy G and object X using the obtained best-fitted Fourier coefficients, $c_\textrm{q}$, $c_\textrm{p}$, and the parameters for the smooth model. Considering the beam size of the VLBA observation ($1.5\,\textrm{mas}\times 3.5$\,mas) at 5\,GHz
and a possible misalignment between our ALMA and the VLBA map, we assumed a positional error of 5\,mas for the lensed quasar core (or q) and jet components p, r, and s. 

As shown in Table \ref{tab:parameters-for-best-fit}, the reduced $\chi^2$ for the best-fit Type A model based on the ALMA (Cycle 2 + Cycle 4) image reconstructed with $\textit{robust}=0$, magnification weighting and $L=3\farcs 6$ or $L=4\farcs 2$ is $\chi^2_{\textrm{pos}}/\textrm{dof}_\textrm{pos}=22.5/(36 - 8)=0.80$. Thus, the fit to the ALMA image, VLBA position of the VLBA core component q, and HST positions of the centroids of G and X is good even without explicit modelling of object Y (Figure \ref{fig:VLBA-fit}). For the four Type A models we analyzed, the rms of convergence perturbation $\delta \kappa_{\textrm{rms}}^q$ and that of shear perturbation $\delta \gamma_{\textrm{rms}}^q$ at the positions of the quadruple image of q are $0.017\pm 0.005$ and $0.010 \pm 0.002$, respectively. Our result suggests that the rms convergence perturbation is significantly larger than the rms shear perturbation on the \ti{positions of the quadruple images}. The Type B models fitted the data slightly better than the Type A models. Inclusion of Fourier modes yielded good fits without large ellipticity for Y (see \citet{inoue2017}). Compared with the Type A models, the a posteriori fit to the VLBA positions was slightly improved.   

We also conducted a similar analysis using the ALMA image reconstructed with $\textit{robust}=0.5$, but it yielded slightly worse fitting for the VLBA positions with $\chi^2_{\textrm{pos}}/\textrm{dof}_\textrm{pos}=49.5/(36 - 8)=1.8$ and $\beta = 0.456$ in Type A models. Therefore, we used only the ALMA image reconstructed with $\textit{robust}=0$ in the subsequent analysis.

As shown in the middle panel in Figure \ref{fig:dif-source-plane}, the differences in the weighted mean de-lensed peak subtracted source images with the same parity were significantly reduced in the perturbed model with $L=3\farcs6$ and $c_\textrm{q}+c_\textrm{p}=0.95$ compared with the unperturbed model (left panel in Figure \ref{fig:dif-source-plane}).  The difference between the unperturbed and perturbed source images 
depicted two brightest spots, which can be interpreted as anomalies in astrometric shifts (right panel in Figure \ref{fig:dif-source-plane}). The position of r is not perfectly aligned with one side of the bipolar structure in the unperturbed model (left panel in Figure \ref{fig:model-best-fit}), but aligned in the perturbed model (middle panel in Figure \ref{fig:model-best-fit}). The misalignment may be due to errors in the fitted gravitational potential. Compared with the unperturbed model, a caustic that crosses a jet component r  (triangle in Figure \ref{fig:model-best-fit}) and  corresponds to a closed critical curve in the vicinity of object X (right panel in Figure \ref{fig:model-best-fit}) shifted towards a jet component r by $\sim 5\,$mas. The shift suggests that the perturbed model has a larger core for object X with a fainter fifth and sixth images around it. Since our ALMA images did not indicate a presence of a bright fifth and sixth images, the perturbed model is considered to be a reasonable model.      

Both the unperturbed and perturbed peak-subtracted de-lensed images show a brightest spot centered at q. The spot may be associated with dust emission from a circumnuclear disk around the quasar core. However, we cannot exclude the possibility of synchrotron emission from the quasar core due to insufficient subtraction of a point-like source component \citep{inoue2020}. Note that the position of the brightest spot obtained in our previous 
analysis that uses the HST positions of lensed core images deviates from q by $\sim 10\,$mas in the source plane (see also Figure \ref{fig:mock-source}). Since the accuracy of VLBA positions ($\lesssim 2\,$mas) used in this new analysis is better than that of the HST positions ($\sim 3$\,mas), we expect that our new de-lensed images yielded much better description of the source structure.

If the obtained model gives a better fit to the data, we expect that the fit to the data in the lens plane is also improved if the fluctuation scale of intensity is larger than the synthesized beam in the lens plane\footnote{If the fluctuation scale of intensity is smaller than the synthesized beam, the fit may become worse due to shrinking of lensed image}. To observe this effect, we smoothed the intensity of the lensed de-lensed image within a pixel with a side length of $0\farcs 05$ and subtracted it from the ALMA image. As shown in Figure \ref{fig:dif-lens-plane} (left and middle), the differences in the residuals between unperturbed and perturbed models is limited to regions near the critical curve, at which the magnification is large (Figure \ref{fig:dif-lens-plane}, right). In the perturbed model, the fit in the lensed arc image A2 was improved at regions near the critical curve (Figure \ref{fig:dif-lens-plane}, middle). The result is expected as astrometric shifts of $\sim 0\farcs 005$ along the lensed arc can be enhanced as $\sim 0\farcs 005 \times \mu$, where $\mu$ is the magnification. In our best-fitted models, the typical magnification at image A1 and A2 are $\sim 15$. Therefore, we expect shifts of $0\farcs 05-0\farcs 1$ in the vicinity of A1 and A2, which is larger than the beam size.

In the middle panels of Figure \ref{fig:real-space-best-fit-full}, we can see a positive peak in the convergence perturbation near image A2. In the models using the magnification weighting, the negative peak in the potential that corresponds to the positive peak in the convergence is slightly shifted from the position of object Y. However, we cannot exclude the possibility of perturbation by object Y because accurately measuring the matter distribution outside the lensed arcs is difficult. As shown in the right panels of Figure \ref{fig:real-space-best-fit-full}, images A1 and A2 are perturbed by shear possibly due to a pair of clumps or a trough.

Our method can probe both the potential perturbation due to subhalos and LOS structures and distortion in the potential of the primary lens. Since the effect is degenerate with that of low-frequency potential perturbation due to subhalos and LOS structures, extracting only the distortion in the potential of the primary lens is difficult. Although it is somewhat ad-hoc to determine the range of the 'low' frequency affected by the primary lens, we adopted the lowest 6 modes with angular wavelength of $2\arcsec \sim 7\arcsec$ as the 'low frequency' modes\footnote{In the limit of $L \rightarrow \infty$, these lowest modes correspond to multipoles $m=0,1,2$ in the polar coordinates.}. As shown in Figure \ref{fig:real-space-best-fit-low}, The convergence perturbation $\delta \kappa $ due to the low frequency modes are less than $0.002$, corresponding to $\sim 0.4$ percent of the background convergence of $\kappa \sim 0.5$ in the effective Einstein radius. The result is expected as low frequency modes whose fluctuation scales are larger than the effective Einstein radius of the primary lens cannot perturb quadruple images \ti{independently}. Thus, it is likely that our results for intermediate and high frequency modes are not very sensitive on the selected low frequency modes (including deviation from SIE) that are partly associated with the potential of the primary lens. However, note that our result cannot exclude the possibility of contribution from very small angular scale modes $<1\farcs 0$, which we did not take into account.   
  
We also studied perturbed Type B models in which object Y is modeled by an SIE at a redshift $z_\textrm{Y}=0.661$ in the smooth model (Figure \ref{fig:convergence-objY.pdf}). We found that the ratio of the difference of convergence contribution $\delta \kappa$ between A2 and B to A1 and B as $(\delta \kappa ({\rm A}2)-\delta \kappa ({\rm B}))/(\delta \kappa ({\rm A}1)-\delta \kappa ({\rm B}))=2.8$. The value is close to the ratio of the differential extinction of A2 to that of A1, $\Delta A_{V}({\rm A}2)/\Delta A_{V}({\rm A}1)\approx 3.4$\citep{inoue2017}. Therefore, the large difference in the differential extinction of A2 relative to A1 can be naturally explained by our model if the convergence is proportional to the dust column density. Owing to the presence of small clumps in the vicinity of B and A1, the perturbed Type B models fit the MIR flux ratios better than the unperturbed background Type B model. Compared with the perturbed Type A models, the perturbed Type B models fit the VLBA positions of jet components much better (see Table \ref{tab:parameters-for-best-fit}). In Figure \ref{fig:object-Y-perturbation.pdf}, we show the best-fitted convergence perturbations in the Type B models. One can visually confirm that the fluctuation patterns are similar to the ones in Type A models with the same parameters (cf. Figure \ref{fig:real-space-best-fit-full}). This suggests that the small scale potential fluctuations are independent of the presence or absence of object Y. 

To estimate the lensing power spectra towards MG\,J0414+0534, we used the homogeneous weighting, which was observed to be more suitable than the magnification weighting (Sec. 5.4). Similar to our mock analysis, the constant $\beta$ was chosen to satisfy $\chi^2/\textrm{dof}=2.8$ in the initial model without any perturbation. The $1\,\sigma$ errors of the powers were calculated using random Gaussian potentials that give $\varDelta \chi \le 1$ with the smoothing term $R_s/\textrm{dof}$ smaller than that for the best-fitted model. We used the two models (36 modes with $L=3\farcs6$ and $L=4\farcs2$) to estimate the lensing power spectra. Table \ref{tab:power-parameters-ALMA} shows the lensing power spectra obtained from the 22 intermediate frequency modes as was calculated in the mock analysis. The lensing power spectra of potential perturbation $\varDelta \psi$ and astrometric shift perturbation $\varDelta \alpha$ did not depend much on the side length $L$. In contrast, the convergence perturbations $\varDelta \psi$ showed a weak dependence on $L$: a model with smaller angular scale has a slightly larger convergence perturbation. This tendency was observed in the lensing power spectra with much smaller bins (see Figure \ref{fig:powerspectrum-result.pdf}). Modes with smaller angular scales have a larger power. This tendency is more apparent for convergence perturbation than astrometric shift and potential perturbations. On the smallest angular scale, the difference in powers of Type A and those of B models becomes smaller. The result suggests that the lensing powers on the smallest angular scale obtained from the 7 high-frequency modes are not so sensitive to the details of object Y. Therefore, we adopt the lensing powers on the smallest angular scale as the robust ones.

We summarize our results obtained from the source plane fit to the ALMA images as follows:
\begin{enumerate}
 \item 
Using a discrete Fourier expansion of potential perturbation $\delta \psi$, we were 
able to fit the VLBA positions of lensed images of a radio core, HST positions of the centroids of the primary lensing galaxy G and object X (object Y), peak-subtracted ALMA continuum image at $340$\,GHz, and MIR flux ratios observed with the Subaru and Keck telescopes. Using the obtained model parameters, we were able to fit the VLBA positions of lensed images of radio jet components using the traditional lens plane fit without changing the model parameters.
\\
\item
The de-lensed source images show a bright spot centered at a radio core with a bi-polar structure in the vicinity of the VLBA jet components. 
\\
\item
The best-fitted models show a complex mass distribution with four clumps near the quadruple images. One clump near image A2 may be associated with object Y.  The detailed structure of the mass distribution depends on the choice of assumed model parameters that were fixed in fitting.   
\\
\item
The contribution to convergence  
from low frequency modes are less than $\sim 0.4$ percent of the background value. 
\\
\item
The range of measured convergence power within $1\,\sigma$ of the mean values at the two smallest angular scales $l= (1.11\pm 0.07) \times 10^6$ and $l= (1.30\pm 0.08) \times 10^6$ was $\Delta_{\kappa} =0.021-0.028$. The mean angular scale $l \sim 1.2 \times 10^6$ of the two measurements corresponds to the effective Einstein radius $b_\textrm{G}\sim 1\farcs 1$ of G. The power is significantly larger than those on larger angular scales $>1\farcs 1$. The ranges of measured astrometric shift and potential powers within $1\,\sigma$ of the two angular scales were $\varDelta_\alpha =7-9\,$mas and $\varDelta_\psi=1.2-1.6\,$$\textrm{mas}^2$, respectively. The measured lensing powers at $l \sim 1.2 \times 10^6$ were not so sensitive to the presence or absence of object Y.

\end{enumerate}
\vspace*{0.cm}
\subsection{Lens Plane Fit to ALMA Visibilities}
\label{sec:6.3}
As we have discussed, the effects of systematic errors due to limited sample of visibility data are significantly reduced in our partially non-parametric model fitting. However, residuals of systematic errors caused by the CLEAN deconvolution process or phase corruption may still affect the fitted models significantly. Therefore, we check one of our best-fitted model (Type A) using $\chi^2$ fit in the visibility plane \citep{hezaveh2013, rybak2015, hezaveh2016a, spilker2016, maresca2022}. Direct fitting to the visibility data obtained by our Cycle 2 and 4 observations can test whether the observed astrometric shifts are due to perturbation by LOS structures/subhalos or systematic errors caused by sidelobes and phase corruption.  

Since our fitting formalism cannot determine the absolute fluxes of lensed images, we parameterize the overall amplitudes of the best-fitted point sources (cores p, q) and extended sources (jets r,s and dust) in the lens plane by multiplying constants $a$ and $b$ to the pre-fitted model visibilities of point and extended sources, respectively. In what follows, we use a Type A model that was fitted to the ALMA data with the magnification weighting ($36$ modes and $L=3\farcs 6$). We describe the detailed procedure of fitting visibilities to the ALMA data in Appendix.

\begin{figure}
\hspace*{0.2cm}
\vspace{0.cm}
\epsscale{0.6}
\plotone{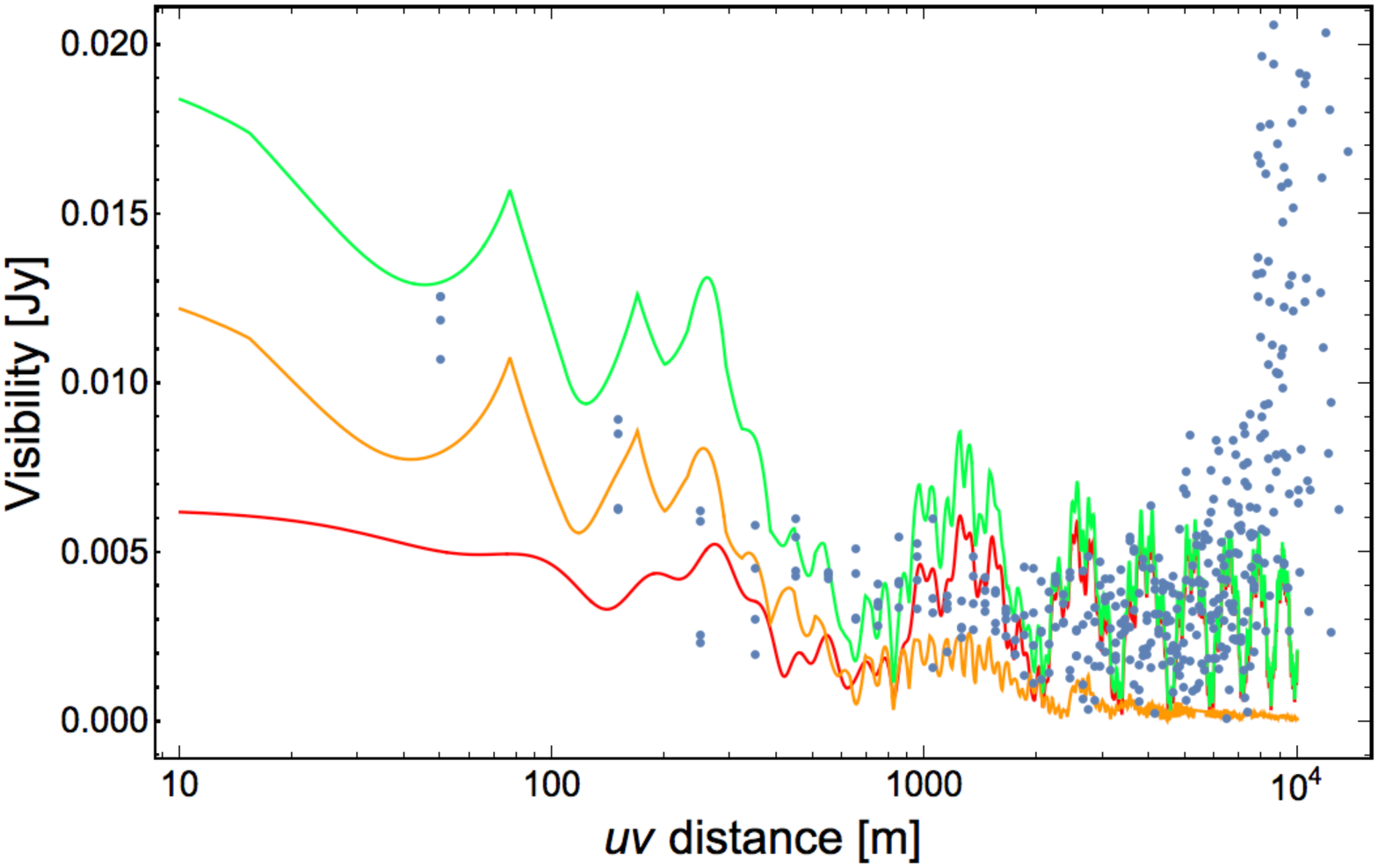}
\vspace{0.cm}
\caption{Comparison between observed and model (Type A) visibilities. The blue dots show the amplitude of visibilities on circular grids observed at 335.3\,GHz versus the uv distance $(=\sqrt{u^2+v^2})$. The radial separation of adjacent grids is 100\,m and the azimuthal separation is $\delta \theta=90^\circ$. The green, orange, and red curves show best-fitted Type A model visibilities at $v=0\,$m for the total, extended source, point source components, respectively.  }
\label{fig:vis-comparison-theta90-spw0.pdf}
\end{figure}

\begin{figure}
\hspace*{-0.2cm}
\vspace{0.cm}
\epsscale{0.7}
\plotone{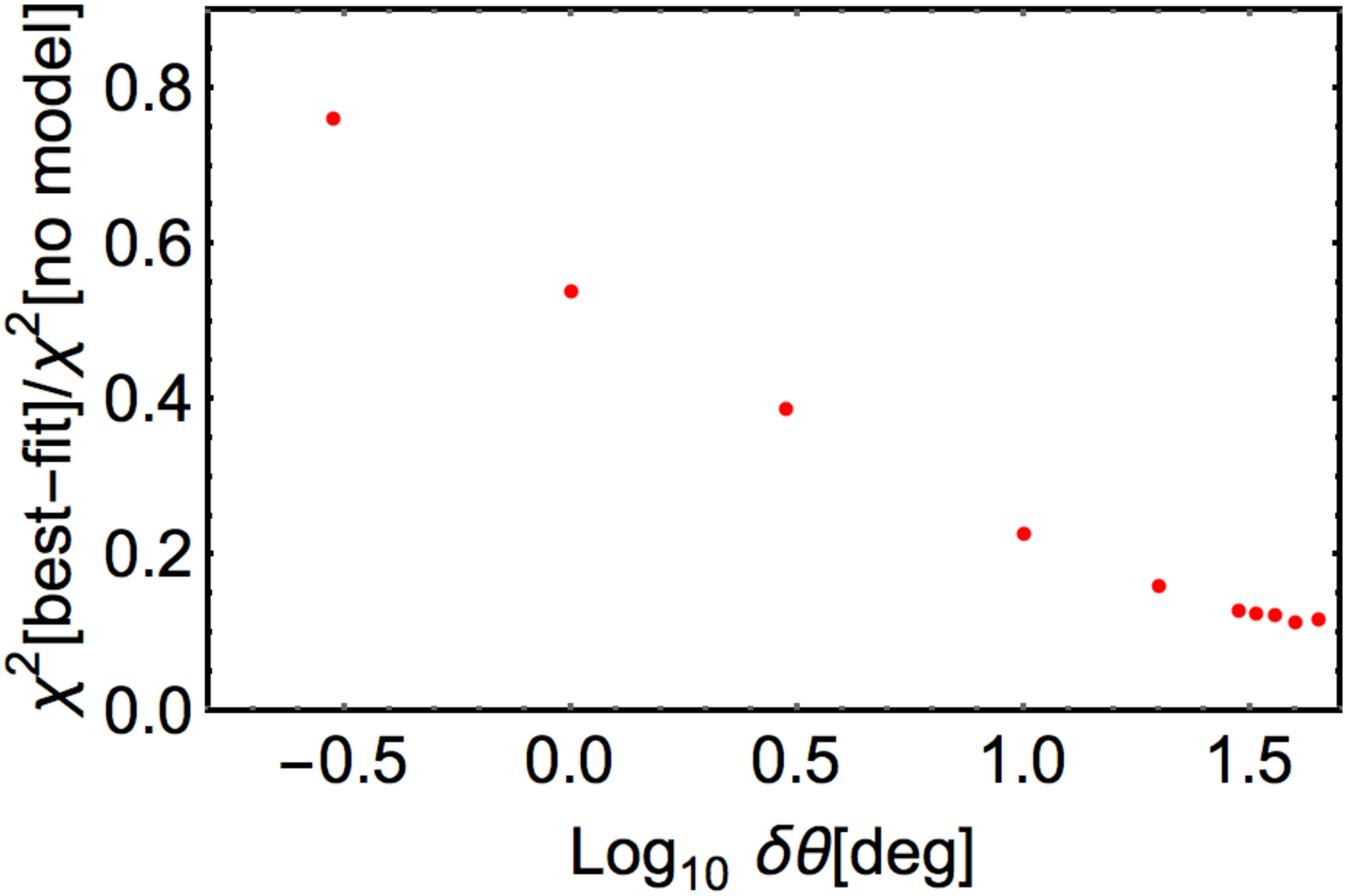}
\vspace{0.cm}
\caption{Improvement in fitting with respect to no (null) model. The red filled circles show the reduced $\chi^2$ in a perturbed Type A model divided by that in no (null) model. $\delta \theta$ is the angular separation between the centers of adjacent grids. }
\label{fig:chi2-perturb-no-model.pdf}
\end{figure}
\begin{figure}
\hspace*{-0.5cm}
\vspace{0.cm}
\epsscale{0.7}
\plotone{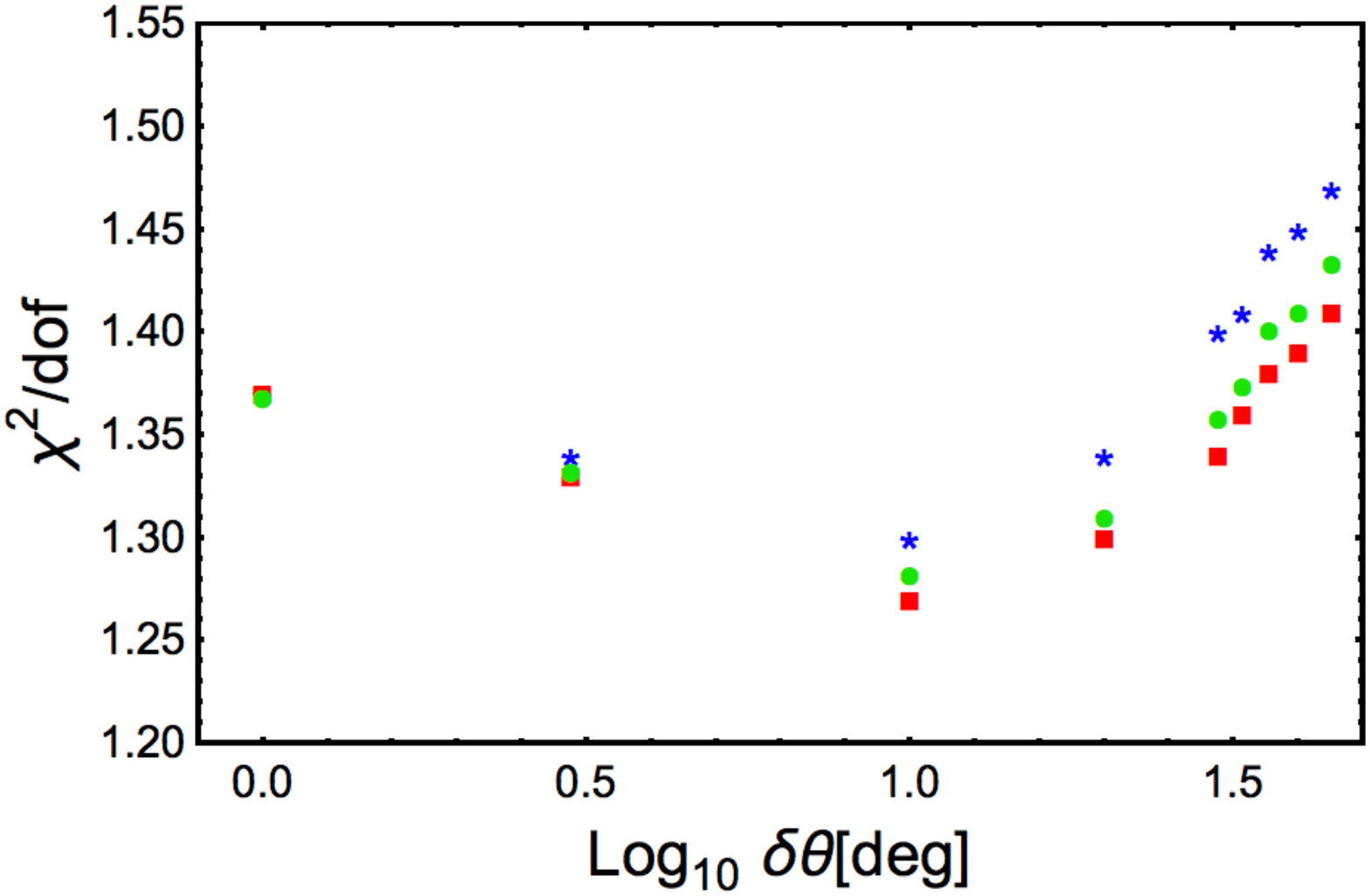}
\vspace{0.cm}
\caption{Comparison between perturbed and unperturbed Type A models. The red boxes and the blue stars show the reduced $\chi^2$ for the perturbed and unperturbed background Type A models, respectively. The green filled circles show the reduced $\chi^2$ for the perturbed Type A model in which the point source flux ratios of A1 to A2 is artificially fixed to be unity. $\delta \theta$ is the angular separation between the centers of adjacent grids.}
\label{fig:chi2-perturb-background.pdf}
\end{figure}

As described in Sec. 6.2, perturbation by LOS structures/subhalos produces thin tangential arcs across a critical line. Therefore, it is likely that such a perturbation preserves the radial structure of extended source visibility components (Figure \ref{fig:model-visibility.pdf}). Moreover, the amplitude of point source visibilities are invariant with respect to rotation in the visibility plane. Thus we consider that grids defined in the polar coordinates are suitable to compress the data to analyze the effects of potential perturbation efficiently.      

As one can see in Figure \ref{fig:vis-comparison-theta90-spw0.pdf}, the ALMA visibilities averaged with polar grids with azimuthal separation $\delta \theta=90^\circ$ are dominated by errors at $uv$ distance $(=\sqrt{u^2+v^2})$ larger than  $\sim 10000\,$m that corresponds to the size of the synthesized beam. The extended components contribute to the visibilities more than the point source components at distances smaller than $\sim 800\,$m whereas the point source components dominate the signal at distances larger than $\sim 800\,$m. Although the amplitudes of the total fitted model visibilities are comparable to the observed ones at distances $\sim 400 - 5000\,$m,  we observe a slight excess in model visibilities at small distances $\lesssim 400\,$m possibly due to systematics in our CLEAN deconvolution process. Improvement in fitting with respect to no (null) model is more conspicuous for larger angular separations $\delta \theta$ in which each grid has a larger number of samples (Figure \ref{fig:chi2-perturb-no-model.pdf}). If $\delta \theta$ is too small, much of information of phase is lost, leading to a worse fit. 

For $1^\circ \lesssim \delta \theta \lesssim 90^\circ$, our visibility fitting showed that the perturbed model fits the ALMA data better than the corresponding unperturbed smooth model (Figure \ref{fig:chi2-perturb-background.pdf}). We found that the optimal value of azimuthal separation that gives the best-fit is $\delta \theta =10^\circ$. The result suggests that the improvement by including Fourier modes of potential perturbation is not caused by systematics due to sidelobes and phase corruption. We also found that the improvement is caused by both the extended and point source components. In order to have the evidence of improvement due to the extended source components, we fitted model visibilities in which the point source flux ratio of A1 to A2 is artificially fixed to be unity while the extended source components were intact. If the improvement is not caused by extended components, then we can expect that the model visibilities will not show any improvements. However, as shown in Figure \ref{fig:chi2-perturb-background.pdf}, we found that the fit was improved to some extent. Therefore, we conclude that the fit to the extended source components was indeed improved.  

At $\delta \theta =10^\circ$, the coefficients of the point-source and extended source components in the perturbed model were $a=1.8\pm 1.5\,$mJy and $b=0.42\pm 0.39$. At $\delta \theta =90^\circ$, the corresponding values were $a=1.8\pm 0.8\,$mJy and $b=0.44\pm 0.19$. The errors were obtained from a condition $\Delta \chi^2/\textrm{dof}<1$. Thus the best-fitted values correspond to $\sim 20$ per cent and $\sim 60$ per cent decrease in overall amplitude with respect to the originally modeled intensity distribution of the point-source and extended source components, respectively. The decrease in $a$ is not statistically significant but the significance of a decrease in $b$ is at the $\sim 3\,\sigma$ level. 

As we have seen in Section \ref{sec:5.2}, our CLEAN deconvolution process gave a systematic boost in overall amplitudes of flux on large angular scales. Therefore, it can affect the model fit on large angular scales $\gtrsim 0\farcs 7$ (corresponding to a distance of $\sim 400\,$m). However, on small angular scales, especially for signals with large S/N ratio, such systematic effects are expected to be small. The improvement of $\chi^2$ fit due to the extended source components implies that the perturbation effect caused by LOS structures and subhalos are limited to intensity fluctuations on relatively small angular scales $ \lesssim 0\farcs 7$. The size of the image residual in the lens plane (Figure \ref{fig:dif-lens-plane}) in Section \ref{sec:6.2} supports this interpretation.  

Thus we conclude that our results in visibility fitting and mock analysis suggest that observed astrometric shifts are due to perturbation by LOS structures/subhalos rather than systematic errors caused by sidelobes and phase corruption.

\vspace{1cm}
\subsection{Lens Plane Fit to VLBA Positions}
\label{sec:6.4}
\begin{table*}
\vspace{0cm}
\hspace{-4cm}
\caption{Parameters, minimized $\chi^2$, flux ratios (estimated at the fitted positions of q), and rms perturbations in Type A models best-fitted to the MIR flux ratios, the HST positions of galaxies and the VLBA positions of four jet components p, q, r, and s. Only the 22 intermediate frequency modes with translation were used for fitting. The parameters in the smooth model were fixed except for the position of an SIE+ES.  }
\hspace{-3.45cm}
\scriptsize{
\setlength{\tabcolsep}{2pt}
\begin{tabular}{cccccccccccccccc}
\hline
\hline
 $L$ & $\delta \psi_0$ & $\delta \alpha_0$ &  $\delta \kappa_0$ &    $\chi^2/\textrm{dof}$ & dof & A2/A1 & B/A1 & C/A1 &  $\chi^2_{\textrm{flux}}/\textrm{dof}$ & $\chi^2_{\textrm{VLBA}}/\textrm{dof}$ & $\chi^2_{\textrm{X}}/\textrm{dof}$   & $\chi^2_{\textrm{G}}/\textrm{dof}$ & $\langle (\delta \psi)^2\rangle^{1/2} $ & $\langle (\delta \alpha)^2\rangle^{1/2} $ & $\langle (\delta \kappa)^2\rangle^{1/2} $ 
\\
\hline  
3.6  & $0.0014$ & $0.0053$ & $0.0090$ & $2.0$ & $7$ & $0.916$ & $0.355$ & $0.157$ & $0.29$ & $1.13$  & $0.06$ & $0.47$ & $0.00097$ & $0.0041$ & $0.0092$ 
\\
\hline
3.6 &  $\infty$  &$\infty$&  $\infty$  &  $0.65$ & $7$ & $0.920$ & $0.345$ & $0.151$ & $0.10$ & $0.30$ & $0.06$ & $0.19$ & $0.0031$ & $0.0144$ & $0.0346$
\\
\hline
4.2 & $0.00082$ & $0.0051$ & $0.0048$ &  $2.1$ & $7$ & $0.918$ & $0.355$ & $0.158$ & $0.33$ & $1.22$  & $0.06$ & $0.44$& $0.0011$ & $0.0042$ & $0.0084$ 
\\
 \hline
4.2 & $\infty$ &  $\infty$ &  $\infty$  &  $0.99$ & $7$ & $0.915$ & $0.355$ & $0.132$ & $0.095$ & $0.676$ & $0.06$ & $0.16$ & $0.0028$ & $0.0113$ & $0.0238$
\\
\hline
\label{tab:parameters-for-best-fit-VLBAonly}
\end{tabular}
}
\vspace{-0.5cm}
\small{
\flushleft{Note. $L$ is the side length of a square at which the Dirichlet boundary condition is imposed. $\delta \psi_0$, $\delta \alpha_0$, and $\delta \kappa_0$ are smoothing parameters that control the rms values defined in the square region. $\chi^2$ is the sum of $\chi^2_{\textrm{flux}}$ for the MIR flux ratios, $\chi^2_{\textrm{X}}$ for the HST position of X,  $\chi^2_{\textrm{G}}$ for the HST position of G, $\chi^2_{\textrm{VLBA}}$ for the VLBA positions of four jet components. The rms perturbations $\langle (\delta \psi)^2\rangle^{1/2} $, $\langle (\delta \alpha)^2\rangle^{1/2}$, and $\langle (\delta \kappa)^2\rangle^{1/2} $ on the square region were obtained from the coefficients for the best-fitted 22 Fourier modes.  }}
\end{table*}
\begin{figure}
\vspace*{-1cm}
\hspace*{-0.2cm}
\epsscale{0.7}
\plotone{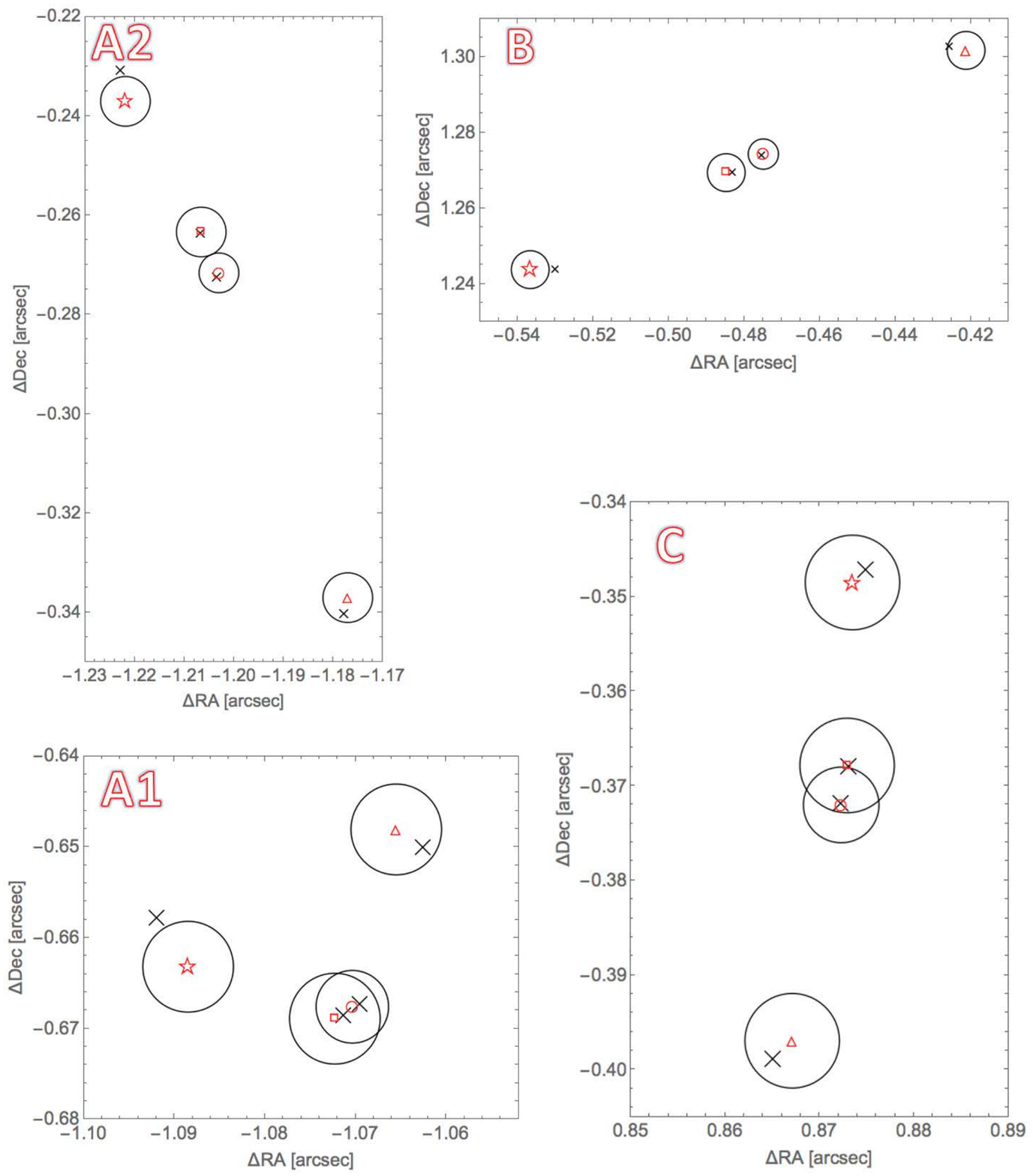}
\vspace{-1.3cm}
\caption{The best-fitted VLBA positions of four jet components in the lens plane using the MIR flux ratios, HST positions of galaxies and smoothing parameters that were used to fit the ALMA data. For the fitting, a Type A smooth model was used and additional 22 intermediate frequency modes were adjusted with smoothing constraints $R_s^2$. The red symbols represent the VLBA positions of p, q, r, and s observed at 5\, GHz as in Figure \ref{fig:alma-lens-image} and circles indicate their $1\,\sigma$ errors. The black X's represent the best-fitted positions in a model with $L=3\farcs 6$. The coordinates are J2000 centered at the centroid of the primary lensing galaxy G. }
\label{fig:VLBA-fit-VLBA-only.pdf}
\end{figure}

%
% Taken from ``Parameters-VLBA-G-X-fit-cy24-LL4.2M-MIR-pq0.95-flux-28modes(Most-Important-result).nb''
% ``Parameters-VLBA-G-X-fit-cy24-LL3.6M-MIR-pq0.95-flux-28modes(Important-result).nb''
% {-0.917945, 0.350472, -0.149318} {0.0013135, 0.00512399, 0.0109973}//{-0.919375, 0.347521, -0.136433}
% {0.00240796, 0.0106728, 0.0249827}

\begin{figure}
\vspace*{-3cm}
\hspace*{-0.2cm}
\epsscale{1.15}
\plotone{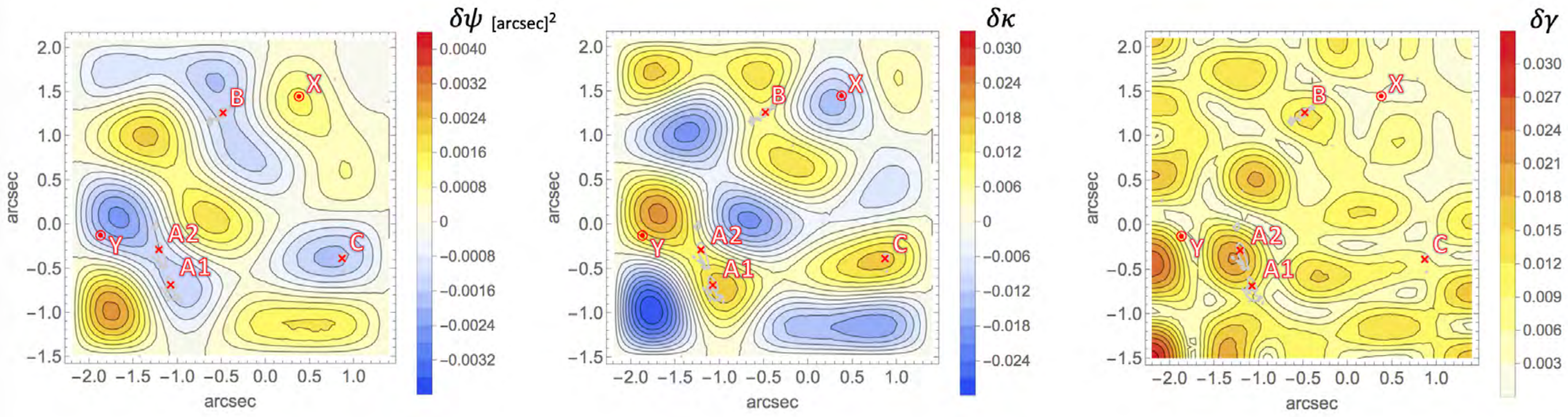}
\vspace{-3.5cm}
\caption{Contour maps of the best-fitted perturbations obtained from the MIR flux ratios, HST positions of galaxies, VLBA positions of four jet components p, q, r, and s and smoothing parameters that were used to fit the ALMA data. For the fitting, a Type A smooth model was used and additional 22 intermediate frequency modes were adjusted with smoothing constraints. The plotted maps show the potential perturbation $\delta \psi$, convergence perturbation $\delta \kappa $, and shear perturbation $\delta \gamma$ reconstructed with $L=3\farcs6$. The gray curves denote the boundaries of regions in which signals are larger than $3\,\sigma$. The red X's denote the positions of lensed jet component q.
The centers of the square region are $(-0.4,0.3)$. The coordinates are J2000 centered at the centroid of the primary lensing galaxy G. }
\label{fig:real-space-best-fit-22modes.pdf}
\end{figure}

Based on the above discussions, the reconstructed potentials in the lens plane exhibit some differences depending on the choice of the center and side length $L$ of a square at which the Dirichlet boundary condition is imposed. In order to scrutinize the effect of perturbation, particularly in the vicinity of object Y, we also performed a $\chi^2$ fitting in the lens plane. We fitted the model to the positions of VLBA jet components p, q, r, and s, centroid of G and object X, and the MIR flux ratios of quadruple images with adjusting coefficients of the 22 intermediate frequency Fourier modes and translation\footnote{We conducted a similar analysis using the 6 high frequency modes and the 22 modes but the fit was worse.}. The positional errors are assumed to be $2\,$mas for q, $5\,$mas for p, r, s, and G, and $0\farcs 1$ for X. The number of observed parameters is $4\times 2\times 4+2\times2 +3=39$ and the number of fitting parameters is $22+2+8=32$. We did not consider rotation of the potential. 

If we conduct $\chi^2$ fitting for the positions, significant discontinuity in the potential perturbation may occur at places farther from the position of lensed jet components as the size of jets are much smaller than the lensed arc observed in the ALMA image. To avoid such discontinuity, we imposed an additional constraint on the smoothness of perturbation. To do so, we minimized $\chi^2$ for the positions of G, X, and the VLBA jet components and MIR flux ratios plus the smoothing term $R_s$ defined in equation (10). We fixed the smoothing parameters $\delta_{\psi 0}$, $\delta_{\alpha 0}$ and $\delta_{\kappa 0}$ to be the same or slightly smaller values obtained from the best-fitted models with the magnification weighting with $L=3\farcs 6$ and $N^2=36$ modes. To test the smoothing effect, we considered two types of $\chi^2$ fitting: with and without constraint on the smoothness of perturbation. 

As shown in Table \ref{tab:parameters-for-best-fit-VLBAonly}, the rms perturbations $\langle (\delta \psi)^2\rangle^{1/2} $, $\langle (\delta \alpha)^2\rangle^{1/2} $, $\langle (\delta \kappa)^2\rangle^{1/2} $\footnote{For the model with $L=3\farcs 6$ and $N^2=36$ reconstructed with the magnification weight, we had
$(\sqrt{\langle \delta \psi^2\rangle}, \sqrt{\langle \delta \alpha ^2\rangle}, \sqrt{\langle (\delta \kappa)^2\rangle})= (0.00084\, \textrm{arcsec}^2, 0.0031\, \textrm{arcsec},  0.0092)$.} for fitting without constraint on smoothness are significantly 
larger than those for fitting with the constraint. Our results indicate that reconstructing the potential beyond the region of jet components is difficult if constraint on smoothness is not taken into account. The fit is slightly better for the model with $L=3\farcs 6$ than that with $L=4\farcs 2$. The obtained rms perturbations in the model with $L=3\farcs 6$ and constraint on smoothness agree with those in the corresponding model with 36 Fourier modes obtained from the previous fits based on the ALMA image within 35 percent. Moreover, as shown in Figure \ref{fig:VLBA-fit-VLBA-only.pdf}, the fit to the VLBA positions of jet components p, q, and r in images A1 and A2 noticeably improved compared with the previous fit in the source plane using the ALMA data dominantly. Therefore, the obtained model may be significantly better in describing the perturbation in the vicinity of A2. As shown in Figure \ref{fig:real-space-best-fit-22modes.pdf}, the feature of potential perturbation $\delta \psi$ is similar to the one obtained with $L=3\farcs6$, $N^2=36 $ modes and the magnification weighting (Figure \ref{fig:real-space-best-fit-full}). 

Interestingly, the best-fitted convergence perturbation $\delta \kappa$ indicates a presence of a clump in the vicinity of object Y as well as three clumps in the vicinity of image A1, B, and C. The distance between the local peak in the convergence and object Y is $\sim 0\farcs3$, which is smaller than the smallest fluctuation scale $\sim 0\farcs 9$ of the fitted Fourier mode functions. Compared with the previous model in \citet{inoue2017}, the shape of the clump in the vicinity of Y is more rounder. In the best-fitted models, a pair of clumps in the vicinity of B and a pair of clumps in the vicinity of A2 cause a shear perturbation $\delta \gamma $ at A2 and B, respectively. They probably result in an improvement in the fit to the MIR flux ratios and VLBA positions of jet components. Our results indicate that potential perturbations consisting of several halos are more realistic and robust compared with perturbations consisting of only one halo. 

\subsection{Object Y}
\label{sec:6.5}
\begin{figure}
\hspace*{-0.3cm}
\epsscale{1.2}
\plotone{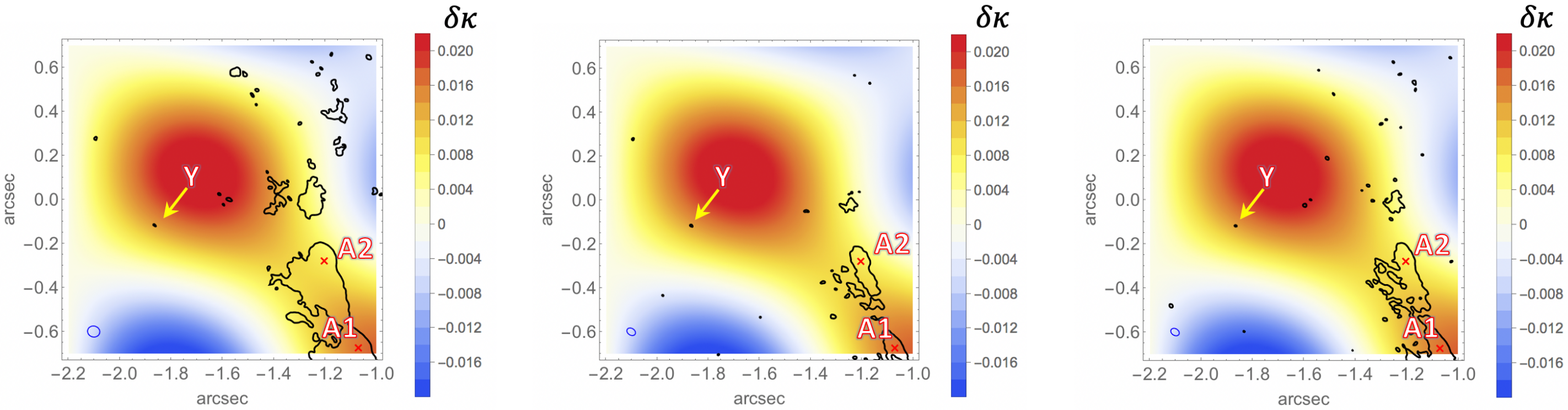}
\vspace*{-4.5cm}
\caption{ALMA images of object Y. The plotted black curves are contours of intensity with $ 3.5\,\sigma$ for line subtracted images with $robust=0$ (left), $robust=0.5$ (middle), and a full band image with $robust=0$ (right) overlaid with the best-fit convergence 
perturbation $\delta \kappa $ obtained from the MIR flux ratios, HST positions of galaxies, VLBA positions of four jet components p, q, r, and s. The reconstruction was conducted with $L=3\farcs6$ and the smoothing parameters that were used to fit the ALMA data. }
\label{fig:objectY.pdf}
\end{figure}
\begin{table}
\caption{Flux of object Y measured in CLEANed images. }
\setlength{\tabcolsep}{2pt}
\hspace{3cm}
\begin{tabular}{lccc}
\hline
\hline
 &Line-free &Line-free & Full-band 
\\
\hline
\ti{Robust} & 0.5 & 0 & 0 
\\
\hline
Rms errors [$\mu \textrm{Jy}\,\textrm{beam}^{-1}$] &22.9  & 30.4  & 22.6
\\
\hline
Peak flux [$\mu \textrm{Jy}\,\textrm{beam}^{-1}$] & 81.8  & 114.5  & 84.2
\\
 \hline
S/N [$\sigma$]  at peak & 3.6 & 3.8 & 3.7 
\\
\hline
flux ($>2\sigma$) [$\mu$Jy] & 36 & 79 & 63
\\
\hline
Peak $\Delta$RA (J2000)  & -1\farcs 865  & -1\farcs 866 & -1\farcs 865
\\
\hline
Peak $\Delta$Dec (J2000)  & -0\farcs116   &  -0\farcs116 
& -0\farcs118 
\\
\hline

\label{tab:fluxes-of-object-Y}
\end{tabular}
\end{table}

\begin{figure}
\vspace*{-0.5cm}
\hspace*{-0.6cm}
\epsscale{0.6}
\plotone{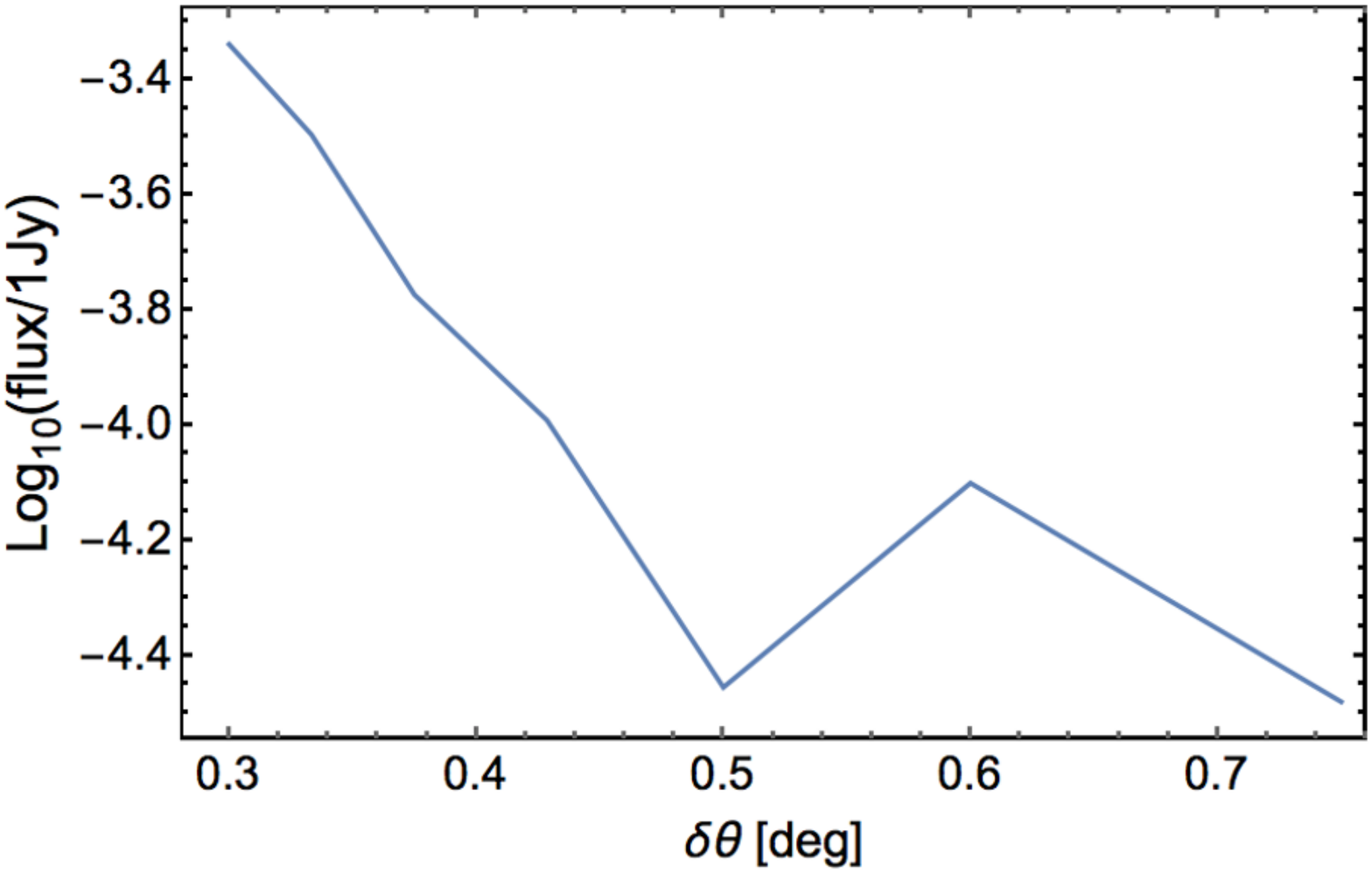}
\vspace*{-0.5cm}
\caption{Result of visibility fitting in which object Y is modeled by a Gaussian source. The blue line shows the best-fitted flux of object Y. $\delta \theta$ is the angular separation between the centers of adjacent grids. We used a perturbed Type A model with the magnification weighting and $L=3\farcs 6$. In the fitting, only the line-free visibility data were used. }
\label{fig:objectY-flux-chi2-lf.pdf}
\end{figure}

\begin{figure}
\vspace*{-2.8cm}
\hspace*{-0.4cm}
\epsscale{1.0}
\plotone{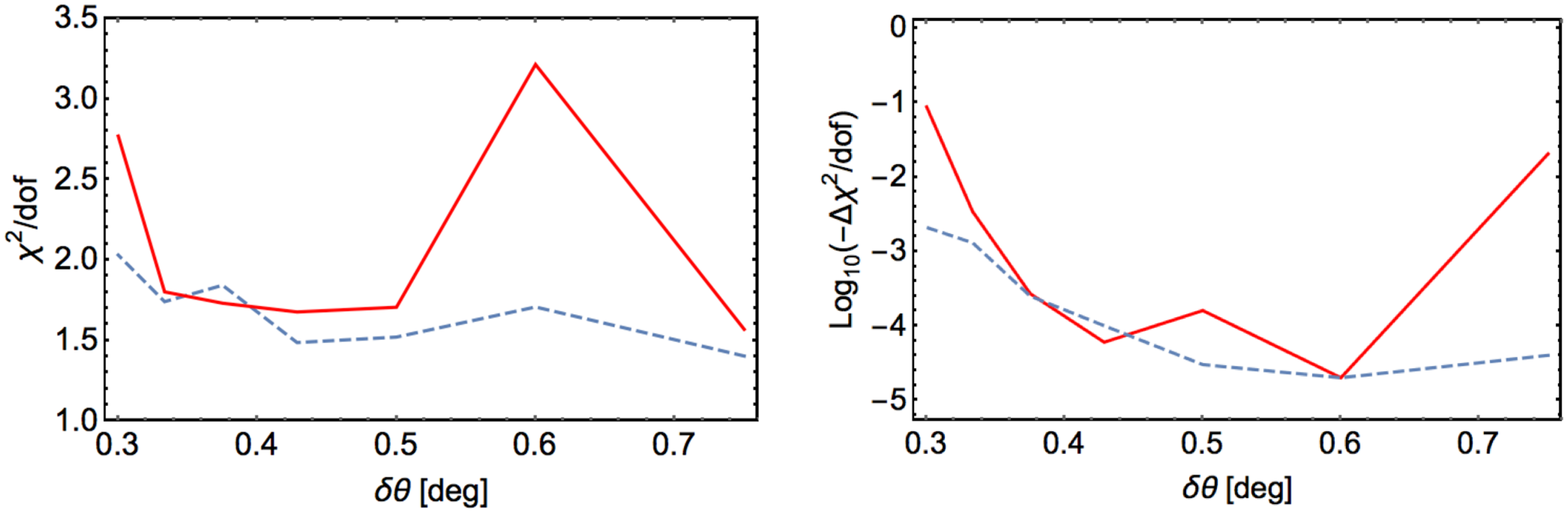}
\vspace*{-2.8cm}
\caption{Comparison of results of visibility fitting. The left and right figures show the reduced $\chi^2$ for models with object Y and the difference in the reduced $\chi^2$ between models with and without object Y $(\Delta \chi^2 \equiv \chi^2(\textrm{with Y})-\chi^2(\textrm{without Y}))$, respectively. In each figure, the blue dashed and red lines correspond to the results fitted to the line-free and full-band visibility data, respectively. }
\label{fig:objectY-lf-full.pdf}
\end{figure}

A faint continuum emission with a flux of 0.2-0.3\,mJy at (RA,Dec)$=(4^\textrm{h}14^\textrm{m}37.818^\textrm{s}, 5^\circ 34\arcmin 42\farcs 810)$, in the north east of A2, was first reported by \citet{inoue2017}. We called it object Y. Using the same ALMA Cycle 2 data, \citet{stacey2018} pointed out that the emission disappears after self-calibration. To check the result, we added the data of our Cycle 4 (high resolution) observations to the Cycle 2 (low resolution) data. For each data set of Cycle 2 and 4 data, we subtracted off the components of line emission and carried out phase-only self calibration. Then we combined both the data sets to obtain CLEANed images with various $robust$ parameters. As shown in Table \ref{tab:fluxes-of-object-Y}, with $robust=0, 0.5$, we detected a faint emission with the peak flux density of $81.8, 114.5\,\mu \textrm{Jy}\,\textrm{beam}^{-1}$ at the position of object Y. The statistical significances were $3.6, 3.8\,\sigma$. To increase S/N, we also carried out imaging with $robust=0$ using the full-band data (self-calibrated line-free and line channels) and detected faint emission with a peak flux density of $84.2\,\mu \textrm{Jy} \,\textrm{beam}^{-1}$ at the position of object Y. The statistical significance is $3.7\,\sigma$. Although the significances are not sufficiently high, as shown in Figure \ref{fig:objectY.pdf}, the signals seem to be stable. In contrast, other faint spots with intensity $<3.5\,\sigma $ are not stable. Their peak flux depend on the selected $robust$ parameters and frequency. Note that the significance of object Y decreased to $<3.0\,\sigma$ for the natural weighting. Thus, we cannot conclude the significance of the object Y. 

%The size of the (dust) emitting region at object Y may be more smaller than the% synthesized beam size $\sim 0\farcs 05$ for the natural weighting if the signa%l is indeed from a dusty dwarf galaxy. 
 
In order to verify the weak emission from object Y, we performed visibility fitting. We used a Type A model that was best fitted to the ALMA, HST, Subaru and VLBA data using the magnification weighting with $L=3\farcs6$. We added a Gaussian component at $(-1\farcs 865, -0\farcs116)$\footnote{The position corresponds to the local peak in intensity in the CLEANed image obtained from the self-calibrated line-free components of our Cycle 2 and 4 data. } in the J2000 coordinates in which the centroid of galaxy G is at (0,0). In the fitting, the flux of the Gaussian component was used as a free parameter as well as $a$ and $b$ that control the overall amplitudes of the point-like and extended sources, but the position of the Gaussian source was fixed at the central position of object Y. The distance between object Y and the phase center (in the vicinity of q in A1) is $\sim 1''$, the corresponding baseline scale at $\sim 340\,$GHz is $\sim 40\,$m. The scale of the synthesized beam is $0\farcs 03$, which corresponds to the baseline scale of $\sim 1500\,$m. Therefore, in order to retain the phase information of object Y, we needed to restrict the azimuthal separation of polar grids to $ \delta \theta \lesssim \arctan{(40/1500)}=1\fdg 5$. We plotted our result in Figure \ref{fig:objectY-flux-chi2-lf.pdf}. The fitted flux values were $40-400\,\mu$Jy and always positive. We also fitted our model to the full-band visibility data. Our fitting result showed that the fitting to the full-band data is worse than the line-free data (left in Figure \ref{fig:objectY-lf-full.pdf}). Since we did not include line components in the model, the result is not surprising. Interestingly, compared to the line-free data, the full-band data showed an enhanced improvement in fitting compared to the model without object Y (right in Figure \ref{fig:objectY-lf-full.pdf}). Since the S/N of emission from object Y is expected to be higher in the full-band data, the result is again not surprising. Thus, we were not able to exclude the possibility that the weak continuum emission is associated with object Y rather than systematic errors. 

However, it should be noted that the origin of the weak emission is still not confirmed. In order to confirm the weak emission, we need to carry out new ALMA observations at frequency $\gtrsim 340$\,GHz with angular resolution $ \lesssim0\farcs1$ with sensitivity better than $\sim 10 \mu $\,Jy/beam.

\section{Consistency with CDM Model}
As shown in Section \ref{sec:6.2}, the range of measured amplitude of convergence perturbation within $1\,\sigma$ at an 
angular scale of $\sim 1\farcs 1$ ($l \sim 1.2 \times 10^6$) was $\varDelta_\kappa=0.021-0.028$. The result was obtained using the 7 high frequency modes. Assuming that the 7 modes are independent each other and obey a Gaussian distribution, the cosmic variance is at most $\varDelta \varDelta_\kappa^2/\varDelta_\kappa^2= \sqrt{2/7}\sim 0.53$. Therefore, the measured values including the error due to the cosmic variance is $\varDelta_\kappa=0.015-0.032$. Assuming a constant convergence power over a range of angular wavenumber $\Delta l\sim l$, the obtained value seems to be at odds with theoretical prediction $\delta \kappa_{\tr{rms}}=0.0059^{+0.0006}_{-0.0004}$ \citep{inoue2016} at quadruple images with separation angle of $\sim 1''$ in which only CDM subhalos in the primary lensing galaxy are taken into account. If we add contribution from LOS structures to that from CDM subhalos, the CDM prediction yields $\delta \kappa_{\tr{rms}}=0.018\pm 0.007$ \citep{inoue2016} assuming that object X is not residing in the primary lens plane \citep{takahashi-inoue2014} and baryons in the primary lensing galaxy do not significantly affect the lensed image \citep{hsueh2018}. Therefore, our result is consistent with the CDM prediction. Although, we cannot deny the possibility of chance alignment of subhalos with the quadruple images, our results support the idea that anomalies in the flux ratios and astrometric shifts in galaxy-scale quadruple lenses at $z_\textrm{s}>2$ are primarily caused by LOS structures rather than subhalos. In other words, our results are consistent with the predicted abundance of intergalactic halos in CDM models on $\sim 10\,$kpc scale. 

Expressing convergence perturbations in terms of mass scale might be helpful to galactic astronomers. The mean of the measured convergence perturbation $\Delta \kappa=0.025$ for a source redshift $z_s=2.639$ and a lens redshift $z_l=0.9584$ is equal to $9.3 \times 10^8 \ms/\textrm{arcsec}^2$ if the total mass resides at $z_l$ \citep{takahashi-inoue2014}. Assuming that the convergence power is proportional to the proper length to the source \citep{inoue-takahashi2012}, our result suggests that a convergence power $5 \times 10^8 \ms$ per $5\,$kpc scale for a source redshift $z_s=0.982$ and a lens redshift $z_l=0.28$, which is slightly larger than the recent upper bound of convergence power $3\times 10^8 \ms$ for a proper length scale of $3\,$kpc obtained from the HST observations of SDSS\,J0252+0039 \citep{bayer2023}. Taking into account the cosmic variance, our result is consistent with the HST result. 
\newpage

\section{Conclusion and Discussion}
% Discussion for the obtained lensing power spectrum
We obtained the first lower and upper bounds of the lensing power spectra on $\sim 10\,$kpc (on the primary lens plane) scale toward MG\,J0414+0534 using our new partially non-parametric method. Based on Type A models, the range of measured convergence, astrometric shift, and potential powers within $1\,\sigma$ at an angular scale of $\sim 1\farcs 1$ (corresponding to an angular wave number a number of waves per 360 degrees) of $l \sim 1.2\times 10^6$ or $\sim 9\,$kpc in the primary lens plane) were $\varDelta_\kappa=0.021-0.028$, $\varDelta_\alpha =7-9\,$mas, and $\varDelta_\psi=1.2-1.6\,$$\textrm{mas}^2$, respectively. Our simple estimate suggests that the obtained values are consistent with the CDM prediction. To measure the power spectra, we conducted ALMA observations with high angular resolution $0\farcs02-0\farcs04$ towards the anomalous quadruply lensed quasar MG\,J 0414+0534 at $z_\textrm{s}=2.639$. We also used the MIR ratios of quadruple images, VLBA positions of jet components, and HST positions of galaxies.

% Discussion for the obtained potential and astrometric shift power spectra
 In a mock analysis, we observed that the residual errors of measured power spectra for potential and astrometric shift perturbations are smaller than that for convergence perturbation. The result is expected as interferometers such as VLBA and ALMA can measure astrometric shifts with small systematic errors. To date, astrometric weak lensing effect has been measured only in the near universe \citep{mondino2020}. Our first measurement of power spectra with upper and lower bounds) for potential and astrometric perturbations in the far universe creates a basis for constraining cosmological models with a better accuracy. 

% Object Y
%Our lens models suggest a presence of a dark halo in the vicinity of object Y. %Although much %fainter than the previous report, we also detected weak c%ontinuum emission possibly from obje%ct Y at the $4\,\sigma$ level in our combined (Cycle2 and Cycle 4) ALMA data. Our visibility f%itting analysis suggests that the signal is not related with systematic errors. However, we de%finit%ely need further evidence to confirm the continuum emission.

% Why our analysis is considered to be 'the best'? VLBA+ALMA+HST
Our method can directly measure lensing power spectra on certain angular scales and provides more accurate results than the previous methods based on only flux ratios or only astrometric shifts. To estimate the rms convergence perturbation from flux ratios of lensed images, we must assume the amplitudes of shear perturbation. For completely isotropic perturbations, the rms convergence perturbation is equal to the rms shear perturbation. Since the number of available fluxes that can be used to estimate convergence perturbation is limited, such an assumption would cause some errors (though not systematic) in each lens system. If we use only astrometric shifts to reconstruct the perturbation, our numerical analysis demonstrated that the convergence tends to become excessively large in regions beyond the lensed arcs in which the astrometric shifts are measured. Since astrometric shifts are related with the first spatial derivatives of the projected potential, not the second one, such a behavior can be naturally expected. Inclusion of ALMA data is essential as it provides information of radial profile of mass that cannot be obtained using VLBA observations. Using positions of jet components and extended dust emission on scales $>1\,$kpc as well as MIR flux ratios, which are microlensing free, our multi-wavelength method provides us with a very effective tool for probing matter fluctuations on scales $\lesssim 10\,$kpc in the universe.  

% Why source-plane chi^2 is useful? What about visibility plane fit?  
Our method based on decomposition with discrete Fourier modes enables us to measure lensing power spectra directly from a single lensing system based on source plane $\chi^2 +\alpha$ evaluation. Note that noise cancellation due to synthesized multiple lensed images makes our method suitable for systems with extended images obtained from interferometers. Moreover, our method can be applied to systems with a source with a complex structure as constraints on the source intensity are not stringent. 

Our test using visibility plane fitting suggested that the improvement of fit achieved by our algorithm is not likely caused by systematics due to sidelobes and phase corruption even if the overall intensity is not preserved by imaging using CLEAN. Gravitational perturbation causes tiny 'arc-like' changes in the lens plane. Therefore, the effect of perturbation is very weak in the visibility plane as tiny arc-like structures are FFTed to large arcs. On the other hand, the signal in the source plane is enhanced due to a linear combination of distant multiple images. Therefore, it is likely that our method has an advantage in detecting potential perturbation due to LOS structures/subhalos if properly applied to the lens system. Our visibility fitting method is useful for calibrating the absolute flux of point-like and extended sources.  

% Why is the homogeneous weighting good to estimate lensing power spectra? 
In our mock analysis, to obtain the lensing power spectra towards MG\,J 0414+0534, the homogeneous weighting seems to be more suitable than the magnification weighting. Since the magnification weighting puts large weights on very bright regions, the improvement in fitting is limited to regions
with a large intensity. On the other hand, the homogeneous weighting treats multiple images equally. Therefore, we expect that it retains the information of fluctuations outside the bright regions, leading to a robust estimate of powers. However, in some systems, a certain 'robust' weighting that lies in the middle of the both weighting may be optimal. We must further study this de-lensing problem using various lens systems. 

% Why our constraints on smoothness are reasonable ? 
In our algorithm, we included a term that controls the smoothness of perturbation beyond the lensed images to $\chi^2$. To verify such constraints, we may require to assume that the lensing power spectra are sufficiently homogeneous inside the square region. Such an assumption is natural if the dominant contribution is from LOS structures. The cosmological principle states that there is no special direction in the universe. If the lensing power spectra are homogeneous, then the smoothness of potential, astrometric shift, convergence due to LOS structures are determined by the lensing power spectra or equivalently the corresponding rms values in the lens plane to some extent. However, higher order correlations such as bispectra and trispectra may be necessary for describing non-Gaussian and non-linear perturbations. If the expected perturbation is strongly non-Gaussian, we may require to consider more inhomogeneous, unsmooth potential perturbations. From our previous analysis based on $N$-body simulations, such effects seem insignificant if the positions of quadruply lensed images can be well-fitted by a smooth potential such as an SIE. Our mock analysis in this paper also supports the result. Nevertheless, extension of our formalism to higher order statistics is one of our next immediate scientific goal.  

% Why we can neglect very small scale <1arcsec perturbations ?
In our analysis, we neglected perturbations whose fluctuation scale is less than $\sim 1\,\arcsec$ due to lack of S/N and angular resolution to resolve tiny distortion. We consider that such restriction is not 
a significant problem. First, our models using the source plane $\chi^2$ successfully fit the MIR flux ratios of lensed images, the VLBA positions of jets and the HST positions of lensing galaxies simultaneously. In our formalism, the strength of astrometric shifts becomes larger for smaller angular wavenumbers. Therefore, even if the MIR flux ratios can be fit with a perturbation whose fluctuation scale is less than $\sim 1\,\arcsec$, it would be difficult to simultaneously fit the VLBA positions of jets and the HST positions of lensing galaxies. Second, in CDM models, the convergence power spectrum moderately decreases as a function of the angular scale for $\lesssim 1\,\arcsec$ (corresponding to a wavenumber $k\gtrsim 100\,h/\textrm{Mpc}$ at $z \sim 1$) \citep{inoue-etal2015}. Therefore, we expect that $\sim 1\,\arcsec$ is the angular scale that contributes dominantly to the power. 

% What about baryonic effects? 
The simulations conducted to date to estimate the non-linear power spectra on small scales at redshifts $z<3$ are for dark matter only. Inclusion of baryonic physics may enhance small scale powers to some extent. Because of the uncertainty in baryonic physics, our results should be tested by future observations with improved S/N and angular resolution.

% Discussion for object Y

% Discussion for time delay

% Future prospect 

\newpage
\section{Acknowledgments}
KTI would like to thank Eiji Akiyama, Misato Fukagawa, Fumi Egusa, and Kazuya Saigo for their support on data reduction, Yuichi Higuchi for valuable discussion, and anonymous referees for their valuable comments, and acknowledge supports from NAOJ ALMA Scientific Grant Number 2018-07A, ALMA Japan Research Grant of NAOJ ALMA Project, NAOJ-ALMA-256, and JSPS KAKENHI Grant Number 17H02868. SM is supported by the Ministry of Science and Technology (MoST)  of Taiwan, MoST 103-2112-M-001-032-MY3, 106-2112-M-001-011, and 107-2119-M-001-020.  
KN is supported by JSPS KAKENHI Grant Number 19K03937.
This paper makes use of the following ALMA data:
ADS/JAO.ALMA$\#$2013.1.01110.S., 2016.1.00281.S. ALMA is a partnership of ESO (representing its member states), NSF (USA) and NINS (Japan), together with NRC (Canada), MOST 
and ASIAA (Taiwan), and KASI (Republic of Korea), in
cooperation with the Republic of Chile. The Joint ALMA Observatory is
operated by ESO, AUI/NRAO, and NAOJ.

\section{Data Availability}
The ALMA data used in this work can be downloaded from the ALMA archive:
\\ 
https://almascience.nao.ac.jp/aq/. The corresponding project codes are ADS/JAO.ALMA$\#$2013.1.01
\\ 110.S. and 2016.1.00281.S. 
\bibliographystyle{mnras}

\bibliography{ALMA-powerspectra2021}
\appendix

\section{Lensing power spectra}
Using discrete Fourier modes, a real-valued scalar perturbation $\delta X(\THE)$ of $X$ defined on a square with an area of $A=L'^2$\footnote{If we impose the Dirichlet boundary condition on a square with a side length of $L$, the corresponding area is $A=4L^2$ and $L'=2L$.} in the real space $\THE$ can be decomposed as  
\BEA
\delta X(\THE)&=&\sum_{\K}\hat{\delta} X(\K)e^{-i \K \cdot \THE},
\nonumber
\\
\hat{\delta} X(\K) &=& A^{-1} \int \delta X(\THE) e^{i \K \cdot \THE} d^2 \theta,
\EEA
where $\hat{\delta} X(\K)$ is a Fourier coefficient for a wave vector $\K$, which 
can be expressed in terms of a set of non-negative integers $(n_1, n_2)$ as  
\BE
\K=\biggl(\frac{2 \pi n_1}{L'}, \frac{2 \pi n_1}{L'}\biggr ).
\EE
The power spectrum 
$P_X(k)$ of an ensemble of perturbations $\delta X$ is defined as  
\BE
P_X(k)\equiv \frac{A}{(2 \pi)^2}\langle  |\hat{\delta} X(\K) |^2  \rangle, 
\EE
where $\langle \rangle $ is an ensemble average for a fixed $k=|\K|$. The dimensionless 
power spectrum $\varDelta^2_X$ as a function of $k$ is defined as
\BE
\varDelta^2_X(k)\equiv 2 \pi k^2 P_X(k).
\EE 
We express the square root of the spectrum as $\varDelta_X\equiv \sqrt{\varDelta^2_X}$.
From an observed image, we can measure the estimated dimensionless 
power spectrum as 
\BE
\tilde{\varDelta}^2_X(k)= 2 \pi n^2 \langle |\hat{\delta} X(\tilde{\K}) |^2   \rangle_{\textrm{angle}},
\EE
where $\langle \rangle_{\textrm{angle}}$ denotes an ensemble average over 
various directions of wave vectors $\tilde{\K}$ that have $|\tilde{\K}|\sim k$ and $n=\sqrt{n_1^2+n_2^2}$, where $n_1$ and $n_2$ are the mode numbers that give $|\tilde{\K}(n_1,n_2)|\sim k$. A lensing power spectrum is defined as a dimensionless power spectrum $\varDelta^2_X(k)$ in which $X$ is either a projected gravitational potential $\psi$, strength of astrometric shift $\alpha$ or convergence $\kappa$. The ensemble average of squared perturbation of $X$ for a range of angular wavenumbers
$k-\varDelta k/2 <k<k+\varDelta k/2$ can be estimated as  
\BE
\langle |\hat{\delta} X(\THE) |^2   \rangle \sim \tilde{\varDelta}^2_X(k) k^{-1} \varDelta k,
\EE
where $|\varDelta k|\ll k$. In terms of real Fourier coefficients $\hat{\varDelta}^2_X(k)$, the 
right hand side of (A6) must be divided by 1/4 (see Appendix B),
\BE
\langle |\hat{\delta} X(\THE) |^2   \rangle \sim \hat{\varDelta}^2_X(k) ({4k})^{-1} \varDelta k.
\EE

The angular wave number can be also expressed as the number of waves per 360 degrees as
$l=180\,k/\pi$. For an angular wavelength of $\delta \theta$ in units of arcsec, the corresponding angular wave number is
\BE
l=\frac{1.296\times 10^6}{(\delta \theta/1'')}.
\EE 
\section{Two dimensional real and complex Fourier Series}
Since we do not consider any zero modes, a real scalar function $f(\R)$ defined in a region $-\tilde{L}/2 \le x \le \tilde{L}/2$ and $-\tilde{L}/2  \le y \le \tilde{L}/2$ can be decomposed as a real Fourier series as 
\BEA
f(\R)&=&\sum_{m,n>0} (A_{mn} \cos{k_m x}\cos{k_n y}+B_{mn}\cos{k_m x}\sin{k_n y} 
\nonumber
\\
&+& C_{mn} \sin{k_m x}\cos{k_n y}+D_{mn}\sin{k_m x}\sin{k_n y}),
\EEA
where $m$ and $n$ are natural numbers and $k_m=2 \pi m/\tilde{L}$. 

Similarly, it can be decomposed as a complex Fourier series as 
\BE
f(\R)=\sum_{m',n'} E_{m'n'} \exp[-i(k_{m'} x+k_{n'} y)], 
\EE
where $m'$ and $n'$ are non-zero integers. Since $f$ is real, we have $E_{-m,-n}=E^*_{mn}$. Thus, we obtain
\BEA
A_{mn}&=&E_{mn}+E^*_{mn}+E_{m-n}+E^*_{m-n},
\nonumber
\\
B_{mn}&=&i(-E_{mn}+E^*_{mn}+E_{m-n}-E^*_{m-n}),
\nonumber
\\
C_{mn}&=&i(-E_{mn}+E^*_{mn}-E_{m-n}+E^*_{m-n}),
\nonumber
\\
D_{mn}&=&-E_{mn}-E^*_{mn}+E_{m-n}+E^*_{m-n}, 
\EEA
which results in
\BEA
& &|E_{mn}|^2+|E_{m-n}|^2+|E_{-mn}|^2+|E_{-m-n}|^2
\nonumber
\\
&=& \frac{1}{4}(A^2_{mn}+B^2_{mn}+C^2_{mn}+D^2_{mn})
\nonumber
\\
&=& \langle f^2 \rangle.
\EEA

\section{Procedure of fitting visibilities to the ALMA data }

In order to implement fast computation, we need to compress the observed visibilities in time and frequency domains. First, we averaged visibility data in each channel bin with an integration time of 60 seconds using CASA 6.4. Then the average distance between adjacent sampling points in the visibility plane was $\lesssim 10$\,m, which is much shorter than the minimum baseline $\sim 15\,$m that corresponds to the angular scale $\sim 20''$ at $\sim 340$\,GHz, which is sufficeintly larger than the angular size of MG\,J0414+0534. Therefore, the data compression does not significantly affect visibility fitting in our ALMA data.Then, we concatenated Cycle 2 and Cycle 4 data and averaged line-free channels in each of 4 spectral windows (SPWs), which produced 4 visibility data sets centerd at 335.3\,GHz, 336.6\,GHz, 346.7\,GHz, and 349.5\,GHz  with $\sim 1.7\times 10^5$ elements. The four SPWs were numbered as $s=1,2,3,4$ in ascending order. The effective bandwidths of the SPWs were $1-2$\,GHz and the expected systematic reduction in the flux of a point source due to the bandwidth smearing effect in our setting (assuming a square bandpass and circular Gaussian beam taper) was $\lesssim 1$ percent. Since the expected change in the fluxes of compact sources are $\sim 10$ percent, the smearing effect is negligible.      

In order to compare the modeled visibilities to the data, we need to estimate the rms noise. The CASA \textit{weight} in the ALMA Measurement Set cannot be used as an estimated error in measured visibilities because the absolute values are not properly calibrated for some technical reason 
(see https://casaguides.nrao.edu/index.php/DataWeights
\\
AndCombination\#Absolute\_Accuracy\_of\_the\_Data\_Weights)
\footnote{We confirmed that the error values obtained from CASA \textit{weight} were systematically smaller than the values obtained from the sample variance of the observed visibilities.}. In our analysis, the standard deviation  
$\sigma_{sI}$ of a visibility ${\cal{V}}_{sI}$ with an SPW number $s$ and a sample number $I$ in the visibility $(u,v)$ plane was simply estimated as
\BE
\sigma_{sI}^2=\frac{|{\cal{V}}_{sI}-{\cal{V}}_{s+1 I}|^2}{2}. 
\EE
In other words, the variance in the subtracted visibilities in a particular bin were assumed to be twice the true noise variance \citep{hezaveh2016a, dye2018}. To obtain the pre-fitted visibilities of the point-source components, we fitted p and q components in the VLBA data to the ALMA data and we assumed that the flux ratios were given by the MIR values. The tentative absolute flux values of each component were specified by $c_\textrm{q}=0.324$ and $c_\textrm{p}=0.626$ (see Table \ref{tab:parameters-for-best-fit}). As the pre-fitted visibilities of the point-source components, we used the Fourier transformed Dirac delta functions of these components. The obtained visibilities were multiplied by a constant parameter $a$ that has a dimension of Jy. 
\begin{figure}
\vspace{-2.5cm}
\hspace*{0.cm}
\vspace{0.cm}
\epsscale{1.0}
\plotone{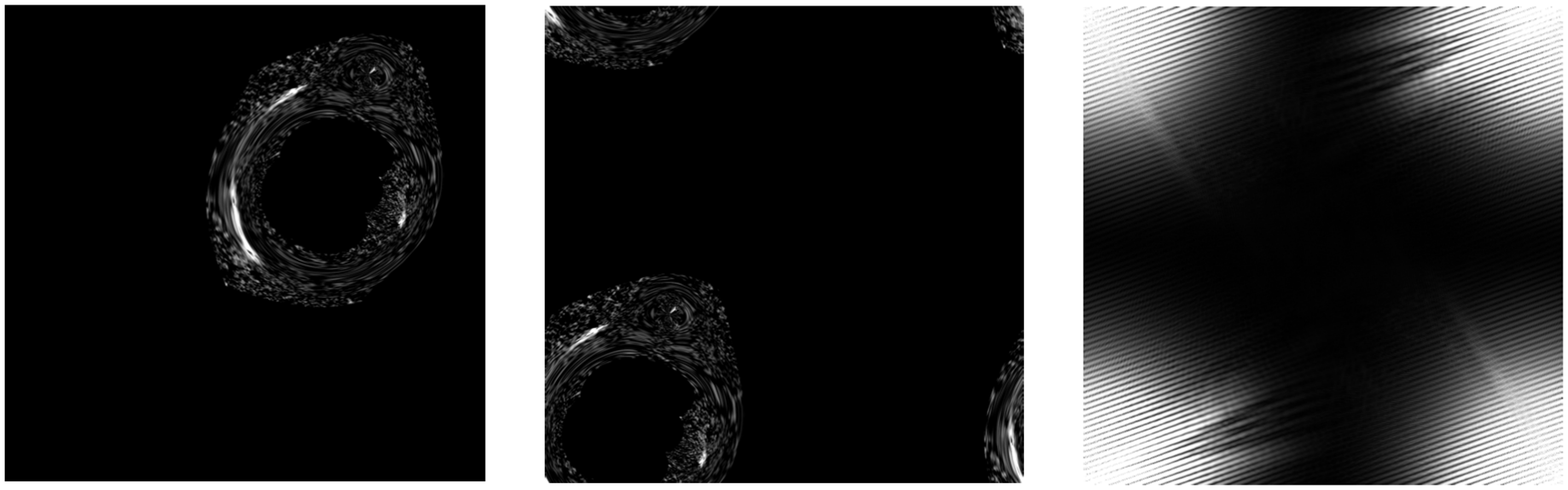}
\vspace{-3cm}
\caption{A pre-fitted pixelized model image (left), the rearranged model image (center), and the FFTed image before rearrangement (right). The image size and the pixel size in the sky are $6'''\times 6''$ and $0\farcs 005$, respectively. The model images with point and extended source components are shown only for illustrative purpose. In our analysis, pixelized images were used only for computing visibilities of extended source components and 0-padding around the image was used to improve the accuracy of visibilities on large angular scales. }
\label{fig:model-image-Fourier.pdf}
\end{figure}
\vspace{2cm}
\begin{figure}
\hspace*{0.cm}
\vspace{0.cm}
\epsscale{1.0}
\plotone{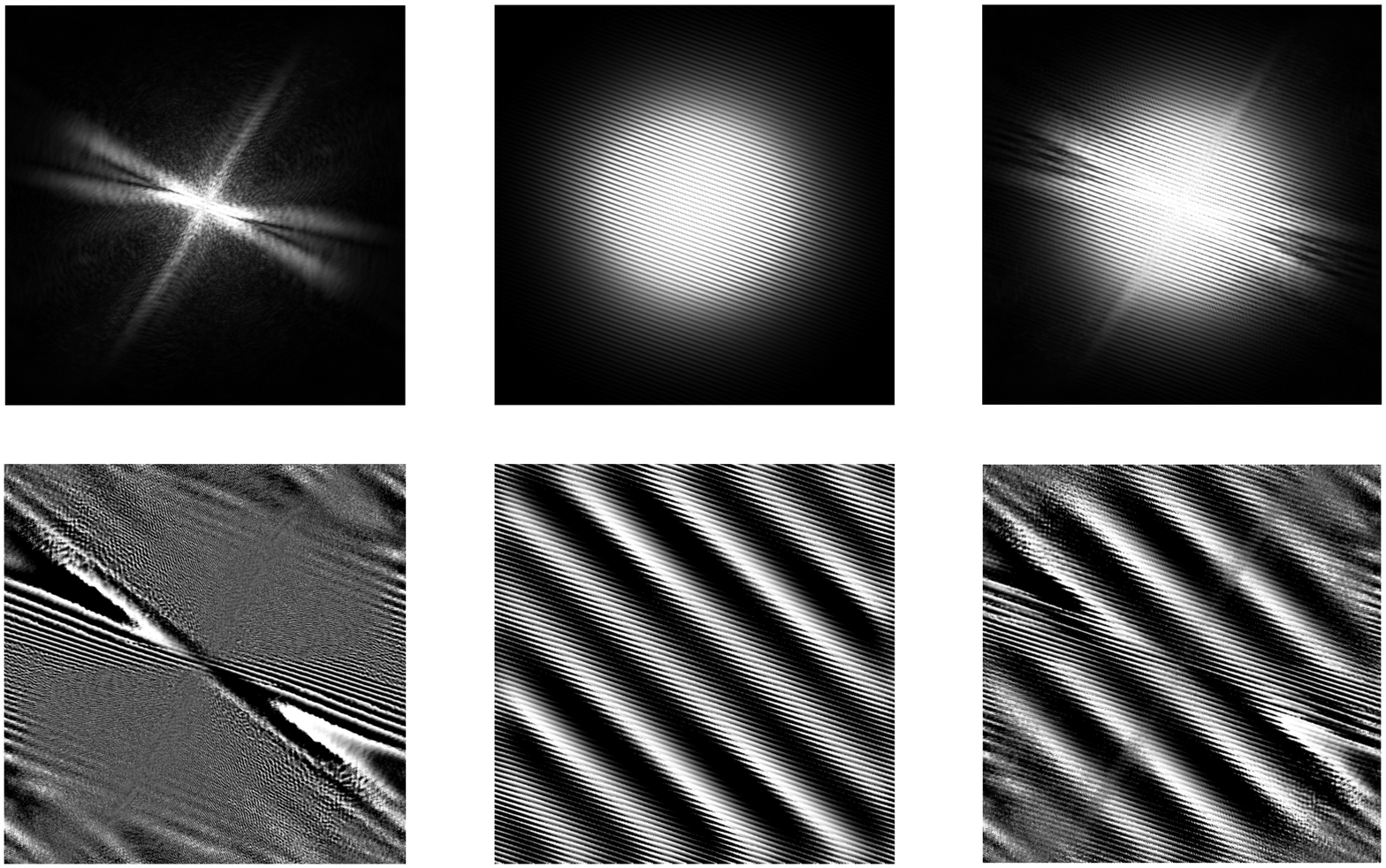}
\vspace{0.cm}
\caption{The amplitudes (top column) and phases (bottom column) of visibilities in the $(u,v)$ plane 
for the pre-fitted model shown in Figure 23. The visibilities (right row) are given by the sum of those of the extended source components (left row) and those of the point source components (center row).   }
\label{fig:model-visibility.pdf}
\end{figure}
\begin{figure}
\hspace*{0.cm}
\vspace{0.cm}
\epsscale{0.7}
\plotone{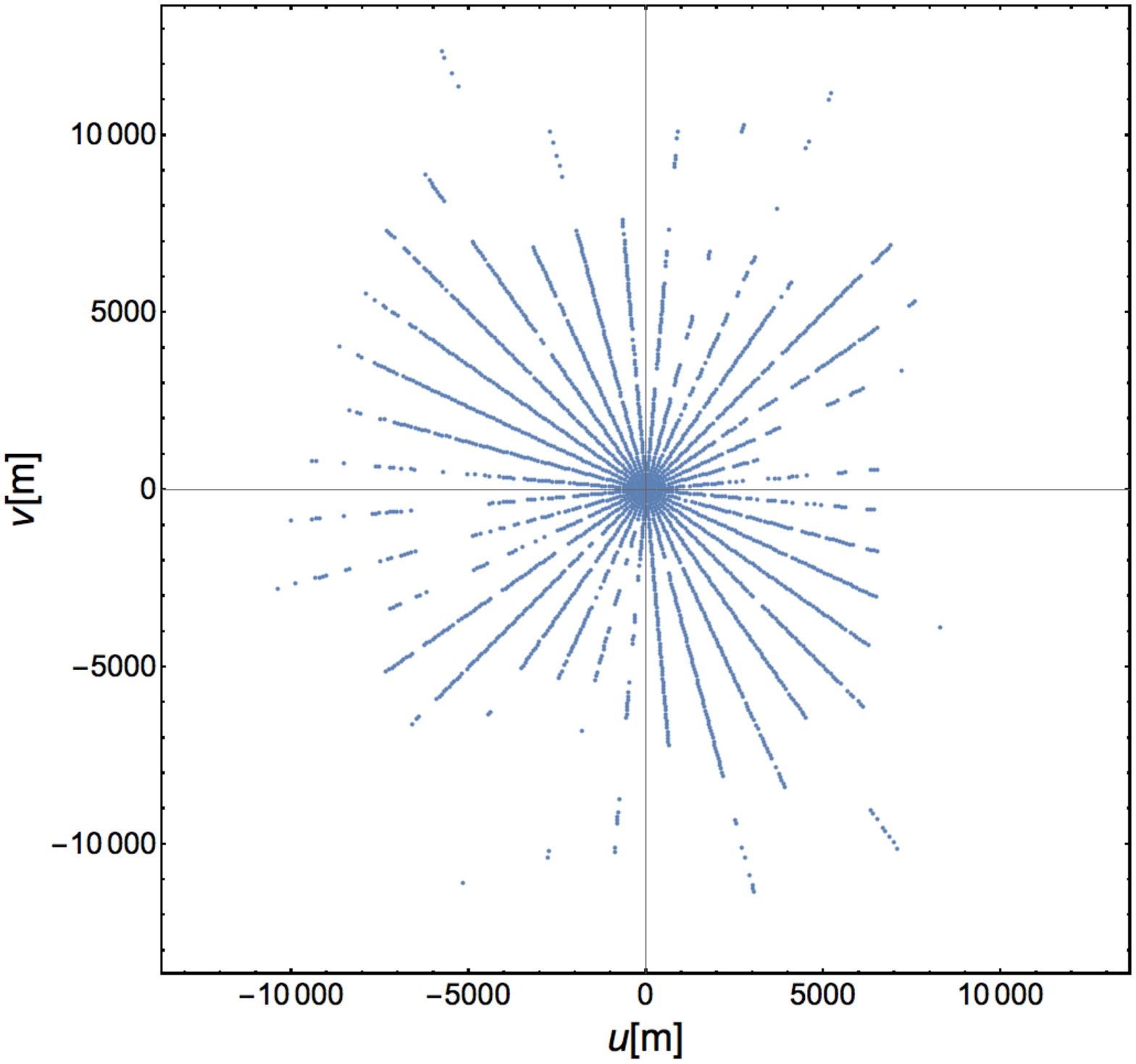}
\vspace{0.cm}
\caption{Distribution of grid centers. The each blue dot shows a center of a grid that contains sampled ALMA visibility data at 335.3\,GHz. The radial and azimuthal angular separations of centers of adjacent grids are $\delta r= 100\,$m and $\delta \theta=10^\circ$, respectively.  }
\label{fig:vis-base-theta10-spw0.pdf}
\end{figure}

\vspace{0cm}
To obtain the pre-fitted visibilities of the extended-source components, we need to compute a matrix that consists of the extended source intensity at pixels in the lens plane. To do so, we used a PSF-subtracted ALMA image with the magnification weighting over a square region with a sidelength of $6\farcs 0$ and a pixel size of $0\farcs 005$ (pixel number $N_c=1200$). Then, we padded 0s around the edge of the matrix (i.e., equivalent to adding a blank field) so that the dimension of the matrix becomes three times larger. The 0-padding significantly improves the accuracy of large scale fluctuations in intensity. In order to perform the fast Fourier transform (FFT), we rearranged the matrix elements so that the center of the extended square region with a sidelength of $18\farcs 0$ is shifted to the bottom-left corner. The rearrangement is equivalent to a swap between the first (second) and the third (fourth) quadrant. Second, we performed a two-dimensional FFT on the matrix data of the extended source components and rearranged the FFTed matrix as is done in the original matrix data (see Figure \ref{fig:model-image-Fourier.pdf}). We made a reflection in the RA direction and a translation by a half pixel size in the negative RA and DEC directions\footnote{This translation is related to the difference between the fits convention that a pixel $(N_c/2 + 1, N_c/2 + 1)$ is at the coordinate center $(0,0)$ and the image convention that $(N_c/2 + 1, N_c/2 + 1)$ is at (pixelsize$/2$, pixelsize$/2$).}. The obtained visibilities of the extended-source components (see Figure \ref{fig:model-visibility.pdf}) were multiplied by a constant parameter $b$ that has no dimensions.

Finally, a statistic $\chi_v^2$ ($\chi^2$ in the visibility plane) can be written as 
\BE
\chi_v^2 \equiv \sum_{s,I} \frac{|{\cal{V}}_{sI}-a {\cal{P}}_{sI} - b{\cal{E}}_{sI} |^2}
{\sigma^2_{sI}},
\EE
where ${\cal{P}}_{sI}$ and ${\cal{E}}_{sI}$ are modeled complex visibilities of the point source components and those of extended source components, respectively. It should be noted that the modeled complex visibilities were calculated using \textit{weights} extracted from the observed ALMA Measurement Sets. We used the same sample $(u,v)$ coordinates and \textit{weights} of observed visibilities before averaging at grids.

We found that the obtained $\chi_v^2$ divided by the degree of freedom (i.e., reduced $\chi^2$) was almost equal to $1$ even without taking into account model visibilities. The result suggests that the S/N ratio of each visibility is so small that performing naive $\chi^2$ fitting is very difficult. We also tried to perform $\chi^2$ analysis using visibilities that have an apparent 'high' S/N ratio but it resulted in biased results. Most of visibilities with a 'high' S/N ratio actually have a systematically large error due to the low S/N ratio. Therefore, restricting visibilities with an apparent 'high' S/N ratios leads to a positive bias in the estimated amplitudes. 

In order to circumvent the problem, we need to compress the visibility data in a certain way. In our analysis, we used visibilities that are averaged on grids (denoted by $J$) in the polar coordinates $(r,\theta)$ with a radial bin size of $\delta r=100\,$m and an azimuthal size of $\delta \theta$ in the visibility plane $(u,v)$ (Figure \ref{fig:vis-base-theta10-spw0.pdf}). The error variance of an averaged visibility at the $J$-th grid and an SPW number $s$ was calculated as 
\BE
\sigma^2_{s J}=\sum_{I(J)}{\sigma^2_{sI(J)}}/N^2_{J},
\EE
where $I(J)$ is a sample number within the $J$-th grid and $N_J$ is the total number of such sampled points. The averaged model complex visibilities on grids were calculated using CASA \textit{weight}s extracted from the observed ALMA Measurement Sets, which are proportional to the errors of visibilities. The selected $(u,v)$ coordinates before averaging visibilities were as the same as those used in our ALMA observations.

\end{document}